\documentclass[twocolumn]{aastex63}
\usepackage{multirow}
\usepackage{color}
\usepackage{lineno}

\usepackage{capt-of}
\usepackage{rotating}
\usepackage{longtable}
\usepackage{graphicx}
\usepackage{float}
\usepackage{amsmath,amssymb}
\usepackage{gensymb}
\usepackage{xcolor}
\usepackage{natbib}
\usepackage[hang,flushmargin]{footmisc}
\usepackage{chngcntr}
\usepackage{hyperref}
\usepackage{enumitem}
\setlist{parsep=0pt,listparindent=\parindent}
\usepackage{xtab}
\usepackage{tabularx} 
\usepackage{xltabular} 
\usepackage{soul}
\usepackage{array}

\usepackage{mwe}
\usepackage{afterpage}
\usepackage{rotating}
\usepackage{afterpage}

\usepackage{colortbl}
\usepackage{lipsum}

    {\global\pdfpageattr\expandafter{\the\pdfpageattr/Rotate 90}}%
    {\clearpage\pagebreak[4]\global\pdfpageattr\expandafter{\the\pdfpageattr/Rotate 0}}%

\newcommand{\ghost}{\texttt{GHOST}}

\usepackage{pifont}

\newcommand{\PS}{Pan-STARRS}

\newcommand{\annoy}{\texttt{ANNOY}}
\newcommand{\laiss}{\texttt{LAISS}}

\definecolor{LightCyan}{rgb}{0.88,1,1}

\newcommand{\Cambridge}{Institute of Astronomy and Kavli Institute for Cosmology, Madingley Road, Cambridge, CB3 0HA, UK}

\newcommand{\STScI}{Space Telescope Science Institute, Baltimore, MD 21218.}

\newcommand{\CfA}{Center for Astrophysics $|$ Harvard \& Smithsonian, Cambridge, MA 02138, USA}
\newcommand{\IfA}{Institute for Astronomy, University of Hawaii, 2680 Woodlawn Drive, Honolulu, HI 96822, USA}

\newcommand{\UCSC}{Department of Astronomy and Astrophysics, University of California, Santa Cruz, CA 95064, USA}
\newcommand{\QUB}{Astrophysics Research Centre, School of Mathematics and Physics, Queen's University Belfast, Belfast BT7 1NN, UK}

\newcommand{\DARK}{DARK, Niels Bohr Institute, University of Copenhagen, Jagtvej 128, 2200 Copenhagen, Denmark}
\newcommand{\Illinois}{Department of Astronomy, University of Illinois at Urbana-Champaign, 1002 W. Green St., IL 61801, USA}
\newcommand{\NCSA}{Center for AstroPhysical Surveys, National Center for Supercomputing Applications, Urbana, IL, 61801, USA}

\newcommand{\WSU}{Department of Physics \& Astronomy, Washington State University, Pullman, Washington 99164, USA}

\newcommand{\Melbourne}{School of Physics, The University of Melbourne, VIC 3010, Australia}

\newcommand{\Southhampton}{Department of Physics and Astronomy, University of Southampton, Highfield, Southampton SO17 1BJ, UK}

\newcommand{\PNNL}{Pacific Northwest National Laboratory, 902 Battelle Blvd, Richland, WA 99354, USA}
\newcommand{\CMU}{McWilliams Center for Cosmology, Department of Physics, Carnegie Mellon University, Pittsburgh, PA 15213, USA}
\newcommand{\NOIRLabGemini}{NSF's National Optical-Infrared Astronomy Research Laboratory, Gemini North, 670 N Aohoku Place, Hilo, HI 96720, USA}
\newcommand{\NOIRLabTucson}{NSF's National Optical-Infrared Astronomy Research Laboratory, 950 North Cherry Avenue, Tucson, AZ 85719, USA}
\newcommand{\IAIFI}{The NSF AI Institute for Artificial Intelligence and Fundamental Interactions}
\newcommand{\IfAHilo}{Institute for Astronomy, University of Hawai'i, 640 N. A'ohoku Pl., Hilo, HI 96720, USA}

\begin{document}

\title{Anomaly Detection and Approximate Similarity Searches of Transients in Real-time Data Streams}



\author[0000-0002-6298-1663]{P.~D.~Aleo} 
\affiliation{\Illinois}
\affiliation{\NCSA}


\author[0000-0003-2348-483X]{A.~W.~Engel} 
\affiliation{\PNNL}

\author[0000-0001-6022-0484]{G.~Narayan} 
\affiliation{\Illinois}
\affiliation{\NCSA}

\author[0000-0002-4269-7999]{C.~R.~Angus} 
\affiliation{\DARK}
\affiliation{\QUB}

\author[0000-0001-7179-7406]{K.~Malanchev} 
\affiliation{\Illinois}
\affiliation{\CMU}


\author[0000-0002-4449-9152]{K.~Auchettl} 
\affiliation{\UCSC}
\affiliation{\Melbourne}

\author[0000-0003-4703-7276]{V.~F.~Baldassare} 
\affiliation{\WSU}

\author[0000-0002-5010-441X]{A.~Berres} 
\affiliation{\Illinois}

\author[0000-0001-5486-2747]{T.~J.~L.~de~Boer} 
\affiliation{\IfA}

\author[0000-0002-0622-1117]{B.~M.~Boyd} 
\affiliation{\Cambridge}

\author[0000-0001-6965-7789]{K.~C.~Chambers}  
\affiliation{\IfA}

\author[0000-0002-5680-4660]{K.~W.~Davis} 
\affiliation{\UCSC}

\author[0009-0002-2806-9379]{N.~Esquivel} 
\affiliation{\NOIRLabTucson}

\author[0000-0002-6886-269X]{D.~Farias} 
\affiliation{\DARK}

\author[0000-0002-2445-5275]{R.~J.~Foley}  
\affiliation{\UCSC}

\author[0000-0003-4906-8447]{A.~Gagliano} 
\affiliation{\IAIFI}
\affiliation{\CfA}

\author[0000-0002-8526-3963]{C.~Gall} 
\affiliation{\DARK}

\author[0000-0003-1015-5367]{H.~Gao} 
\affiliation{\IfA}

\author[0000-0001-6395-6702]{S.~Gomez} 
\affiliation{\STScI}

\author[0000-0002-6741-983X]{M.~Grayling} 
\affiliation{\Cambridge}

\author[0000-0002-6230-0151]{D.~O.~Jones} 
\affiliation{\IfAHilo}

\author[0000-0002-7272-5129]{C.-C.~Lin} 
\affiliation{\IfA}

\author[0000-0002-7965-2815]{E.~A.~Magnier} 
\affiliation{\IfA}

\author[0000-0001-9846-4417]{K.~S.~Mandel} 
\affiliation{\Cambridge}

\author[0000-0001-6685-0479]{T.~Matheson} 
\affiliation{\NOIRLabTucson}

\author[0000-0002-6248-398X]{S.~I.~Raimundo} 
\affiliation{\DARK}
\affiliation{\Southhampton}

\author[0009-0009-1590-2318]{V.~G.~Shah} 
\affiliation{\Illinois}

\author[0000-0001-6360-992X]{M.~D.~Soraisam} 
\affiliation{\NOIRLabGemini}

\author[0000-0002-9886-2834]{K.~M.~de~Soto} 
\affiliation{\CfA}

\author[0009-0009-7129-7538]{S.~Vicencio} 
\affiliation{\NOIRLabTucson}

\author[0000-0002-5814-4061]{V.~A.~Villar} 
\affiliation{\CfA}

\author[0000-0002-1341-0952]{R.~J.~Wainscoat} 
\affiliation{\IfA}

\submitjournal{ApJ}

\correspondingauthor{P.~D.~Aleo}
\email{paleo2@illinois.edu}


\begin{abstract}

We present \texttt{LAISS} (Lightcurve Anomaly Identification and Similarity Search), an automated pipeline to detect anomalous astrophysical transients in real-time data streams. We deploy our anomaly detection model on the nightly ZTF Alert Stream via the ANTARES broker, identifying a manageable $\sim$1--5~candidates per night for expert vetting and coordinating follow-up observations. Our method leverages statistical light-curve and contextual host-galaxy features within a random forest classifier, tagging transients of rare classes (\emph{spectroscopic} anomalies), of uncommon host-galaxy environments (\emph{contextual} anomalies), and of peculiar or interaction-powered phenomena (\emph{behavioral} anomalies). Moreover, we demonstrate the power of a low-latency ($\sim$ms) approximate similarity search method to find transient analogs with similar light-curve evolution and host-galaxy environments. We use analogs for data-driven discovery, characterization, (re-)classification, and imputation in retrospective and real-time searches. To date we have identified $\sim$50 previously known and previously missed rare transients from real-time and retrospective searches, including but not limited to: SLSNe, TDEs, SNe~IIn, SNe~IIb, SNe~Ia-CSM, SNe~Ia-91bg-like, SNe~Ib, SNe~Ic, SNe~Ic-BL, and M31 novae. Lastly, we report the discovery of 325 total transients, all observed between 2018-2021 and absent from public catalogs ($\sim$1\% of all ZTF Astronomical Transient reports to the Transient Name Server through 2021). These methods enable a systematic approach to finding the ``needle in the haystack" in large-volume data streams. Because of its integration with the ANTARES broker, \texttt{LAISS} is built to detect exciting transients in Rubin data.
\end{abstract}

\keywords{supernovae: general – astronomical databases: surveys - time-domain astronomy: anomaly detection}


\section{Introduction} 
\label{sec:intro}

Serendindipity has played a disproportionately large role in breakthrough time-domain astronomical discoveries. For example, the discovery of SN~1987A \citep{Kunkel1987} in the nearby Large Magellanic Cloud ushered in a renewed interest in supernova (SN) science---spurring systematic searches in the forms of pencil-beam surveys to discover high-redshift SNe for cosmological study \citep{Perlmutter1997,Schmidt1998} and rolling searches to drastically increase observed events \citep{Barris2004}. The discovery of SN~1998bw, which is the first SN associated with a gamma-ray burst, was found by astronomers who were observing gamma-ray bursts rather than SNe \citep{Galama1998}. Nearby Type Ia SN 2014J in M82 (Cigar Galaxy) at 3.7~Mpc was discovered by chance during an undergraduate telescope training session at the University of London Observatory by Steve Fossey and his students \citep{Fossey2014, McIntosh2014AAS}. Recently, \cite{Itagaki2023ixfTNS} discovered SN~2023ixf in M101, the closest SN to Earth since 2014J, from an untargeted search before automated algorithmic discovery. \par

Beyond those closest and brightest to Earth, transient discoveries of extremely rare origin often arise from astronomers aiming to increase the observable parameter space or find dissimilar objects in ever-increasing data sets \citep{Li2022}. These tasks are worthwhile, because in-depth study of such rare events reveals insights of their progenitors, explosion mechanisms, and diversity---all of which continue to be active areas of research \citep[see, e.g.][]{Margutti2019,Gagliano2022,Galan2022tlf,Perley2022,Kuncarayakti2023,Pierel2023}.  \par 

Astronomers have attempted to automate serendipity in a ``systematic'' manner (see, e.g., \citealt{Giles2019}) to increase rare transient discovery. This task is analogous to finding ``the needle in the haystack" \citep[See, e.g.,][]{Villar2019}, which becomes not only increasingly difficult in the era of large, data driven surveys (where human vetting for the vast majority of objects is infeasible), but increasingly important due to the limited spectroscopic resources and multi-wavelength follow-up observations. It is estimated $\sim$1\% of the approximate ten million Vera C. Rubin Observatory \citep{Ivezic2019} Legacy Survey of Space and Time (LSST) transients will be selected for spectroscopic follow-up \citep{hambleton2022rubin}, down from the $\sim$10\% of all optical transients classified spectroscopically today. Thus, automated algorithms that identify rare or interesting transients must do so with high purity as to be judicious with our resources. Moreover, because it is all but guaranteed that LSST will discover entirely new phenomena, theorized or not \citep{Li2022, hambleton2022rubin}, we need adept algorithms that are able to identify events with currently unknown feature distributions. \par

The astronomical community has been hard at work developing anomaly detection methodologies. Some have focused on static data sets, such as images \citep{Reyes2020arXiv, Storey-Fisher2021, Etsebeth2023} or spectra \citep{Liang2023, Bohm2023}, or time-series light curve data \citep{Rebbapragada2009,Nun2014,Nun2016,Solarz2017,Giles2019,Pruzhinskaya2019,Soraisam2020,Webb2020,Villar2021,Galarza2021,Lochner2021,Ishida2021,Malanchev2021,Perez-Carrasco2023,Cui2023}. The added complexity in detecting anomalies in time-series data, combined with strong interest in SN photometric classification (e.g., see \citealt{Lochner2016,Muthukrishna2019,Moller2020,Villar2020,Qu2021,2022Gagliano_CCA,Aleo2023} and references therein), is a key reason why most efforts have concentrated on identifying anomalous light curves. \par

To that end, most of the effort has been dedicated to full phase anomaly detection, after the entire light curve has been observed. Only recently has the field focused on real-time anomaly detection \citep{Soraisam2020, Villar2021, Muthukrishna2022, Perez-Carrasco2023, Gupta2024}, with current efforts combining the SN light curve with contextual information (e.g., host galaxy, spectra, etc.), as seen in \cite{Perez-Carrasco2023} and this work. \par

In conjunction, instead of discovering transients and isolating anomalies independently, recent effort has been devoted to finding \emph{analogs} of a given object or transient event. \cite{Giles2019} utilized a similarity score ascribed to the t-SNE (t-distributed Stochastic Neighborhood Embedding; \citealt{vanderMaaten2008}) representation of Kepler light curve features to search for outliers in cluster distributions. Their follow-up work \citep{Giles2020} demonstrated the effectiveness of k-Nearest Neighbor distance in feature space to efficiently identify anomalous light curves. \cite{Galarza2021} expanded on this work by running an Unsupervised Random Forest \citep{Shi2006} on the joint space of Kepler light curves and power spectra with two manifold-learning algorithms (t-SNE and UMAP, \citealt{McInnes2018}) to create a low-dimensional embedding. They tested their analog-finding ability by analyzing the anomalies' location and clustering in the embedded feature space, identifying those with shared astrophysical properties. In a different approach, \cite{Aleo2022} simulated bright ZTF SNe, extracted statistical light curve features, applied a brute-force k-D tree algorithm to identify nearest neighbors, and visually inspected the closest matches. They discovered 11 previously unreported transient events in the Zwicky Transient Facility (ZTF; \citealt{Bellm2019}) fourth data release (ZTF DR4) out of 105 manually-vetted objects. Beyond light curves, similarity search has been used to identify similar galaxy images from their low-dimensional representations \citep{Stein2021} and associating galaxy images with their optical spectra via cross-modal contrastive learning \citep{Lanusse2023}, as well as selecting synthetic galaxies matching statistical properties of observational data for simulations \citep{Lokken2023}. \par

\begin{figure*}
    \centering
    \includegraphics[width=18cm]{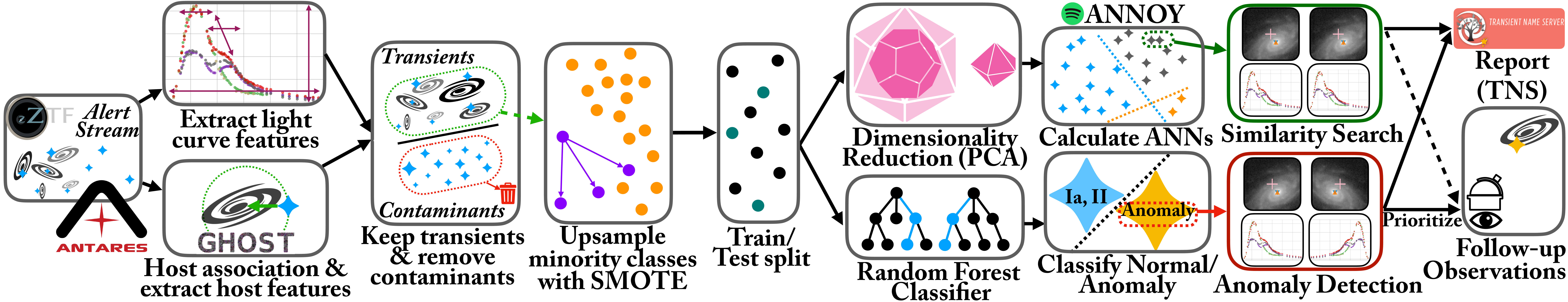}
    \caption{
    The \laiss{} Pipeline. We ingest ZTF Alert Stream data, and extract statistical light curve features with \texttt{light-curve} \citep{Malanchev-LC} and contextual host galaxy features with \texttt{GHOST} \citep{Gagliano2021}. We apply strict preprocessing cuts to remove contaminants and obtain a reference dataset of transient candidates. We apply a 70/30 train/test split, subsequently upsampling our training set with SMOTE \citep{Chawla2002} such that each class is balanced. We train a Random Forest Classifier to perform a binary classification task between ``Normal" and an ``Anomaly" class. We consider \emph{spectroscopic} anomalies, \emph{contextual} anomalies, and \emph{behavioral} anomalies via expert vetting. In parallel, we apply principal component analysis (PCA) followed by approximate nearest neighbors (ANNs) search in feature-space via \texttt{ANNOY} (\citealt{Github:annoy}; Section~\ref{sec:annoy}) to find transient analogs with similar light curve evolution and host galaxy environment. Finally, we deploy our anomaly detection model onto the ANTARES broker \citep{Matheson2021} as a \emph{Filter}. For transients with $P(anom)\geq50$\% at any point in the light curve, we manually vet them and request for follow-up resources. We report to TNS any new or updated classifications from either our anomaly detection or approximate nearest neighbors search.
    } 
    \label{fig:pipeline}
\end{figure*}

Here, we present an approach for simultaneous anomaly detection (real-time and retroactive) and similarity search that uses both SN light curve and contextual information (host galaxy photometry) from the ZTF Alert Stream. We identify transient events that are of rare spectroscopic class (\emph{spectroscopic} anomalies), exhibit peculiar behavior or interaction-powered phenomena (\emph{behavioral} anomalies), or are found in host galaxy environments uncommon to their type (\emph{contextual} anomalies), and categorize these broadly as anomalies. In tandem, we find transient analogs across via an approximate nearest neighbors (ANNs) similarity search. We demonstrate the effectiveness of these methods by identifying archival rare or unique transients with high purity and low-latency, and reclassifying transients after being prompted to investigate their spectra due to their nearest neighbors' class labels. We call our pipeline \laiss{} (Lightcurve Anomaly Identification and Similarity Search). We provide a schematic overview in Figure~\ref{fig:pipeline}. \par

Our paper is structured as follows. In \S\ref{sec:methodology}, we describe our methodology for constructing our database of SNe from the ZTF Alert Stream. In \S\ref{sec:ztf_real_time}, we discuss our real-time Random Forest Classifier (RFC) anomaly detection model and the results of its application to the ZTF Alert Stream via the ANTARES broker \citep{Matheson2021}. In \S\ref{sec:AD_ysedr1}, we provide additional results based on our model's application to the Young Supernova Experiment Data Release 1 \citep[YSE DR1;][]{Aleo2023}. In \S\ref{sec:annoy}, we demonstrate the power of low-latency approximate similarity search of SNe, including finding missed SN candidates and the reclassification of some SNe. In \S\ref{sec:discussion}, we speculate on additional applications and extensions of this work, such as performing a calculation and preliminary analysis of our tagged transients' host galaxy masses and star formation rates. We conclude in \S\ref{sec:conclusion}, and we detail all used features and newly reported SN candidates in the \hyperref[sec:appendix]{Appendix}. \par

The code is publicly available on Github\footnote{\url{https://github.com/patrickaleo/LAISS-local}} and the version of this code used in this work is available on Zenodo\footnote{\url{https://zenodo.org/records/11541806}} \citep{LAISS_Zenodo}.

\section{Constructing our Training Set} \label{sec:methodology}

We are motivated by two Directives: 1) find anomalous transients in real-time within the ZTF Alert Stream with high purity (see Section~\ref{sec:ztf_real_time} for details), and 2) find $k$ ANNs (i.e. analogs) of any transient through our embedding space within $\sim$ms to scale with the rate of LSST transients. Achieving these Directives with current datasets is a strong prototype for eventual application to Rubin transients, where the volume and diversity of transients will demand such infrastructure. \par

First, we need to construct a reference database of transients (hereafter, ``databank"). This Section details this process. \par

\subsection{ZTF Data \& Broker} \label{subsec:data_broker}

ZTF\footnote{\url{http://ztf.caltech.edu}} \citep{Bellm2019} uses the Palomar 48-inch Schmidt telescope equipped with a 47~deg$^{2}$ field-of-view camera and an eight-second readout time to observe the entire northern sky in ZTF-$g$, ZTF-$r$, and ZTF-$i$ passbands.\footnote{We do not use ZTF-$i$ observations for this work because of poor coverage ($\sim$10\% of all ZTF observations) and the 18-month grace period for private survey data before public release, which is not conducive for real-time applications.} \par

Now in Phase II operations, ZTF allocates 50\% of camera time and 50\% of SEDM (Spectral Energy Distribution Machine) spectrograph time to a 2-night cadence public survey of the entire northern sky. SEDM spectra are uploaded daily to the Transient Name Server (TNS)\footnote{\url{https://www.wis-tns.org}}. The Infrared Processing and Analysis Center (IPAC) provides ZTF image reduction and object identification in near real-time, producing transient alerts from raw images in $\sim$4 minutes. These are available to the community in the public ZTF Alert Stream via alert brokers such as ANTARES\footnote{\url{http://antares.noirlab.edu}} \citep[Arizona-NOIRLab Temporal Analysis and Response to Events System;][]{Matheson2021}, ALeRCE\footnote{\url{http://alerce.science}} \citep[Automatic Learning for the Rapid Classification of Events;][]{Forster2021}, Fink\footnote{\url{https://fink-broker.org}} \citep{Moller2021}, \texttt{AMPEL}\footnote{\url{https://github.com/AmpelAstro/Ampel-contrib-sample}} \citep[Alert Management, Photometry and Evaluation of Lightcurves;][]{Nordin2019}, Pitt-Google\footnote{\url{https://github.com/mwvgroup/Pitt-Google-Broker}}, and Lasair\footnote{\url{http://lasair.roe.ac.uk/}}. \par

For this work, we extensively use the ANTARES broker. 
The primary advantage of ANTARES is the allowance of a user-created \emph{Filter}---snippets of Python code which can analyze and tag light curves with numerical or categorical values. These tags can then be used to filter objects of interest, such as optimal candidates for follow-up observations. For example, our full-phase anomaly detection RFC is implemented as a \emph{Filter} on ANTARES, and has been running since 22~August~2023. More details on this \emph{Filter} is presented in Section~\ref{subsec:AD_RFC}. \par

ANTARES has processed millions of ZTF objects (``loci"\footnote{A locus (plural, loci) is a point on the sky where alerts cluster and is roughly equivalent to an astrophysical object.}) with their \emph{Filter} \texttt{lc\_feature\_extractor}. This \emph{Filter} uses the \texttt{light-curve}\footnote{\url{https://github.com/light-curve}} package \citep{Malanchev-LC} to quickly ($\sim$ms) calculate 53 statistical light curve features in each passband. These features and its subset we choose for this work is highlighted in Section~\ref{subsec:feature_select}. 

Before preprocessing, we start with all $\sim$2.5~million loci tagged by the \texttt{lc\_feature\_extractor} \emph{Filter}. The overwhelming majority are variable stars, quasi-stellar objects (QSOs), and active galactic nuclei (AGN), while a small portion are transient in nature like supernovae (SNe) and tidal disruption events (TDEs). In addition to the desired \texttt{light-curve} statistical features, these loci have additional metadata\footnote{\url{https://antares.noirlab.edu/properties}} and data products, from RA/Dec to tags from other science \emph{Filters} and beyond. While no transient loci have associated hosts, we perform our own host associations  (Section~\ref{subsubsec:host_features}) using \texttt{GHOST} \citep{Gagliano2021}, whose features are used downstream in \laiss{}.

\subsection{Preprocessing} \label{subsec:preprocess}

We require a highly pure sample of transients to train, test, and validate our models. Although the methods we present here can be extended to variable stars and other phenomena, for this work we are interested in transients exclusively and thus want to minimize contamination from non-transients. \par

To obtain a pure transient sample, we impose a strict preprocessing pipeline: 
\begin{enumerate}
    \item First, we excise any objects in the galactic plane by requiring $|b|~\textgreater~15\degree$. We do so because variable stars far outnumber extragalactic transients along the galactic plane, and the light curves of longer period variables (e.g., Mira stars) can resemble those of SNe in statistical feature space \citep{Malanchev2021, Aleo2022}. This selection cut leaves $\sim$1~million objects.
    \item We require both ZTF-$g$ and ZTF-$r$ \texttt{light-curve} features, which are only calculated for bands with at least 4 observations. Thus, we remove all objects with fewer than 4~observations in either passband, leaving $\sim$0.5~million objects.
    \item At this stage, we have yet to distinguish between variable stars, asteroids, bogus observations, AGNs, transients, and the like. Because ANTARES has no built-in star/galaxy or variable/transient separator, we outsource this step to existing and proven infrastructure. \par
    We query the ALeRCE light curve classifier \citep{SanchezSaez2021} and their stamp classifier \citep{Carrasco-Davis2021}. We prioritize the results from the light curve classifier, which parses objects into 15 subclasses of variable and transient objects across ``Transient", ``Stochastic", and ``Periodic" sources. We keep only the objects that have a Transient classification from light curve classifier (`SN~Ia', `SN Ibc', `SN II', `SLSN'), or stamp classifier (`SN') as its most probable classification, prioritizing the results from the light curve classifier.\footnote{According to \cite{Carrasco-Davis2021}, the stamp classifier has an overall 87\% SN completeness of SN, and agrees with 78\% of the light curve classifier SN.} After cuts, we have $\sim$10,000 objects, the majority of which are bonafide transients.
    \item From this sample, we obtain host galaxy features using the \ghost{} software\footnote{\url{https://pypi.org/project/astro-ghost/}} \citep{Gagliano2021}. \ghost{} is a database of 16k PanSTARRS (\PS1{}, \citealt{Chambers2016}) spectroscopic supernovae and the catalog-level properties of their host galaxies, equipped with analysis tools for transient host association. 
    A final association is made using the directional light-radius (DLR) method at the catalog level. We do not opt to use the gradient ascent (GA) method at the postage stamp level\footnote{\url{https://ps1images.stsci.edu}} for computational reasons. \par
    Because we require host galaxy features as part of our input data schema, transients with unsuccessful host associations (e.g., hosts not in the \PS1{} host galaxy catalog, transients occurring in faint host galaxies that is either not present in the PS1 catalog or cannot be found during the association etc.) are discarded. From 9402 bonafide transients, 6571 have successful host associations. However, we additionally remove transients whose host galaxies have null values for any of the features presented in Appendix~\ref{appx:host_gal_features}. This requires PS1 photometry in $grizy$ passbands.\footnote{This effectively removes the faintest sources (such as those above the S/N limit in only a few bands) and extremely red ones detected only in the reddest filter ($y$). This includes the faintest host galaxies (in which SLSN are preferentially found) and red sources like brown dwarfs and high-redshift quasars \citep{Magnier2016}. The impact on this selection function is left unquantified and left for future work.} After cuts, we have 5837 remaining objects for which we query TNS for a spectroscopic classification (if it exists).
    \item As a final attempt to increase the purity of bonafide transients in our databank, we query the PS1 point source catalog \citep[PS1-PSC;][]{Tachibana2018}. The PS1-PSC contains $\sim$1.5 billion sources from the PS1 first data release (DR1), and is used within ZTF's real-time extragalactic alert stream to automatically reject stellar sources. \par
    We query the PS1-PSC at two locations per transient candidate: the location of the transient, and the location of the host galaxy, both with a search radius of 1\arcsec. 
    We retain transient candidates whose host galaxy locations match to a resolved extended object with $\textgreater$50\% probability, and whose transient locations either A) also match to resolved extended object with $\textgreater$50\% probability (in the case of a transient occurring at/near the galaxy center) or B) match to no existing counterpart in the PS1-PSC catalog (in the case of a transient occurring far/offset from the galaxy center). After cuts, we have a remaining dataset of 5472 likely bonafide transients.
\end{enumerate}

A table summarizing the counts of loci at each stage of our preprocessing and quality cuts is found in Table~\ref{tab:preprocessing-cuts}.
\begin{table*}
\centering
\caption{Preprocessing Cuts}
\label{tab:preprocessing-cuts}
\begin{tabular}{lr}
\hline
\textbf{Selection Cuts} & \textbf{\# Loci Remaining} \\
\hline
Tagged by \texttt{lc\_feature\_extractor} Filter & 2,536,582 \\
Have $|b|~\textgreater~15\degree$ & 976,842 \\
Require both ZTF-$g$, ZTF-$r$ passbands & 488,300 \\
Have ALeRCE `SN' classifications & 9,420 \\
Have successful \texttt{GHOST} host galaxy association & 6,571 \\
Have non-null host galaxy features & 5,837 \\
Cross-matched to resolved extended source (PS1-PSC) & 5,472 \\
\hline\\[-2.0ex]
\multicolumn{2}{c}{
\begin{minipage}{10cm}
The number of loci (representing potential transient candidates) after selection cuts to construct our databank.
\end{minipage}}
\end{tabular}
\end{table*}

Of our remaining 5472 objects, only 1656 ($\approx$30\%) have a spectroscopic classification available from TNS. This leaves 3816 transients with no spectroscopic classification. The spectroscopic breakdown is found in Table~\ref{tab:dataset_bank}. The subsequent process for generating our ZTF training set from this full databank is explained in Section~\ref{subsec:train_set}.


\begin{table*}
\centering
\caption{Databank Spectroscopic Breakdown (5754 objects)}
\begin{tabular}{lllll}
\hline
\textbf{Misc. (3852)} & \textbf{SN Ia (1160)} & \textbf{SN II (340)} & \textbf{SN Ib/c (75)} & \textbf{Exotic (45)} \\
\hline
Phot (\textbf{3816}) & Ia (\textbf{1098}) & II (\textbf{243}) & Ib (\textbf{25}) & TDE (\textbf{20}) \\
CV (\textbf{8}) & Ia-91T-like (\textbf{37}) & IIn (\textbf{59}) & Ic (\textbf{21}) & SLSN-II (\textbf{14}) \\
AGN (\textbf{7}) & Ia-pec (\textbf{10}) & IIb (\textbf{21}) & Ic-BL (\textbf{14}) & SLSN-I (\textbf{11}) \\
SN~I (\textbf{6}) & Ia-91bg-like (\textbf{6}) & IIP (\textbf{14}) &  Ibn (\textbf{7}) & \nodata \\
SN (\textbf{5}) & Ia-CSM (\textbf{4}) & II-pec (\textbf{2}) & Ib/c (\textbf{5}) & \nodata \\
Other (\textbf{5}) & SN Iax[02cx-like](\textbf{3}) & IIn-pec (\textbf{1}) &  Ib-pec (\textbf{2}) & \nodata \\
Nova (\textbf{3}) & Ia-SC (\textbf{2}) & \nodata & Icn (\textbf{1}) & \nodata \\
Varstar (\textbf{2}) & \nodata & \nodata & \nodata & \nodata \\
\hline\\[-2.5ex]
\multicolumn{5}{c}{
\begin{minipage}{10cm}
The spectroscopic classifications from TNS used in our databank before a train/test split and upsampling.
\end{minipage}}
\end{tabular}
\label{tab:dataset_bank}
\end{table*}

\subsection{Feature Selection} \label{subsec:feature_select}

Here we outline our motivation for feature selection. Instead of using the observed, irregularly sampled light curve, we opt to use derived statistical features for the anomaly detection and similarity search methods (Sections~\ref{subsec:AD_RFC},~\ref{subsec:sim_search_ztf}). A review of the challenges astronomers face when analyzing light curves of astronomical sources and the practices for their characterization via statistical feature extraction is found in \cite{Babu2016}. \par
 
The main advantage of using statistical features is that they transform the sparse, gappy, and heteroscedastic light curves into regularly sampled representations, enabling the use of standard time series methods. These methods do not rely on imputation or interpolation; they are calculated only from the data that is observed. The fast, efficient, and scalable calculation of statistical light curve features features is now ubiquitous in LSST brokers. The \texttt{light-curve} package \citep{Malanchev-LC}, which is used to extract the light curve features for our dataset, is already integrated into the pipelines of ANTARES \citep{Matheson2021}, AMPEL \citep{Nordin2019}, and Fink \citep{Moller2021} broker systems\footnote{As for non-\texttt{light-curve} implementations, ALeRCE has their own in-house feature extractor \citep{Forster2021}.}. Moreover, it has been selected for use with LSST. The Rust version of \texttt{light-curve} package outperforms other implementations by a factor 1.5-50, such that it only takes a few milliseconds per CPU core to extract ``cheap” features for 1,000 light curve observations \citep{lavrukhina2023}. \par

Although relying solely on interpolation-based methods like Gaussian process (GP, \citealt{Rasmussen2005}) is a popular choice (see, e.g., \citealt{Lochner2016, Boone2019, Alves2022}), these methods tend to be slower and are not ubiquitous in broker pipelines. Generally, light curve interpolation, smoothing, and padding is sometimes necessary or at least tends to improve results in a variety of deep learning applications: multilayer perceptrons (MLPs, \citealt{Demianenko2022}), normalizing flows (NF, \citealt{Demianenko2022}), temporal convolutional networks (TCNs, \citealt{Muthukrishna2019}), recurrent neural networks (RNNs, \citealt{Charnock2017, Moller2021, 2022Gagliano_CCA}) convolutional neural networks (CNNs, \citealt{Pasquet2019b, Qu2021, Burhanudin2022}), Bayesian neural networks (BNNs, \citealt{Demianenko2022}), variational autoencoders (VAEs, \citealt{Villar2020, Boone2021, Aleo2023}), and transformers \citep{Donoso-Oliva2023, Moreno-Cartagena2023, Pimentel2023}. Often times, the physical interpretation of such learned features from latent or deep-learning representations can be nebulous. \par

In the literature, there has been proven successes of using manually-selected statistical features for classification (e.g., the ALeRCE light curve classifier, \citealt{SanchezSaez2021}; $dmdt$-mappings, \citealt{Mahabal2017}; anomaly detection $p(dm|dt)$ distributions, \citealt{Soraisam2020}; the \texttt{SNAD Transient Miner}, \citealt{Aleo2022}). The lessons learned from these works inform our feature selection, methods, and approach to follow.  \par

\subsubsection{Statistical Light Curve Features} \label{subsubsec:lc_features}

We start with the suite of 106 light curve features (53 in ZTF-$g$, 53 in ZTF-$r$ bands) automatically extracted (at the last observed epoch; features calculated at earlier phases are overwritten) from the \texttt{lc\_feature\_extractor} \emph{Filter} in ANTARES using the \texttt{light-curve} package \citep{Malanchev-LC}. This ensemble of features describe different aspects of the light curve shape, capturing rise information, error distribution, periodic signals, skewness, etc. Moreover, transients whose feature values are in the tails of their parent feature distribution are deemed outliers and could indicate anomalous activity. 

We know from \cite{Malanchev2021} (their Figure~A3) that many of the calculated light curve features are correlated. To mitigate the effect of correlated features and attempt to partially reduce the dimensionality of our dataset, we drop features that are \emph{not} used in \cite{Aleo2022}, leaving us with 82 light curve features (41 per band). Because \cite{Aleo2022} were able to successfully find previously undiscovered SNe candidates among nearly $\sim$1~million sources (most of which were standard variables) using this feature set, we were confident that if we adopted a similar feature set we would also be successful given that the dataset we work with here has significantly higher purity of transients. 
Lastly, we removed 10 features (per band) that were strongly dependent on the cadence (or time-dependence) of observations. We do this because downstream we want to find analogs of the objects themselves without biasing \emph{how} their light curves are measured. This includes features pertaining to measuring the period or calculating the periodogram.  \par

Our final light curve feature list comprises 62 total: 48 magnitude-based and 14 flux-based features, all of which are brightness-related; see the complete list of features and their description in Appendix~\ref{appx:lc_features}.

\subsubsection{Host Galaxy Features} \label{subsubsec:host_features}

As SN discovery rates grow at a rapid pace, particularly in today's expansive landscape of systematic sky surveys, the correlation between SNe and their global galactic environments has started to crystallize \citep[e.g.,][and references therein]{Li2011, Qin2022}. For instance, the rate of core-collapse (CC)~SNe, whose progenitors are short-lived ($\textless$50~Myr) massive stars, is tightly coupled to the star formation rate of a galaxy \citep{Graur2017}. Similarly, there are unique galactic correlations that can be gleaned for many SN demographics, and this information should be considered when quantifying the `anomalous' nature of a transient. Consider the following: all but a single SN~Iax to date have been discovered in late-type galaxies \citep{Lyman2018}. Thus, a new discovery of a second SN~Iax in an early type galaxy could be considered \emph{more} anomalous than the same SN~Iax (i.e., with the same light curve evolution and derived properties) in a late-type galaxy. \par

Our work, which incorporates derived global host galaxy properties, lays the groundwork for such future analyses of varying SNe types. Thus, the object's anomaly score is affected by not only its light curve behavior, but its global host galaxy properties as well. We use approximately a 1:1 ratio of light curve features to host galaxy features.\footnote{In practice, this weighting does not hold exactly (e.g., see Figure~\ref{fig:RFC_feat_importance}, where 8 of the top 10 most important features used in our anomaly detection task are light curve features, and the remaining 2 are host galaxy features).} Ultimately, global host galactic environments of the SNe are considered to be meaningful, though weighted less when compared to unusual light curve behavior. As for local correlations, we conduct no formal investigation because of the general lack of spatially resolved explosion site data. 

Incorporating contextual information into SN identification and classification tasks has steadily increased in the last decade. \cite{Foley2013} demonstrated that host-galaxy morphology and color were sufficient to construct photometric SN~Ia samples at the purity level of then-current light curve methods. \cite{Baldeschi2020} used photometric host features (colors and moments) from the PS1~DR2 catalog to build a RFC and assign probabilities between low and highly star-forming host galaxies, subsequently to help distinguish SNe~II from SNe~Ia. ALeRCE developed a postage-stamp classifier \citep{Carrasco-Davis2021}\footnote{Note that the postage stamp was centered on the detection, thereby encoding \emph{some} local/offset information} to disambiguate SNe from non-SNe (AGN, variable stars, asteroids, bogus detections) with a single detection image, in part relying on the fact that many SN detections contain a visible host galaxy in both science and reference images, but only the science and difference images contain the flux from the SN. \cite{Gagliano2021}, which debuted \texttt{GHOST}, found that photometric features of transient-hosting galaxies alone was sufficient in classifying SNe II and SNe Ia with $\sim$70\% accuracy when fed into a RFC. \cite{2022Gagliano_CCA} extended such work to combine light curve and host galaxy information for adaptive, real-time photometric classification, with emphasis on early-time phases, achieving an accuracy of 82\% $\pm$ 2\% within 3 days of an event's discovery. 

Moreover, publicly accessible SN data products have been keen to include host galaxy information: the YSE DR1 \citep{Aleo2023} included the associated PS1 host galaxies of 1975 SNe, and the simulated Extended LSST Astronomical Time-Series Classification Challenge (ELAsTiCC; \citealt{Narayan2023ELAsTiCC}) produced LSST alert packets with both transient photometry and host information. In the LSST era, where transient spectroscopic classification capabilities will be capped at approximately 1\% of all transients \citep{hambleton2022rubin}, any additional information to augment the sparse transient photometry will be required to fully characterize the SNe and its environment. It is sensible for host galaxy information to provide such a link. \par

With the laundry list of works which were successful in SN identification and classification tasks using host galaxy information, as well as its prevalence in data products, it is a natural extension to use it in the context of anomaly detection and similarity search. \par

\texttt{GHOST} \citep{Gagliano2021} natively connects our transients to a host in the PS1 galaxy catalog and its hundreds of catalog-level properties\footnote{\url{https://outerspace.stsci.edu/display/PANSTARRS/PS1+Database+object+and+detection+tables}}. All host galaxy features per transient are retrieved from the PS1 catalog or explicitly calculated with one host association calculation, and are reused downstream in our anomaly detection and similarity search methods (we make the reasonable assumption that the host association and derived host properties of each SNe is constant in time). \par

The selection of global host features include the \cite{Kron1980} flux from $g,r,i,z,y$ filter stack detections, the normalized offset of the SNe from its host with respect to the Directional Light Radius (DLR), and the extendedness measure for the $g,r,i,z,y$ filter stack detections based on the deviation between PSF and \cite{Kron1980} magnitudes normalized by the PSF magnitude uncertainty, to name a few. Our final host galaxy feature list comprises 58 total. See the full list and a brief description of each host galaxy feature used in this work in Appendix~\ref{appx:host_gal_features}, and discussion of our feature correlations in Appendix~\ref{subsec:APP_feat_corr}. \par

\subsection{ZTF Training Set} \label{subsec:train_set}

We build our training set using our databank presented in Table~\ref{tab:dataset_bank}. Naturally, due to pre-existing follow-up strategies and relative SN rates, our spectroscopic dataset is heavily class-imbalanced; our majority classes are vastly overrepresented, and our minority classes are vastly underrepresented. Moreover, any algorithm to demarcate anomalies from normal SNe trained on such a class-imbalanced dataset would incur a bias toward the majority classes. \par

Because we want an algorithm to learn the full feature space of these different SN classes, as well as to mitigate any bias incurred from class-imbalance, we will upsample all classes. However, before upsampling, we restrict our SNe classes to those that have at least 14 members (before train/test split). This strikes a balance between having several rare SNe classes in our training set while retaining enough members in each class to properly upsample in feature space (via interpolation between known examples) after a train/test split. In this way, the feature space can be satisfactorily representative of the full, unknown feature distribution to achieve sufficient performance. \par

There are 3816 photometric objects without a spectroscopic classification (`Phot' class). Those that do have spectroscopic classifications breakdown into: 1098 SN~Ia, 243 SN~II, 59 SN~IIn, 37 SN Ia-91T-like, 25 SN~Ib, 21 SN~IIb, 21 SN~Ic, 20 TDE, 14 SN~Ic-BL, 14 SN~IIP, and 14 SLSN-II. We perform a standard 70/30 train/test split. The training set before upsampling breaks down into: 2656 photometric-only (`Phot' class), 777 SN~Ia, 175 SN~II, 44 SN~IIn, 28 SN Ia-91T-like, 19 SN~Ib, 15 SN~IIb, 12 SN~Ic, 12 TDE, 9 SN~Ic-BL, 9 SN~IIP, and 11 SLSN-II. From this class-imbalanced training set, we upsample to a class-balanced training set using the Synthetic Minority Over-sampling Technique (SMOTE; \citealt{Chawla2002}), such that each minority classes has the same number of samples as the majority class (2656), totalling 31872 objects. Thousands of labeled samples per class well encapsulate the feature space needed by the RFC for training. \par

SMOTE works by drawing random samples along vectors joining every grouping of $k$ objects in feature space until all classes are balanced. We use $k$=8 neighbors; this is the largest value we can use for our training set, because our smallest class to be upsampled for training has 9 members (SN~IIP, SN~Ic-BL). 
Note that these upsampled features are derived from the full light curve (i.e., not partial light curves) and the host galaxy features. Our final training set is detailed in Section~\ref{subsec:AD_RFC}. \par

We leave the spectroscopic test set as is, and do not perform upsampling because we want to evaluate our performance on a representative sample of true events, matching the selection function we impose. Our spectroscopic test set is composed of: 321 SN~Ia, 68 SN~II, 15 SN~IIn, 9 SN Ia-91T-like, 6 SN~Ib, 6 SN~IIb, 9 SN~Ic, 8 TDE, 5 SN~Ic-BL, 5 SN~IIP, and 3 SLSN-II. Our observed photometric set is not used in training or testing, but is used in our approximate nearest neighbors search (Section~\ref{sec:annoy}). \par

Note that at this stage, we do not perform any linear or non-linear transformation of feature values. We do no rescaling for our RFC (Section~\ref{subsec:AD_RFC}), but we do standardize features by removing the mean and scaling to unit variance for PCA before our ANN search with \texttt{ANNOY} (Section~\ref{sec:annoy}). Due to the limited size of our spectroscopically labeled dataset, especially for rare classes with few members, we opt for a train/test set instead of the more commonly used training/test/validation set. \par

SMOTE \citep{Chawla2002} is known to have certain limitations, particularly when dealing with small training sets and high-dimensional feature spaces. It is prone to generating synthetic points that lie along the lines connecting original data points rather than curves. This behavior, while expected, indicates that SMOTE may not fully capture the complex, curved manifolds \citep{Bellinger2016} inherent in our 120-dimensional feature space and the complex phenomena of supernovae light curves and hosts. \par

Given these constraints, we acknowledge that alternative oversampling techniques, such as the Adaptive Synthetic Sampling method (ADASYN, \citealt{He2008}) and normalizing flows \citep{Papamakarios2021}, might better address these issues by considering the local density and distribution of data points, or correlations across features. This is a subject of future work. 

Despite these limitations, our use of SMOTE facilitates direct comparisons with other anomaly detection methods, such as the FLEET algorithm \citep{Gomez2020, Gomez2023SLSN, Gomez2023TDE}, which also employs SMOTE in a similar context.

\section{Anomaly Detection in ZTF Alert Stream} \label{sec:ztf_real_time} \par

With any anomaly detection task, the definition of ``anomaly" and its associated evaluation metric must be clear. For this work, we consider a transient to be an anomaly if it falls into any of the following three categories:
\begin{enumerate}
    \item \textbf{Spectroscopic anomaly}: Any transient whose spectroscopic label is \textit{not} of Type Ia-normal, Ia-91T-like\footnote{Our model performs more competitively when we consider the rare subtype SN~Ia-91T-like SNe as non-anomalous due to overlap in feature-space with SN~Ia-normal. See Section~\ref{subsec:AD_RFC} for details.}, II-normal, and IIP (the ``Normal" type of our RFC). Thus, we consider the remaining transient labels explicitly defined in our training set (e.g., TDE, SLSN-II, SN~IIn, SN~IIb, SN~Ib, SN~Ic, SN~Ic-BL) as well as labels outside our training set (e.g., SN~Ia-91bg-like, SN~Iax, SN~Ibn, SN~Icn, SLSN-I, etc.) as anomalous (the ``Anomaly" type of our RFC). Note that when we consider spectroscopic anomalies alone, we do so using the original TNS classification label without looking at the spectra. However, when we consider spectroscopic anomalies after vetting (grouped with contextual and behavioral anomalies), we determine as such through the re-classification of the transient's spectra with \texttt{SNID} \citep{Blondin2007} (if needed). This is often spurred by ``Anomaly" classifications from our model on objects originally outside of our anomalous classification taxonomy, or objects whose ANNs are instances of the ``Anomaly" class. \par
    \item \textbf{Contextual anomaly}: Any transient whose host environment is peculiar (e.g., an SN~II-normal in an evolved elliptical galaxy\footnote{\cite{Irani2022} concluded approximately 0.3\% of all core-collapse (CC)~SNe occur in elliptical galaxies, as derived from the spectroscopically complete ZTF-BTS.}, or a SN in a rare galaxy type) based on our current understanding of host galaxy correlations. Note that we do not consider SNe~Ia in spiral galaxies to be rare enough alone to be considered anomalous. We identify contextual anomalies through manual vetting of the host galaxy type in conjunction with the known or likely SN type. \par
    \item \textbf{Behavioral anomaly}: Any transient whose light curve behavior exhibits an atypical or peculiar evolution indicative of an underlying physical process (e.g., a prominent second bump in the light curve caused by cirumstellar interaction) and not an observational effect (e.g., missing rise information, color information, imaging artefacts, etc.). We identify behavioral anomalies through manual vetting by an expert of the light curve evolution with respect to the typical evolution of the SN type. \par
\end{enumerate}

Our philosophy in constructing \laiss{} to discover anomalous transients is similar to that of FLEET \citep{Gomez2020} to find SLSNe \citep{Gomez2023SLSN} and TDEs \citep{Gomez2023TDE}, in part due to their proven success. Our guiding principles are as follows: \par
\begin{itemize}
    \item We ``classify" only anomalies with no regard for the classification success of other transients.
    \item We favor anomaly sample purity over sample completeness.
    \item We prioritize low-latency and simple model construction to allow for real-time identification.
    \item We aim for a manageable list of currently active flagged anomalous transients, such that an expert can manually vet the source and select the best targets each night.
    \item We flag anomalies while they are bright to enable photometric and spectroscopic follow-up.
\end{itemize}

\subsection{Full-Phase Anomaly Detection with Random Forest Classifier and Isolation Forest} \label{subsec:AD_RFC}

\subsubsection{Spectroscopic Anomalies (Before Vetting)} \label{subsubsec:spec_anom_only}

With our class-balanced training set, we remove objects with no spectroscopic label (the `Phot' class), leaving us with 29216 labeled spectroscopically-confirmed and SMOTE-generated events. From our testing set we assign SN~Ia, SN~Ia-91T-like, SN~II, SN~IIP as ``Normal" (403 objects), and TDE, SN~IIn, SLSN-II, SN~IIb, SN~Ib, SN~Ic, and SN~Ic-BL as ``Anomaly" (52 objects). \par

Now we set out to ``classify" anomalies. We do not classify by transient class label; rather, we do a binary classification into ``Normal" and ``Anomaly" classes. To discourage our algorithm from tagging normal SNe~Ia as anomalous, we assign SN~Ia-normal and SN~Ia-91T-like as ``Normal", despite the latter being a more rare overluminous subtype. We find that including SN~Ia-91T-like as ``Anomaly" resulted in a higher rate of Ia tagged as anomalous\footnote{We find that including SN~Ia-91T-like as ``Anomaly" resulted in a higher rate of Ia tagged as anomalous, lowering our purity about 10\%.}. This is perhaps due to the large photometric similarity in the light curve profile between SN~Ia-91T-like and SN~Ia-normal (unlike SN~Ia-91bg-like) because distinguishing factors like absolute magnitudes are not explicitly used. Thus, to be consistent throughout this work, we consider SN~Ia-91T-like to be like any other SN~Ia-normal, and state that our model is not well suited to tag SN~Ia-91T-like as ``Anomaly''. \par

We perform many iterations of varying machine learning algorithms for anomaly detection while varying hyperparameters via grid search. Some of these tests included a variational autoencoder, support vector machine, isolation forest, and a random forest classifier. A comparative table summarizing the performance metrics of all tested algorithms is found in Table~\ref{tab:comparison_algs}. We highlight the combination of hyperparameters from the grid search resulting in the best anomaly purity, while balancing both recall and fraction of predicted test set anomalies. Additionally, we note commonalities among truly non-anomalous SNe classified as anomalous (``false-positive"). More in-depth modification of standardized out-of-the-box algorithms (e.g., customizing a weighted reconstruction class loss or anomaly-specific class loss using pairwise distances in latent space) for specialized anomaly detection tasks is a subject of future work.

Ultimately, we achieved the best and most robust performance with the highest sample purity on the spectroscopic test set via the \texttt{sklearn} implementation of a Random Forest Classifier\footnote{\texttt{sklearn.ensemble.RandomForestClassifier}, see \url{https://scikit-learn.org/stable/modules/generated/sklearn.ensemble.RandomForestClassifier.html}}. A RFC is an ensemble learning method constructed of many decision trees (each trained on bootstrap samples using a random subset of features without replacement) which outputs the mode of the classes in a classification task. \par

\begin{table*}[ht]
\footnotesize
\centering
\caption{\textbf{Comparison of AD Algorithms.} Comparing tested anomaly detection algorithms with grid search}
\begin{tabular}{c|ccccc}
\hline
\hline
Algorithm & Hyperparameters & Recall & Purity & Total Anomalies & False Positive \\
 & via grid search & (Anomaly) & (Anomaly) & Predicted (Test Set) & Characteristics \\
\hline
\hline
VAE & batch\_size = \{1, 64, 128, \textbf{256}, 512\} & 6\% & 10\% & 30/455 (7\%) & Bright ($m~\lesssim~18$~mag),  \\
  & epochs = \{5, 25, \textbf{50}, 75, 100\} &  &  & & SN~Ia in spiral/irregular hosts \\
  & intermediate\_dim = \{2, 8, 16, \textbf{64}\} &  &  & & SN~II with $r$ plateau, \\
  & latent\_dim = \{1, \textbf{3}, 5, 10\} &  &  & & linear $g$ decline \\
  & pos\_weight = \{0.75, \textbf{0.9}, 0.95, 1, 2, 5\} &  &  & & \\
  & metrics = \{\textbf{Precision()}, Recall(), Accuracy()\} &  &  & & \\
  & activation = \textbf{``relu", ``sigmoid"} &  &  & & \\
  & optimizer = \textbf{``adam"} &  &  & & \\
\hline
SVM & kernel = \{\textbf{``rbf}", ``linear", ``poly"\}  & 25\% & 19\% & 68/455 (15\%) & Often in edge-on, \\
 & C = \{0.5, \textbf{1}, 5, 10\} &  &  & & faint, or compact hosts \\
 & gamma = \{``scale", \textbf{``auto"}\} &  &  & &\\
& decision\_function\_shape = \{\textbf{``ovo"},  ``ovr"\} \\
 & class\_weight = \textbf{``balanced"} \\
\hline
IF & n\_estimators = \{50, \textbf{100}, 500\} & 23\% & 20\% & 60/455 (13\%) & Very bright ($m~\lesssim~16.5$~mag) \\
  & max\_features = \{5, 15, 25, \textbf{35}, 45\} &  &  & & \\
  & contamination = \textbf{13\%} &  &  & & \\
\hline
RFC & n\_estimators = \{50, \textbf{100}, 500\} & 29\% & 52\% & 29/455 (6\%) & Large gaps in light curve, \\
  & max\_depth = \{5, 15, 25, \textbf{35}, 45\} &  &  & & or long-lived \\
  & max\_features = \{5, 15, 25, \textbf{35}, 45\} &  &  & & \\
  & class\_weight = \textbf{``balanced"} &  &  & & \\
\hline
\hline\\[-0.5ex]
\multicolumn{6}{c}{
\begin{minipage}{16cm}
\vspace{0.3cm}
NOTE: Bold text indicates the hyperparameters chosen for the final algorithms from a grid search.\\ 
NOTE: The ``Total Anomalies Predicted" column indicates how many objects were tagged as ``Anomaly" out of the test set sample of 455 objects for each tested algorithm. The corresponding percentage is shown in parentheses.\\
NOTE: The ``False Positive Characteristics" column indicates common patterns of ``Normal" SNe tagged as anomalous, to reveal common failure modes. \\
After comparing several algorithms, the Random Forest Classifier performed the best, with highest overall anomaly purity balanced with a suitable number of anomalies tagged relative to their estimated population. 
\end{minipage}}
\end{tabular}
\label{tab:comparison_algs}
\end{table*}

We use a RFC with the following hyperparameters: 100 trees (\texttt{n\_estimators=100}), a tree depth of 35 (\texttt{max\_depth=35}), a maximum of 35 features out of the 120 input features per tree (\texttt{max\_features=35}), a contamination level of 13\%\footnote{to match our observed value within a conservative limit to account for potential misclassifications and peculiar behavior} (\texttt{contamination=0.13}), and balance the weighting of normal to anomalous classes (\texttt{class\_weight="balanced"}). Surprisingly, increasing the weighting towards anomalous classes had no statistically significant increased performance, and thus we kept the simpler balanced weighting. We optimized our hyperparameters with a grid search: \texttt{n\_estimators} from 50 to 500 in steps of 50, \texttt{max\_depth} from 5 to 65 in steps of 5, \texttt{max\_features=35} from 5 to 65 in steps of 5, and \texttt{class\_weight} from 1 to 10 in steps of 1. We used the Gini impurity as our measure for the quality of feature split. 
To estimate the classifier's uncertainties, we run each version of the model 25 times using different random seed initializations. Training for each run was performed on an 2 GHz Quad-Core Intel Core i5 (macOS Version 11.7) and finished in $\sim$20~s after grid search. \par

The simplest overview metrics to understand the performance of a classifier are completeness, purity, and accuracy. These metrics are defined for a single class as:
\begin{align}
\label{eqn:metrics}
\begin{split}
 \text{Completeness} &= \frac{TP}{TP+FN}
\\
 \text{Purity} &= \frac{TP}{TP+FP}
\\
 \text{Accuracy} &= \frac{TP+TN}{S}
\end{split}
\end{align}
where TP (FP) is the number of true (false) positives, TN (FN) is the number of true (false) negatives, and S is the total sample size. \par

In this work, completeness (``recall") quantifies the percentage of a true type that is correctly classified (``Normal" or ``Anomaly"). Purity (``precision") quantifies the percentage of a predicted type that is correctly assigned the true type. Accuracy is the overall fraction of events that were correctly classified into their respective groups. \par 

\begin{figure*}
    \centering
    \includegraphics[width=\columnwidth]{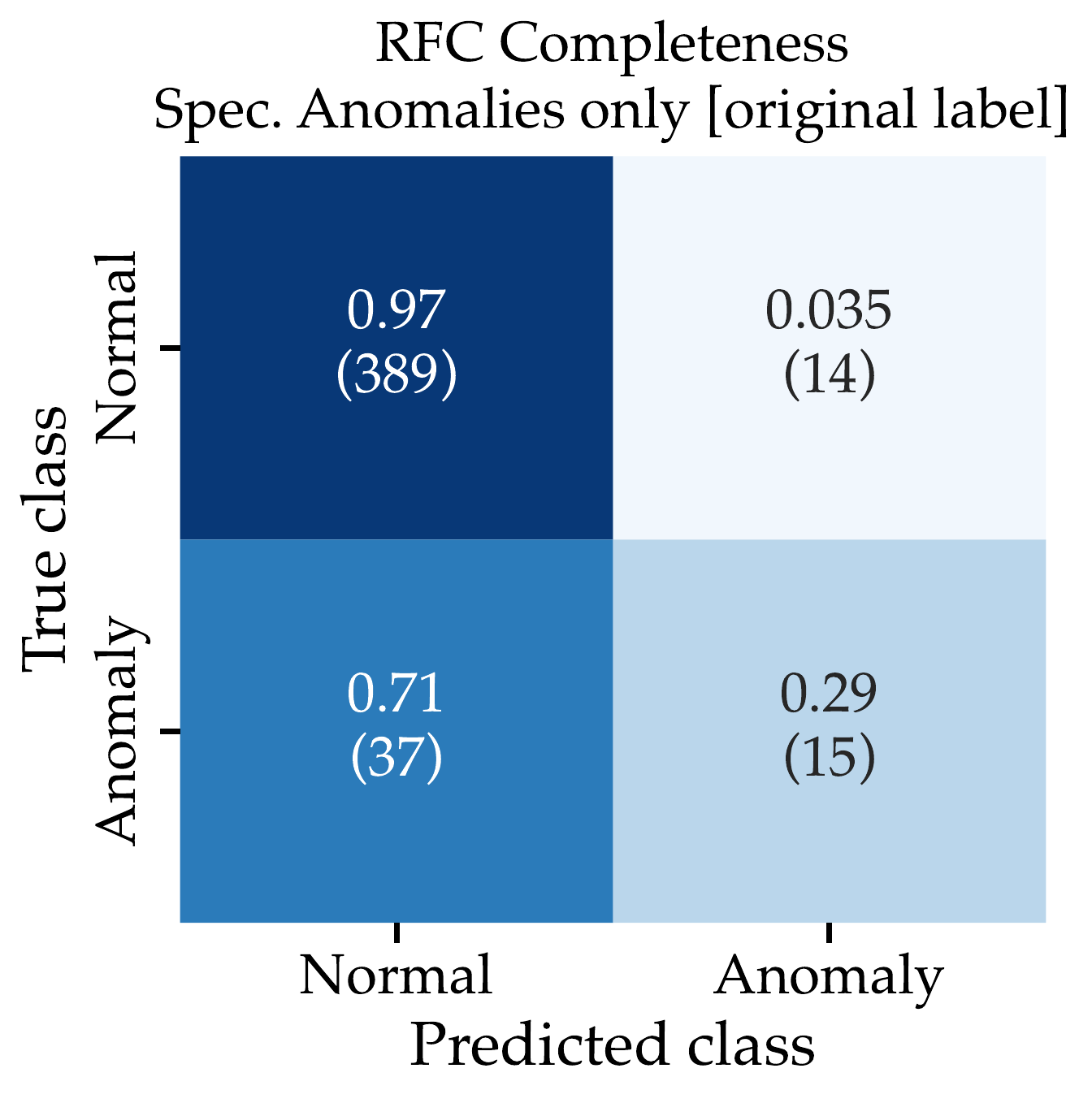}
    \includegraphics[width=\columnwidth]{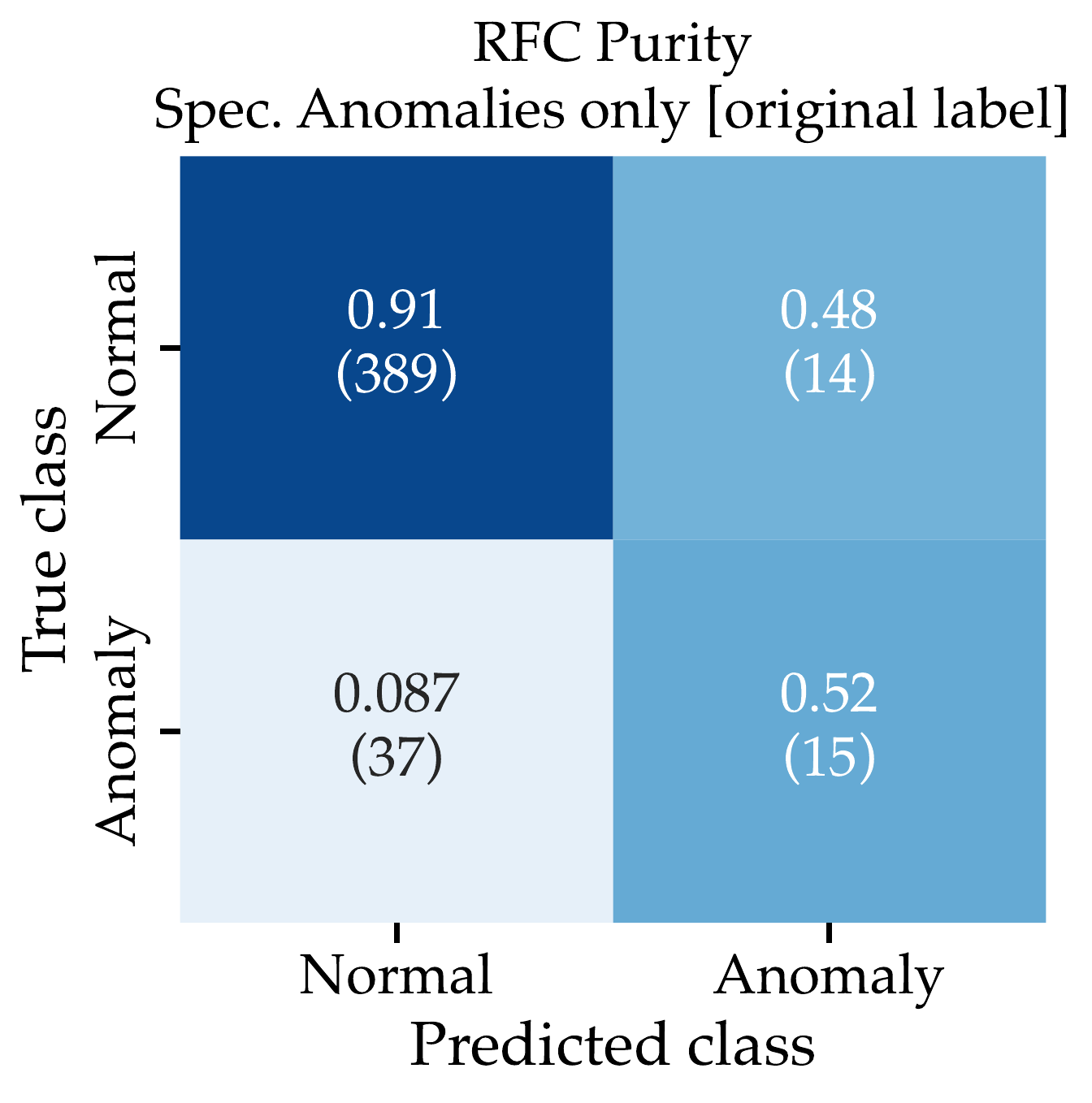}
    \caption{
    Completeness (left panel) and purity (right panel) confusion matrices of our spectroscopic test set for binary  classification performance between ``Normal" SNe (SN~Ia-normal, SN~Ia-91T-like, SN~II-normal, SN~IIP) and ``Anomaly" SNe (TDE, SLSN-II, SN IIn, SN IIb, SN Ib, SN Ic, SN Ic-BL), based on the original TNS spectroscopic label only (i.e., \emph{spectroscopic} anomalies before vetting). We achieve 52\% purity of detecting \emph{spectroscopic} anomalies alone.
    } 
    \label{fig:cm_spec_anomalies}
\end{figure*}

\begin{figure*}
    \centering
    \includegraphics[width=\columnwidth]{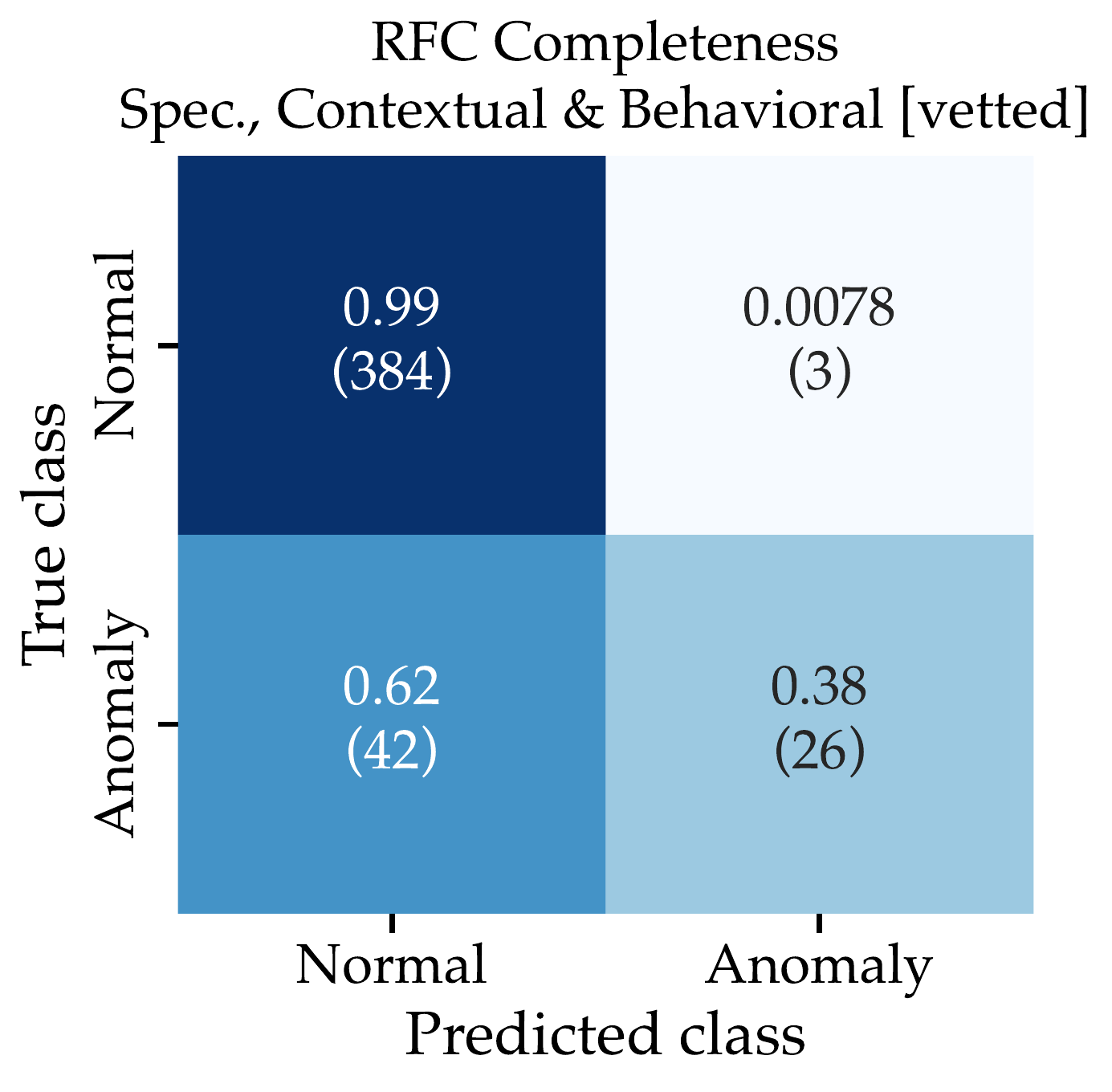}
    \includegraphics[width=\columnwidth]{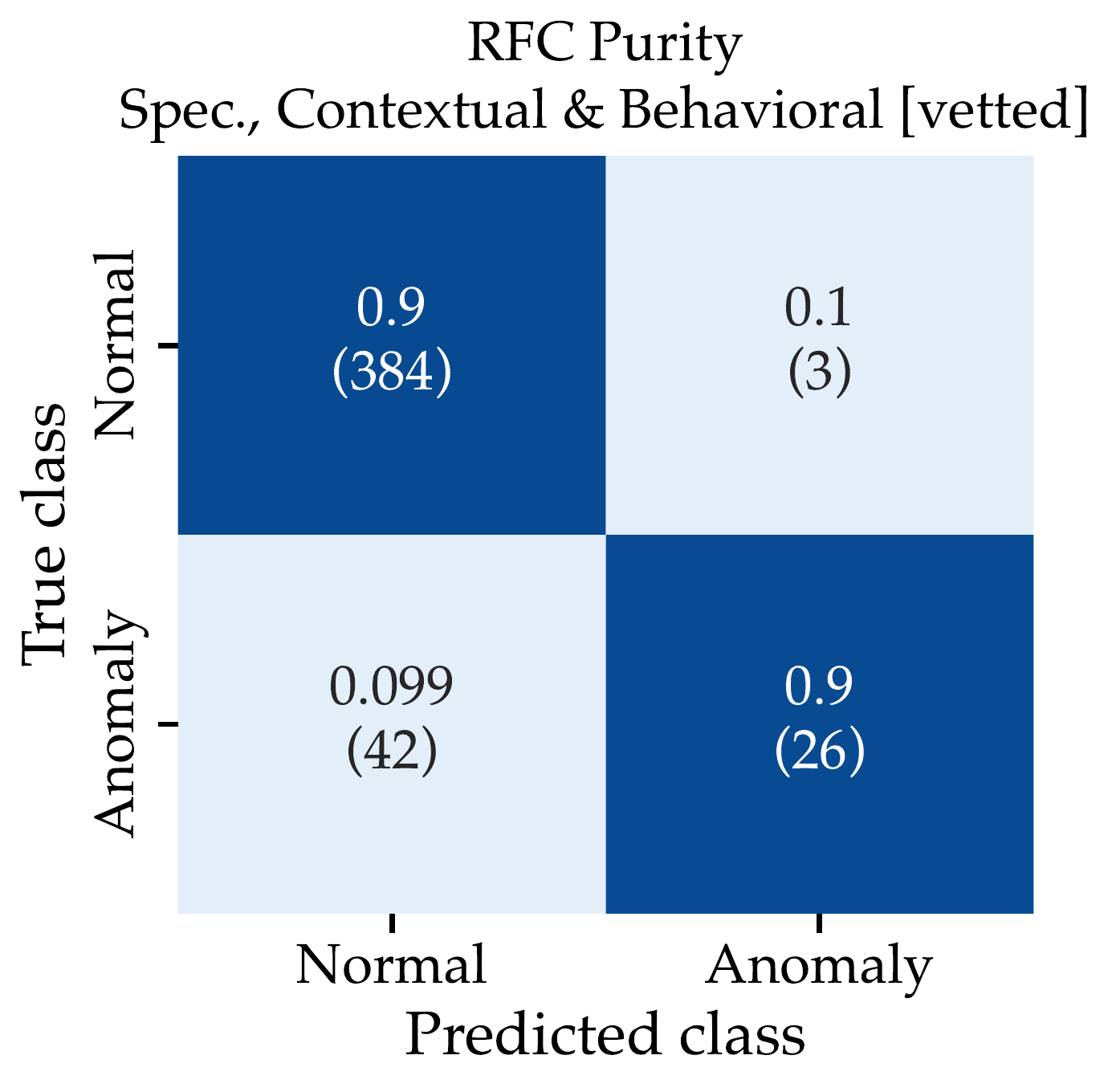}
    \caption{
    Same as Figure~\ref{fig:cm_spec_anomalies}, but including additional \emph{spectroscopic}, \emph{contextual}, and \emph{behavioral} anomalies after expert vetting of the classification spectra, host galaxy environment, and light curve evolution, respectively. We achieve 90\% effective purity of detecting \emph{spectroscopic}, \emph{contextual}, and \emph{behavioral} anomalies combined. We find this effective purity score is consistent with the spectroscopic subset of transients identified as anomalous in our real-time deployment on the nightly ZTF Alert Stream via ANTARES ($\sim$83\%, see Table~\ref{tab:real_time_any_phase}).
    } 
    \label{fig:cm_context_behavior}
\end{figure*}

The performance of this RFC on the spectroscopic test set is shown as completeness (left) and purity (right) confusion matrices in Figure~\ref{fig:cm_spec_anomalies}. Note that for these confusion matrices we only consider the value of the original spectroscopic label from TNS alone, and do not yet consider expert vetting for updated classifications, peculiar behavior, or atypical host galaxy environment; such analysis can be found later in this Section. \par

Our RFC achieves an overall accuracy of 89\%~$\pm$~1\% with a 97\%~$\pm$~1\% ``Normal" completeness and 91\%~$\pm$~1\% ``Normal" purity, with 29\%~$\pm$~3\% ``Anomaly" completeness and 52\%~$\pm$~5\% ``Anomaly" purity (with $1\sigma$ uncertainty derived from the different random seed iterations). \par

In a magnitude-limited survey like ZTF, we can expect about 70\% of observed SNe to be normal SNe~Ia, and about 20\% to be normal SNe~II, leaving about 10\% to be of a rare/anomalous nature (for ZTF~BTS, the measured relative rates are $\sim$72\% SNe~Ia, $\sim$20.5\% SNe II, $\sim$5.5\% SNe Ib/c, $\sim$2.5\% SLSNe; see \citealt{Fremling2020}). Indeed, in our spectroscopic test sample of 455 SNe, 52 (11\%) are considered anomalous. To have a completeness of 29\% and a purity of 52\% on the remaining 11\% of our test set implies that our RFC model significantly outperforms random selection. Because our test set is heavily imbalanced, the probability of randomly selecting ``Anomaly" should be proportional to its prevalence in the dataset. Thus, if we expect the observed sample to be roughly 10\% ``Anomaly", we would expect to successfully recover roughly 10\% anomalous events by random selection. Moreover, at 29\% completeness, our RFC still manages to identify roughly 1/3 true anomalies, a significant recovery fraction.  \par

We also find that our RFC model vastly outperforms a more sophisticated baseline model commonly used for anomaly detection (see Table~\ref{tab:comparison_algs}): an Isolation Forest (IF, \citealt{Liu2012}). An IF is an unsupervised decision-tree based method that isolates outliers by randomly selecting a feature from an array of features, and randomly selecting a threshold value (between the max and min) to split. This random partitioning of features and values produces shorter paths in trees for outlier/anomalous datapoints when aggregated in a tree ensemble. We use the same hyperparameter setup when available (\texttt{n\_estimators=100}, \texttt{max\_features=35}, \texttt{random\_state=11}) and a contamination level of 13\% (\texttt{contamination=0.13}). If we compare the 29 objects tagged ``Anomaly" (where 15 are correctly recovered, according to the spectroscopic label only) to the 29 objects with the highest anomaly score from the IF, we find the IF only successfully recovers 7 events. Moreover, 27/29 objects deemed most anomalous from the IF are brighter than 18.5 mag, with 13/29 exceeding $\sim$16.5~mag in brightness. Of those, 11/13 are simply nearby, normal SNe~Ia, and only 2/11 are what we consider anomalous, SNe~Ic. In essence, the most anomalous objects identified by the IF are simply bright, nearby, normal SNe that can be found with a selection cut. However, this result may be unsurprising given the disparity between a supervised approach like our RFC model compared to an unsupervised IF model. \par

A table highlighting all 29 test set objects tagged anomalous by the RFC model can be found in Table~\ref{tab:spec_test_set}. First we consider the resulting distributions from the spectroscopic label alone before expert vetting and potential reclassifications. The objects of ``Normal'' spectroscopic classes are 9 SN~II (13.2\% of all SN~II in test set) and 5 SN~Ia (1.5\% of all SN~Ia in test set). Similarly, for our anomalous classes, our model tagged 6 TDE (75\%), 3 SN~IIn (20.0\%), 2 SLSN-II (66.7\%), 1 SN~IIb (16.7\%), 1 SN~Ib (16.7\%), 1 SN~Ic (11.1\%), and 1 SN Ic-BL (20.0\%). However, because of small number statistics for the anomalous classes, we cannot say that our algorithm will detect 75\% of all TDE, for example. \par

\begin{table*}[ht]
\footnotesize
\centering
\caption{The 29 events from our spectroscopic test set with $P(anom)\geq50\%$ at \emph{full} light curve phase, ordered by $P(anom)$.}
\begin{tabular}{lllcccl}
\hline
\hline
IAU Name & TNS Class & New Class? & $z$ & Peak $M_{abs}$ & $P(anom)$ & Remarks \\
\hline
\textbf{2020yue} & SLSN-II & TDE\footnote{The updated classification was made by \cite{Yao2023} after our training process.} & 0.204 & $-21.8$ & 91 & Visible for $\sim$220~d. Blue, nuclear. \\
\textbf{2021uzt} & SN~II\footnote{Spectra displays narrow H$_{\alpha}$ at early times, then normal width at late times with broadened Lorentzian H$_{\alpha}$ profile (no clear P-cygni). Coupled with slow rise time and bright peak, we favor a IIn classification.} & SN~IIn & 0.061 & $-19.4$ & 89 & No visible host (host association incorrect). \\
\textbf{2018iih} & TDE & \nodata & 0.212 & $-21.3$ & 76 & Visible for $\sim$1000~d. No data at peak. \\
\textbf{2021iui} & SN~IIn & \nodata & 0.106 & $-19.7$ & 76 & Faint host. \\
\textbf{2020aeuh} & SN~Ia & SN~Ia-CSM\footnote{CSM interaction has been posited as the the likely driver for the secondary bump seen in this light curve \citep{Soraisam2022}.} & 0.126 & $-19.8$ & 72 & Second bump (likely CSM interaction). Visible for $\sim$175~d. \\
\textbf{2018khn} & SN~IIn & \nodata & 0.091 & $-19.4$ & 71 & Faint host. Visible for $\sim$450~d. \\
\textbf{2021bxq} & SN~IIn & \nodata & 0.09 & $-19.0$ & 70 & Faint host. Visible for $\sim$225~d. \\
\textbf{2021gje} & TDE & \nodata & 0.358 & $-21.9$ & 69 & Visible for $\sim$50~d. Blue, nuclear. \\
\textbf{2021aazw} & SN~Ic-BL & \nodata & 0.024 & $-16.6$ & 68 & Visible for $\sim$50~d. \\
2021our & SN~Ia & \nodata & 0.039 & $-18.2$ & 66 & Normal SN Ia. Tagged because underluminous? \\
2021aff & SN~II & \nodata & 0.051 & $-17.3$ & 64 & Incorrect host association. No decline. Visible for $\sim$150~d. \\
\textbf{2021fyp} & SN~II & \nodata & 0.053 & $-18.6$ & 64 & Peculiar light curve. Spectrum has He, weak H (possible IIb). \\
\textbf{2021cpi} & SN~II & SN~IIb & 0.055 & $-18.2$ & 64 & Lack of strong features in 1st peak spectrum (shock cooling?). \\
\textbf{2021ackd} & SN~Ia & SN~Ia-91bg-like & 0.050 & $-18.1$ & 63 & Matches to SN~1991bg, 2007ax, 1986G. \\
\textbf{2021ckb} & SN~II & SLSN (He-rich) & 0.070 & $-19.1$ & 62 & Narrow H$_{\alpha}$ emission from host. Match to PTF10hgi.\footnote{See \cite{Quimby2018, Gal-Yam2019ARA} for details.} \\
\textbf{2021nwa} & TDE & \nodata & 0.047 & $-18.2$ & 61 & Blue, nuclear. Visible for $\sim$150~d. \\
2020tnq & SN~Ia & \nodata & 0.033 & \textless$-18.3$ & 61 & First observations around +26~d after peak (from spectrum). \\
\textbf{2021M} & SN~IIb & \nodata & 0.011 & $-15.2$ & 60 & Edge-on host galaxy. Visible for $\sim$55~d. \\
\textbf{2021mhg} & TDE & \nodata & 0.073 & \textless$-19.5$ & 60 & Re-brightening/ second transient event at location.\footnote{See \cite{Munoz-Arancibia2023} for details.} \\
\textbf{2020abah} & SN~II & \nodata & 0.03 & $-16.5$ & 60 & Known member of long-rising SN~II class at $\sim$90~d.\footnote{$\approx$1.4\% core-collapse~SNe. See \cite{Sit2022} for details.} \\
\textbf{2021adpx} & SN~Ia & SN~Ia-91bg-like & 0.047 & \textless$-18.7$ & 59 & $\sim$20~d $r$-band gap. Spiral host. Matches to SN~2002cf, SN~2006gt. \\
\textbf{2020ywo} & SN~II & SN~Ia-CSM & 0.0475 & \textless$-18.8$ & 58 & Asymmetric H$_{\alpha}$ hints IIn, but redwards fit match Ia. \\
\textbf{2020rmk} & SN~II & \nodata & 0.025 & $-16.9$ & 57 & Candidate member of long-rising SN~II class. \\
\textbf{2020acka} & TDE\footnote{\cite{Frederick2021} this object to be a Trakhtenbrot AGN flare \citep{Trakhtenbrot2019}.} & \nodata & 0.338 & $-23.2$ & 57 & Blue, nuclear. Visible for $\sim$300~d. \\
\textbf{2020scb} & SN~Ic & \nodata & 0.018 & $-17.7$ & 56 & Visible for $\sim$130~d. \\
\textbf{2021zj} & SN~II & \nodata & 0.046 & $-18.6$ & 56 & Flash ionization followed by CSM interaction.\footnote{See \cite{Jacobson-Galan2024}} \\
\textbf{2020acty} & SN~Ib & \nodata & 0.047 & $-17.9$ & 55 & Visible for $\sim$70~d. \\
\textbf{2018lnb} & SLSN-II & \nodata & 0.222 & $-22.0$ & 52 & Incorrect host association. Visible for $\sim$175~d. \\
\textbf{2021axu} & TDE & \nodata & 0.192 & $-21.4$ & 51 & Incorrect host association.\footnote{Association is correct in current \texttt{GHOST} version.} Visible for $\sim$160~d. \\
\hline\\[-1.5ex]
\multicolumn{7}{l}{
\begin{minipage}{16cm}
NOTE: The bolded text designates a transient event that is likely anomalous, and we consider successfully tagged. We consider 26/29 (90\%) objects as anomalies due to their (updated) classification, or peculiar light curve behavior and/or host galaxy, as opposed to the 15/29 originally considered anomalous based solely on the original TNS classification labels.
\end{minipage}}
\end{tabular}
\label{tab:spec_test_set}
\end{table*}

When considering other distributions, such as redshift, we find that our anomaly detection algorithm (which does not use redshift as input), is agnostic. Unlike the IF model, our RFC model tags objects that encompass the entire redshift range of ZTF. We find $z=[0.011, 0.338]$, a mean redshift of $\tilde{z}=0.087$, a median redshift of $z_{med}=0.052$, and a standard deviation of $\sigma_{z}=0.079$. The bounds and standard deviation of the tagged redshift distribution implies that our model has learned to preferentially tag intrinsically bright (correlated with high redshift due to Malmquist bias, which favor SLSNe, TDEs) and intrinsically dim (overrepresented in nearby, low redshift galaxies, which favor SESNe) events. Meanwhile, the median redshift aligns with the median redshift of ZTF~BTS (See Figure~4 of \citealt{Fremling2020}), which is correlated with the redshifts of the majority of ``Normal" objects. Using redshift measurements, we can transform apparent magnitudes into absolute magnitudes using a flat $\Lambda$CDM cosmology with $H_0 = 70~\text{km s}^{-1} \text{Mpc}^{-1}$, and $\Omega_M$ = 0.3. We find that the tagged objects span the gamut of extragalactic transient peak absolute magnitudes: $M \in [-23.2, -15.2]$~mag, evidence that our model is able to find objects at all intrinsic brightness scales \citep{Richardson2014}. \par

\subsubsection{Spectroscopic, Contextual, and Behavioral Anomalies (After Vetting)} \label{subsubsec:spec_cont_beh_anom}

As we describe at the beginning of Section~\ref{sec:ztf_real_time}, we consider a transient to be an anomaly if it falls into any one (or more) of the three categories: spectroscopic anomaly, contextual anomaly, and behavioral anomaly. The former can be obtained via classification of the SN spectrum and without careful inspection of the light curve or host galaxy, except in the case of re-classification. Additionally, the latter two require such expert vetting to identify peculiar behavior (which can occur, e.g., after the spectrum was taken) or host galaxy environment. When we consider the results after expert vetting, our purity is greater than what we were led to believe from the TNS spectroscopic label alone. Objects that fall into any one or more of these three categories after careful manual vetting by an expert \textit{post facto} are in bold text in Table~\ref{tab:spec_test_set}. We consider all such objects a successfully tagged event by our model. Those without bold text are likely to be non-anomalous, and thus represent a ``misclassification". We offer several insights into common failings of our model in Section~\ref{sec:discussion}. \par

For completeness, we carefully inspect the light curves, spectra, and host galaxies of all 455 test set objects. This was done consulting photometry, spectra, host galaxy postage stamps, redshift estimates, catalog information, and other resources available on the TNS, ANTARES \citep{Matheson2021} and ALeRCE \citep{Forster2021} brokers, and YSE's transient survey management platform \texttt{YSE-PZ} \citep{Coulter2022_YSEPZ}. We find 16 transients originally categorized as ``Normal" that are updated to the anomalous ``Anomaly" classification, resulting in 68 total and a 31\% overall increase in anomalies. Of these 16 new anomalies, 11 (or $\sim$69\%) were presently tagged by our model as anomalous, of which we argue eight (SN~2021uzt, SN~2020aeuh, SN~2021fyp, SN~2021cpi, SN~2021ackd, SN~2021ckb, SN~2021adpx, and SN~2020ywo) were previously misclassified on TNS (and therefore, should be considered spectroscopic anomalies) and three can be characterized as behavioral anomalies: two (SN~2020abah, SN~2020rmk) as members of the rare, long-rising ($\textgreater$40~d) SN~II class ($\approx$1.4\% core-collapse~SNe. See \cite{Sit2022} for details), and one (SN~2021zj) with two peaks separated by 100~d, and exhibits flash ionization followed by CSM interaction \citep{Jacobson-Galan2024}. Note that all reclassifications were made with the original classification spectrum uploaded to TNS; we did not use any additional proprietary spectra. It follows that there remain five objects that were incorrectly not tagged as anomalous but appear to be of an anomalous nature. Specifically, SN~2021wun is also a known member of the long-rising SN~II class \citep{Sit2022}; SN~2020eyj has previously been described as having late-time CSM interaction \citep{Fremling2020}; SN~2021yfi has a peak absolute magnitude of $-21.3$~mag (but lacks narrow emission lines which prevented a superluminous~IIn classification\footnote{See classification report from \url{https://www.wis-tns.org/object/2021yfi}}); SN~2019bcv is uncommonly bright ($\sim-19.5$~mag), red ($g-r\approx1$~mag), and visible for $\sim$600~d; and SN~2021ttg is reclassified from SN~Ia to SN~Ia-91bg-like in this work, prompted by an approximate nearest neighbors search and re-vetting of the spectrum (see Table~\ref{tab:reclassification_SNe} and Section~\ref{subsec:reclass_sne}). \par

Now, \textit{post facto}, we find 26/29 objects successfully tagged anomalous, resulting in an effective purity of $\sim$90\%. Note that of the three remaining objects considered ``Normal", one (33\%) has an incorrect host association (SN~2021aff), which is a possible explanation for its misclassification.\footnote{We remind the reader that the host association is performed upstream, and as in the real-time data processing scenario, an expert cannot \textit{a priori} know the correctness of the association of a flagged anomaly before human-on-the-loop vetting. Erroneous associations (estimated to be $\sim$5\% by \citealt{Gagliano2021}) can be identified and dismissed by an expert, underscoring the indispensable role of human intervention in the analysis pipeline.} The second (SN~2020tnq) has no observations until approximately +26~d after peak (estimated from the evolution of the spectrum), possibly tricking our model into inferring the object is intrinsically fainter (and thus more likely to be of anomalous SESNe classes) than it is. The third (SN~2021our) is a normal SN~Ia from the spectrum in a standard elliptical galaxy, but is underluminous based on peak absolute magnitude calculations ($M_{abs}\sim-18.3$~mag), the strongest evidence for the misclassification. \par

After vetting and reclassifications, we update the per-type percentage of objects tagged by the AD model according to the updated spectroscopic label. The tagged objects of ``Normal'' spectroscopic classes drops from 9 SN~II to 5~SN~II (7.3\% of all SN~II in test set) and from 5 SN~Ia to 2 SN~Ia (0.06\% of all SN~Ia in test set). Similarly, for our anomalous classes, our model tagged 7 TDE (77.8\%), 3 SN~IIn (26.7\%), 2 SLSN-II (66.7\%), 2 SN~IIb (33.3\%), 1 SN~Ib (16.7\%), 1 SN~Ic (11.1\%), and 1 SN Ic-BL (20.0\%). Our model additionally tags 2/3 SN~Ia-91bg-like (66.7\%) and 2/3 SN~Ia-CSM (66.7\%), both classes that were identified through reclassification and not included originally in our training set. \par

Subsequently we show updated confusion matrices in Figure~\ref{fig:cm_context_behavior}, which describe the effectiveness of our model at classifying \emph{spectroscopic}, \emph{contextual}, and \emph{behavioral} anomalies within the observed ZTF Alert Stream. Now, we record an accuracy of 90\%. We score a ``Normal" completeness of 99\% at a ``Normal" purity of 90\% while achieving an ``Anomaly" completeness of 38\% at an ``Anomaly" purity of 90\%. As designed, our algorithm indeed prioritizes anomaly purity over sample recall, and is effective at identifying SNe with peculiar attributes in regards to the spectroscopic label and beyond. \par

\subsubsection{Additional Performance Evaluation}
\label{subsubsec:add_performance}
Beyond a confusion matrix, there are other methods to evaluate the performance of a binary classifier. With a RFC, the final prediction is an aggregate of the final prediction of each tree in the forest, and so the fraction of votes belonging to either ``Normal" or ``Anomaly" can be conceptualized as a classification probability. Traditionally, the final classification is assigned as a simple majority, where 50\% is the decision threshold, but this choice can be changed to reflect different aspects of the model. It follows that a higher decision threshold considers only events with near-unanimous decision by the trees in the forest, whereas a lower decision threshold is the opposite case; such a trade-off is construed by a Receiver Operating Characteristic (ROC) curve. The ROC curve measures the rate of true positives and false positives as a function of the decision threshold from 0 to 1, and the model accuracy represents a single point along the curve. Meanwhile, the Area Under the Curve (AUC) quantifies the separability of our two classes, where in the limit of perfect classification the AUC approaches unity. A high AUC indicates a high true positive rate and a low false positive rate, whereas a low AUC indicates a low true positive rate and a high false positive rate. \par

\begin{figure*}
    \centering
    \includegraphics[width=\columnwidth]{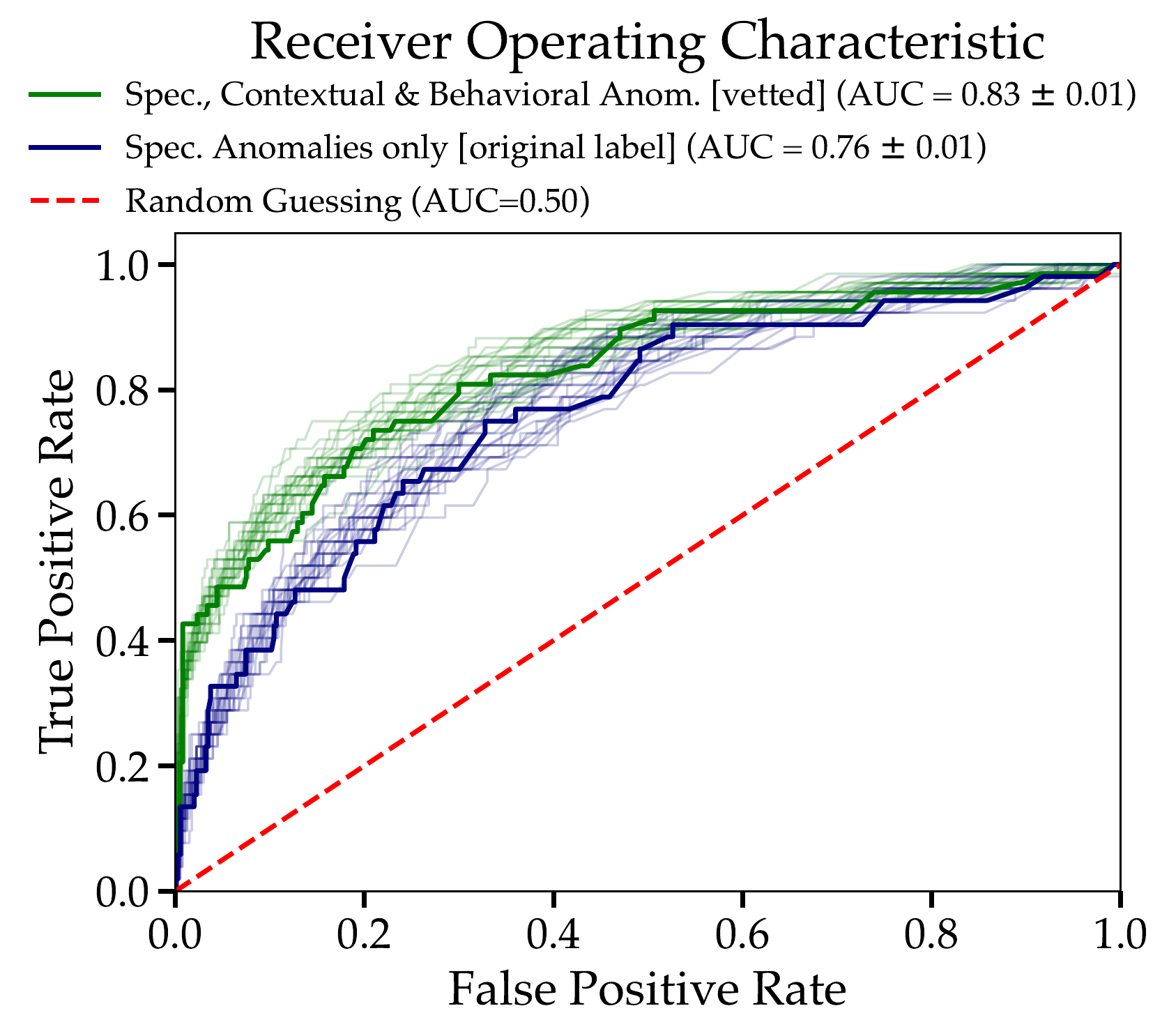}
    \includegraphics[width=\columnwidth]{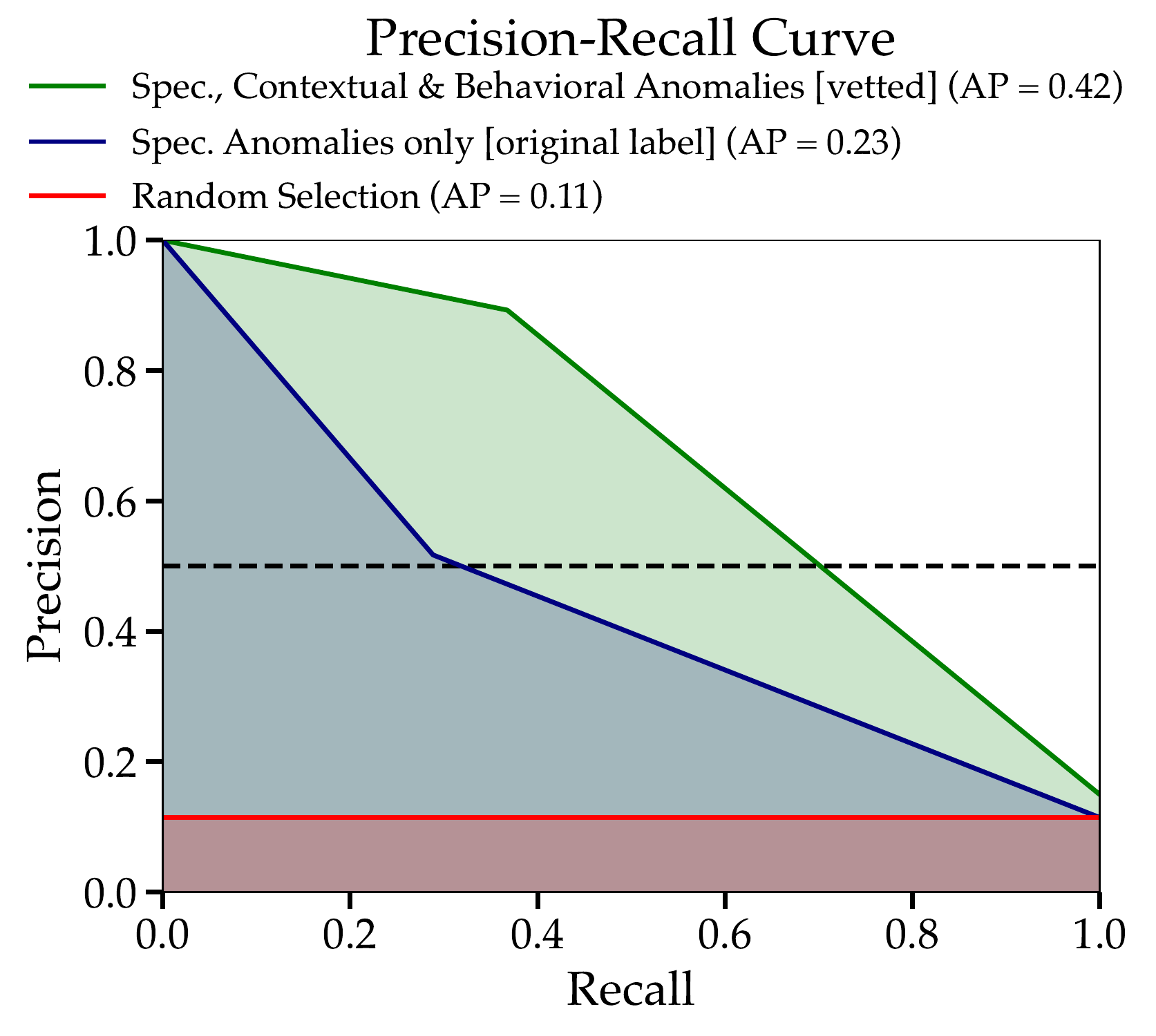}
    \caption{
    \textbf{\textit{Left:}} Receiver operator characteristic (ROC) curves for \emph{spectroscopic}, \emph{contextual}, and \emph{behavioral} anomalies after vetting (green solid line), \emph{spectroscopic} anomalies only before vetting (blue solid line), and by random selection (red dashed line). Each curve is a different initialization from one of 25 different random seeds, where the bold line is the chosen model's performance. The area under the curve (AUC) for the ``Anomaly" classification is listed along with its standard deviation across the 25 different random seeds. 
    \textbf{\textit{Right:}}
    Precision-Recall curves for \emph{spectroscopic}, \emph{contextual}, and \emph{behavioral} anomalies after vetting (green solid line), \emph{spectroscopic} anomalies only before vetting (blue solid line), and by random selection (red solid line), where any object with $P(anom)\textless50\%$ is cast to 0, and $P(anom)\geq50\%$ is cast to 1. The average precision (AP) score is shown in parentheses, and a 50\% purity is shown as a dashed black line. Our anomaly detection model significantly outperforms random selection, and can achieve high levels of purity ($\geq$50\%) at a range of recall thresholds ($\leq$70\% after vetting, $\leq$32\% before vetting).
    } 
    \label{fig:AUROC_PRcurves}
\end{figure*}

In the left panel of Figure~\ref{fig:AUROC_PRcurves}, we show the ROC curve for our model, reporting the AUC for only the anomalous ``Anomaly" class in the cases of random guessing (red), \emph{spectroscopic} anomalies only before vetting (blue), and any of \emph{spectroscopic}, \emph{contextual}, \emph{behavioral} anomalies after vetting (green). We also report the AUC standard deviation across our 25 different random seed iterations (shown as faded lines, with the bold line denoting our final model). We find by the spectroscopic label alone before vetting we achieve an AUC of $76\% \pm 1\%$, and for all anomaly categories after vetting we achieve an AUC of $83\%~\pm~1\%$. The greatest separation between these two anomaly criteria occur at the low true positive and low false positive rate regime, which indicates that the model's improvement primarily benefits in the detection of true positives (anomalies) whilst maintaining low false positives when vetted anomalies are considered. \par

In the right panel of Figure~\ref{fig:AUROC_PRcurves}, we show a Precision-Recall curve for our model performance. We consider any object with $P(anom)~\textless~50\%$ as ``Normal" and $P(anom)\geq50\%$ as ``Anomaly". At a $P(anom)=50$\% threshold, we achieve a purity of \emph{spectroscopic} anomalies of approximately 50\% (52\%), which optimizes the trade-off between precision and recall for our use-case---we achieve the maximum recall for which we identify more anomalies than non-anomalies. For this threshold visualized in our Precision-Recall curve, we cast any object with $P(anom)\textless50\%$ to $P(anom)=0\%$ and any object with $P(anom)\geq50\%$ to $P(anom)=100\%$. This manifests as an inflection point, occurring at the ``Anomaly" recall values as shown in the confusion matrices (29\% before vetting, 38\% after vetting). We show the the average precision (AP)\footnote{\url{https://scikit-learn.org/stable/modules/generated/sklearn.metrics.average_precision_score.html}} from prediction scores in parentheses, denoted as the weighted mean of precision values achieved at each threshold, using the increase in recall from the previous threshold as the weight. Our anomaly detection model significantly outperforms random selection, and can achieve high levels of purity ($\geq$50\%) at a range of recall thresholds: $\leq$32\% for \emph{spectroscopic} anomalies only before vetting, and $\leq$70\% for vetted \emph{spectroscopic}, \emph{contextual}, and \emph{behavioral} anomalies combined. For even higher levels of purity ($\geq$75\%), these recall thresholds are $\leq$15\% for \emph{spectroscopic} anomalies only before vetting, and $\leq$49\% for vetted \emph{spectroscopic}, \emph{contextual}, and \emph{behavioral} anomalies combined. \par

\begin{figure*}
    \centering
    \includegraphics[width=\columnwidth]{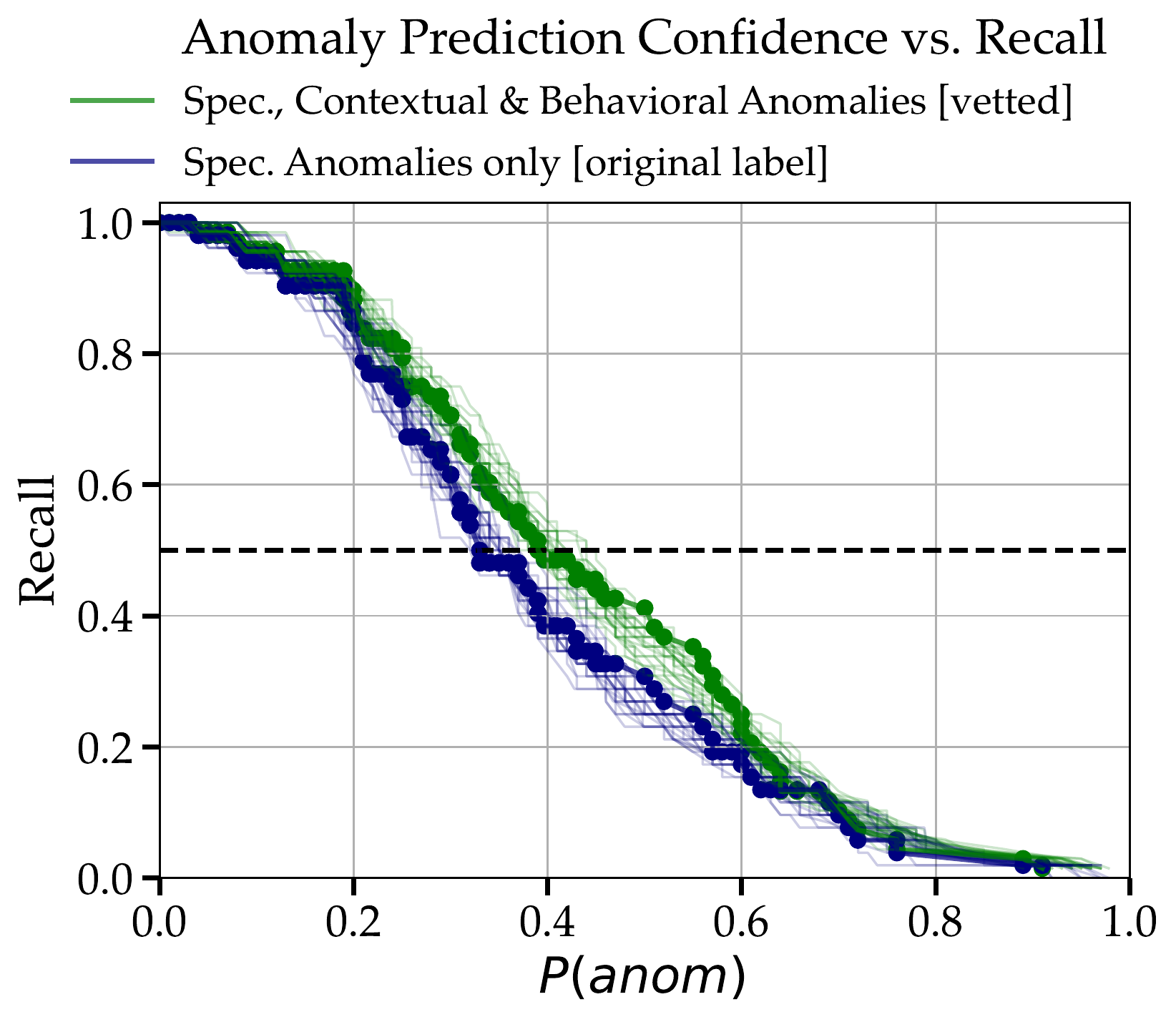}
    \includegraphics[width=\columnwidth]{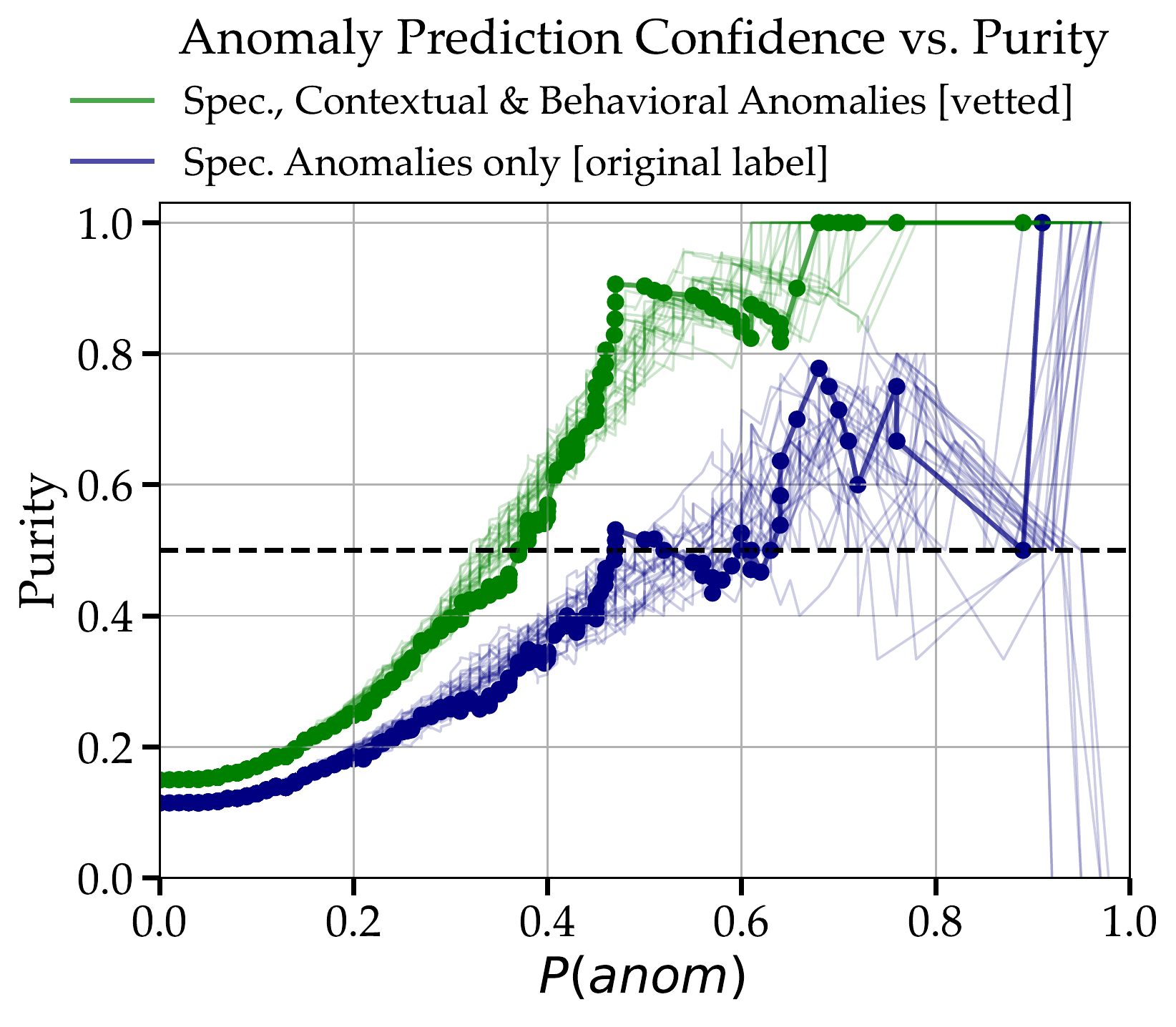}
    \caption{
    The observed completeness (left panel) and observed purity (right panel) of our model as a function of anomaly classification confidence, $P(anom)$, calculated at \emph{full} light curve phase. Individual object confidence scores are represented as circles. Each curve is a different initialization from one of 25 different random seeds, where the bold line is the chosen model's performance. At our chosen thresholds of $P(anom)=50\%$ for this work, we achieve an overall purity score of 52\% and a completeness of 29\% for \emph{spectroscopic anomalies} only before vetting (blue) and an overall purity score of 90\% and a completeness of 38\% for \emph{spectroscopic}, \emph{contextual}, and \emph{behavioral} anomalies after vetting (green), reflected in the confusion matrices of Figures~\ref{fig:cm_spec_anomalies},~\ref{fig:cm_context_behavior}, respectively.
    } 
    \label{fig:Panom_vs_PR}
\end{figure*}

Ultimately, we want to better understand on a granular scale how reliably we can trust the model output at varying confidence scores, particularly to investigate if higher confidence correlates in the increased likelihood of an object being anomalous. In Figure~\ref{fig:Panom_vs_PR}, we show the observed completeness (left panel) and observed purity (right panel) of our model as a function of anomaly classification confidence, $P(Anomaly)$ (hereafter $P(anom)$). Because of our hyperparameter selection, there are fewer high confidence scores ($P(anom)\geq50\%$) than low ones ($P(anom)~\textless~50\%$), and of those which have $P(anom)\geq50\%$, the majority are grouped within the $50\%-70\%$ range, with only a few 70+\% (see Column~6 of Table~\ref{tab:spec_test_set}). \par

In our test set before vetting, every object with $P(anom)\geq70\%$ is anomalous except one, SN~2021uzt, which has the overall second highest anomaly score at $P(anom)=89\%$. A possible driver of the high anomaly score is the likely incorrect host association for this object (the real host is not visible). SN~2021uzt was originally classified as an SN~IIn \citep{TuckerTNS2021uzt}, but later classified as an SN~II \citep{ChuTNS2021uzt}. Due to this, SN~2021uzt is represented as the sharp dip shown in the right panel in blue, and is the sole reason for the decline in purity at high confidence scores. However, upon vetting and re-evaluating the two spectra, there is evidence of a IIn-like Lorentzian H$_{\alpha}$ profile and none of P-cygni. Moreover, SN~2021uzt is bright at its peak ($M_{abs}\sim-19.4$~mag) with a long rise (\textgreater30~d). This evidence suggests the original IIn classification is the best characterization, and we adopt it for this work. Thus, after vetting (green), all objects with $P(anom)\geq70\%$ are anomalous. \par 

Averaged over the 25 different seed iterations, the purity of vetted anomalies steadily increases with rising confidence score from 0\% to a peak around $P(anom)\approx45\%$, plateaus until $P(anom)\approx65\%$, then rises to a perfect purity at $P(anom)\approx70\%$ and sustains it until a maximum classification confidence\footnote{As we will see in Table~\ref{tab:ysedr1_real_time_any_phase}, $P(anom)\geq70\%$ does not guarantee perfect purity; however, in the case of Table~\ref{tab:ysedr1_real_time_any_phase}, our model is not trained on features extracted from YSE photometry and we consider the score at \emph{any} light curve phase instead of at the end of the full phase.}. If $P(anom)\geq47\%$, the purity achieved is $\geq80\%$ (the nearby $P(anom)=50\%$ is reflected in the purity confusion matrix of Figure~\ref{fig:cm_context_behavior}). This trend shares similarities when considering only \emph{spectroscopic} anomalies (before vetting) except the difference that stems from SN~2021uzt and that the maximum purity that can be achieved is overall lower; if $P(anom)\approx47\%$, the purity achieved before vetting is nearly half that compared to vetted anomaly candidates, achieving $\sim40\%$ at worst (but usually $\geq50\%$). Perhaps the most interesting insight gleaned from Figure~\ref{fig:Panom_vs_PR} is that if we vet our anomaly candidates, we achieve a 50\% purity score at a lower $P(anom)$ threshold (38\%) than if we do not (47\%). Fluctuations in the model from random seed iterations impact the outcome of the anomalies before vetting more severely than after vetting, likely due the fact that more objects with high anomaly scores are deemed anomalous after vetting than vice-versa. \par

In a similar manner, the recall steadily decreases with increasing confidence score across the range of recall scores, alluding to the fact that the majority of objects have low anomaly confidence scores (because most are truly non-anomalous), and a minority have high anomaly scores. This behavior reflects the distribution of anomalies we observe. There is little difference in the recall score before and after vetting at the $P(anom)$ margins; the greatest gain ($\approx10\%$ in recall) comes around $P(anom)\sim50\%$, though we observe that this large offset is in part attributed to fluctuations from the random seeding. \par

For our chosen model we achieve a maximum purity of 100\% at a recall of 13\% given a confidence score of $P(anom)=68\%$. But the best balance achieved between purity and recall is at $P(anom)=50\%$, which is reflected throughout this work. \par

\subsection{Feature Importance} \label{subsec:feature_importance}

\begin{figure*}
    \centering
    \includegraphics[width=16cm]{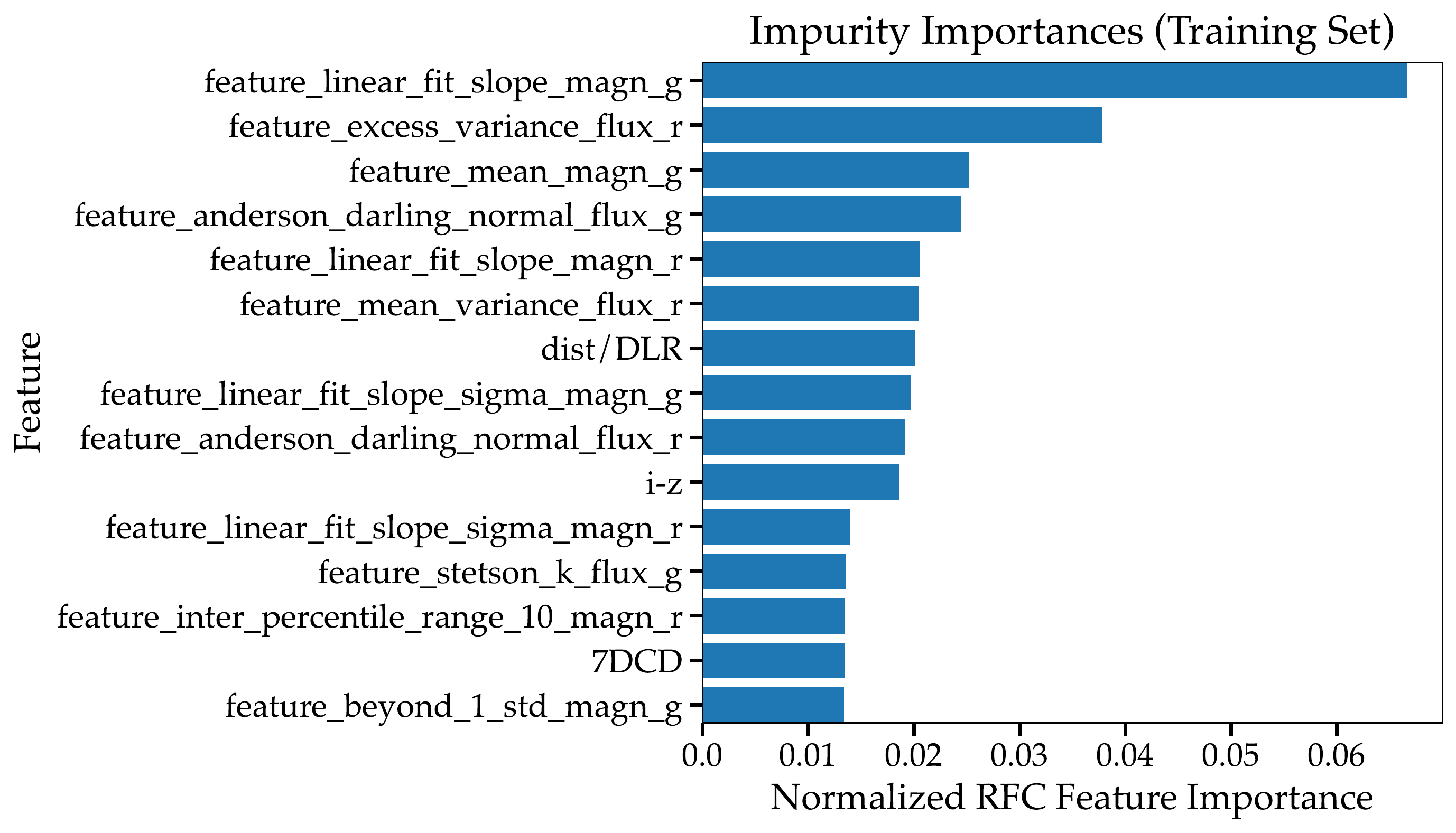}
    \caption{
    The top 15 light curve and host galaxy features, ordered by the greatest normalized impurity-based feature importance in our binary ``Normal" vs. ``Anomaly" task as determined by our final Random Forest Classifier model. The features with highest importance can be broadly categorized by light curve slope fit and error, $g$-band brightness and its variance, transient radial separation, and host galaxy color. The high importance of such host galaxy features is consistent with findings from \cite{Gomez2020} and \cite{Gagliano2021} despite their different classification tasks (SLSNe vs. non-SLSNe, and SNe~Ia vs. CC~SNe, respectively). 
    } 
    \label{fig:RFC_feat_importance}
\end{figure*}

To better understand our model, we desire to determine which light curve and host galaxy features (see the complete list in Appendix~\ref{appx:lc_features}) are most valuable for our anomaly detection task. A common method is to calculate impurity-based feature importances, computed as the (normalized) total reduction of the criterion brought by that feature (Gini importance). We use the \texttt{feature\_importances\_} attribute from \texttt{sklearn.ensemble.RandomForestClassifier}, and display the results in Figure~\ref{fig:RFC_feat_importance}. As is standard procedure, we use the normalized feature importance, defined as the percentage of times the feature is used as a split, as our metric for feature significance. \par

Under this method, we find that the most important light curve features are the light curve slope in a least squares fit of the linear stochastic model with Gaussian noise described by observation errors \{$\delta_{i}$\} (\texttt{feature\_linear\_fit\_slope\_magn\_\{g,r\}}) and its error in $g$-band (\texttt{feature\_linear\_fit\_slope\_sigma\_magn\_g}), the mean and excess variance $r$-band flux (\texttt{feature\_excess\_variance\_flux\_r}, \\ \texttt{feature\_mean\_variance\_flux\_r}), mean magnitude in $g$-band (\texttt{feature\_mean\_magn\_g}), and the unbiased Anderson–Darling normality test statistic for flux (\texttt{feature\_anderson\_darling\_normal\_flux\_\{g,r\}}). We suspect that the \texttt{feature\_linear\_fit\_slope\_magn\_g} at 0.067 is the most important feature overall at nearly twice the contribution of the second most important feature (0.038; \texttt{feature\_excess\_variance\_flux\_r}) because many rarer classes of transients like TDEs show consistently strong blue colors or SESNe which tend to exhibit constantly weak blue colors relative to red throughout the light curve evolution. \par

Interestingly, like FLEET \citep{Gomez2020, Gomez2023SLSN, Gomez2023TDE} which is designed to find SLSNe and TDEs with a RFC, we find the most important host galaxy feature is the projected angular separation between the transient and host galaxy normalized by host radius (\texttt{dist/DLR}, denoted as $\theta$/$d_{DLR}$ in the original \texttt{GHOST} paper; \citealt{Gagliano2021}). This is an indication that SN classes, particularly rare ones, may preferentially occur at different locations throughout their host galaxies than more normal SNe. For example, TDEs are nuclear events, and would predominantly have a small \texttt{dist/DLR} value, making this potentially a powerful indicator for separating TDEs from non-TDEs. \par

The second most important host feature is the aperture magnitude $i-z$ color difference (\texttt{i-z}), followed by \texttt{4DCD}, the 4-dimensional color distance in $g-r$, $r-i$, $i-z$, and $z-y$ from the PS1 stellar locus \citep{Tonry2012}. This is the path traced by stars in color-color space (see Section~2.2 of \cite{Gagliano2021} for details). Color-derived features encode information about the metallicity, mass, and star formation rate of host galaxies. These characteristics are known to be associated with the type of SN (e.g., \citealt{Hansson2012}), including host galaxy property correlations with SNe~Ia \citep[e.g.,][]{Johansson2013, Henne2017, Kelsey2023}. Furthermore, the color of galaxies can effectively distinguish between early and late-type galaxies \citep{Strateva2001, Nair2010}. This suggests that valuable host galaxy features for classifying SNe, and potentially anomalous SNe, are related to previously established galaxy correlations.  \par

\begin{figure*}
    \centering
    \includegraphics[width=16cm]{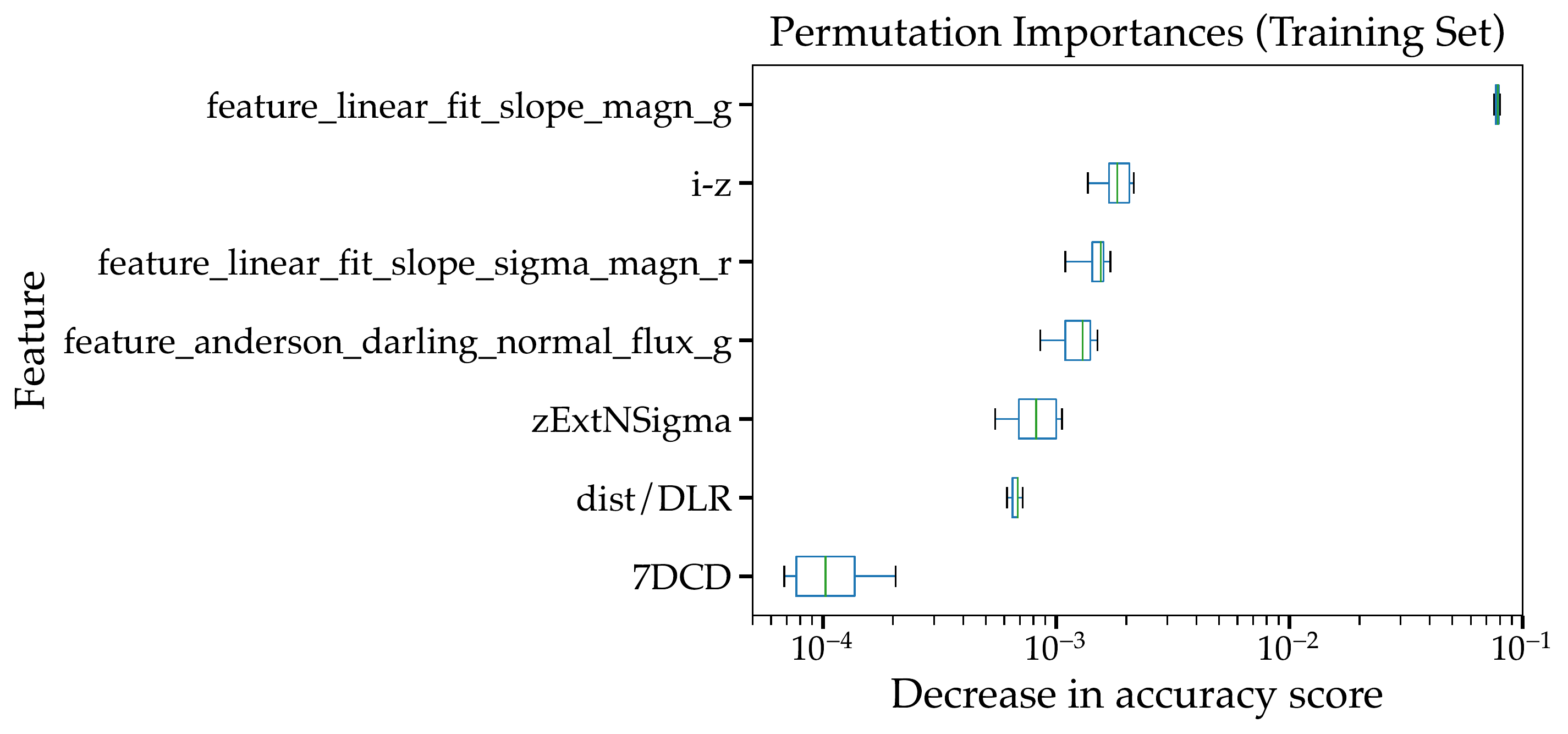}
    \caption{
    A boxplot of the top 7 most important features from our training set with the highest permutation importance using the method from \cite{Breiman2001}. We use 10 permutations per feature, finding the features that most negatively impact accuracy. All features except \texttt{zExtNSigma} are shown as among the most important identified by the impurity-based feature importance metric (see Figure~\ref{fig:RFC_feat_importance}). The light curve feature \texttt{feature\_linear\_fit\_slope\_magn\_g} remains the singular most important feature for identifying anomalies for this work.
    } 
    \label{fig:RFC_feat_permutation}
\end{figure*}

Moreover, our host galaxy feature importance results align very well with \cite{Gagliano2021} when considering radial offset and color-derived features. They used a gradient boosting model to classify SN~Ia vs. core-collapse SNe from host galaxy information alone derived from \texttt{GHOST} host-galaxy associations, finding their 10 most important features for classification were of the same three main categories: radial offset, including \texttt{$\theta$} and \texttt{$\theta$/d$_{DLR}$} (their \#1 and \#2 most important features, respectively); color-derived features, including \texttt{4DCD}, \texttt{g-r}, \texttt{g-i}, \texttt{r-i}, and \texttt{i-z} (their \#7, \#3, \#4, \#5, and \#9 most important features, respectively); and morphological features, including \texttt{momentXX} in $g$ and $i$ and \texttt{ExtNSigma} in $g$ (their \#10, \#8, \#6 most important features, respectively). \par

In this work we do not use \texttt{$\theta$}, \texttt{g-r}, \texttt{g-i}, \texttt{r-i}, but we do use \texttt{dist/DLR} (\texttt{$\theta$/d$_{DLR}$}), \texttt{i-z}, \texttt{4DCD}, \texttt{\{g,r\}momentXX}, and \texttt{gExtNSigma}. For us, \texttt{dist/DLR} is our most important host galaxy feature, and \texttt{4DCD} is our second most important, with \texttt{rmomentXX}, and \texttt{gExtNSigma} being in the top half of host galaxy features, and \texttt{gmomentXX} being in the lower half of important host features. Thus, even though our classification task is different from that of \cite{Gagliano2021}, there is great overlap in the most important features chosen in regards to radial offset (\texttt{dist/DLR}) and color-derived features (\texttt{i-z}, \texttt{4DCD}). Further exploration and quantization of the usefulness of these features is an area of future work. \par

However, it is known that impurity-based feature importances can be misleading for high cardinality features and artificially inflate numerical feature importance\footnote{See \url{https://scikit-learn.org/stable/modules/generated/sklearn.ensemble.RandomForestClassifier.html\#sklearn.ensemble.RandomForestClassifier.feature\_importances\_}}. Because of the strong correlations between the majority of host galaxy features, the variable importances returned by our RFC model may not reveal the most valuable features for supernova classification, or paint the entire picture. Thus, we opt to additionally investigate feature importance using the the permutation importance method described in \cite{Breiman2001}. Permutation importance measures the impact of each feature on the model's performance by evaluating how much the model's accuracy (or another evaluation metric) degrades when the values of that feature are randomly shuffled. Features that have a high impact on the model's accuracy will result in a significant decrease in accuracy when permuted. Therefore, it provides a measure of the feature's importance in making accurate predictions. We display the top 7 features with greatest permutation importance in Figure~\ref{fig:RFC_feat_permutation}. \par

Six of the seven top features identified by permutation importance align well with those identified by impurity-based importance. Specifically, \texttt{feature\_linear\_fit\_slope\_magn\_g} remains the most important by a wide margin over every other feature, indicating that particular $g$-band slope profiles are the best indicator of anomalous activity. Light curve features \texttt{feature\_linear\_fit\_slope\_sigma\_magn\_r} and  \texttt{feature\_anderson\_darling\_normal\_flux\_g} remain as among the most important. As before, radial offset (\texttt{dist/DLR}) and color-derived features (\texttt{i-z}, \texttt{4DCD}) host galaxy features are vital. Meanwhile, the only feature flagged as important by permutation and not by impurity methods is \texttt{zExtNSigma}. Because of our highly dimensional feature space, the individual impact of any one feature on the final anomaly classification (besides these top few) is small. \par

\subsection{Filter Deployment, Real-time Anomaly Detection \& Follow Up} \label{subsec:follow_up}


After demonstrating the success of our anomaly detection model, we deploy it in a real-time scenario: the ZTF Alert Stream. We compose a \emph{Filter} on the ANTARES broker, which first requires objects to have been tagged successfully upstream by the \texttt{lc\_feature\_extractor} \emph{Filter}. Then, we perform the same pre-processing cuts to initially create our databank as outlined in Section~\ref{subsec:preprocess}, except that we do not query PS1-PSC to save on computational time in a real-time scenario. 
We then apply our trained RFC model. Calculated light curve features, host galaxy features, and anomaly score are saved to the tagged loci as Locus Properties. \par

All objects that successfully pass through the \emph{Filter} are tagged as ``\texttt{LAISS\_RFC\_AD\_filter}" regardless of anomaly score. However, those with $P(anom)\geq50\%$ (or any threshold of a user's choosing) at any point during their light curve evolution are processed by our Slack API App\footnote{\url{https://api.slack.com}} as a bot that sends out a notification of the IAU name, ANTARES Locus ID, TNS spectroscopic classification (if it exists), and the current anomaly score. An expert then manually inspects each anomaly candidate. Our bot queries the ANTARES broker using the \texttt{antares\_client} package \citep{Gitlab:antares-client}, and collates all objects tagged by \texttt{LAISS\_RFC\_AD\_filter} within any number of days of the expert's choosing. From there, we apply Wide-field Infrared Survey Explorer (WISE, \citealt{Wright2010}) color selection criteria for active galactic nuclei (AGNs) from \cite{Jarrett2011} and \cite{Stern2012} to remove likely AGN candidates with high anomaly scores. We typically run the bot once a day to vet the previous nights' up-to-date and active anomaly candidates. \par

Our \emph{Filter} was deployed on 2023~August~22. The data cutoff for this work for our real-time deployment is 2023~October~22, two months since deployment. In that short time, \laiss{} has successfully processed $\sim$1,200 loci, of which 45 have achieved an anomalous classification of $P(anom)\geq50\%$ and are listed in Table~\ref{tab:real_time_any_phase}, ordered by maximum anomaly score, $max(P(anom))$. In total, 39 are transients (21 with spectroscopic classifications) and 6 are AGN or AGN candidates (1 spectroscopically-confirmed). \par

\begin{table*}[ht]
\footnotesize
\centering
\caption{\textbf{Real-time search:} \emph{Transient-only} loci with $max(P(anom))\geq50\%$ at \emph{any} light curve phase}
\begin{tabular}{lllccl}
\multicolumn{6}{c}{45 total tagged loci, \texttt{LAISS\_RFC\_AD\_filter} Filter, ordered by $max(P(anom))$.} \\
\hline
\hline
\multicolumn{6}{c}{Spectroscopic} \\
\hline
\hline
ZTF ID & IAU Name & Spec. Class & $z$ & $max(P(anom))$ & Remarks \\
\hline
\textbf{ZTF22aadesap} & 2022fpx & TDE & 0.073 & 0.92 & Tagged anomalous 39~d before SGLF's spectrum. \\
\textbf{ZTF23aatzhso} & 2023oom & Nova & --- & 0.87 & In M31. \\
\textbf{ZTF23aaspcfl} & 2023nlj & SN~Ib & 0.03 & 0.83 & Tagged anomalous 16~d before ZTF's spectrum. \\
\textbf{ZTF23aanptpp} & 2023koq & SLSN-II & 0.104 & 0.77 & Tagged anomalous at peak. \\
\textbf{ZTF23aarktow} & 2023myo & SN~IIb\footnote{We reclassify from Type II to IIb.} & 0.035 & 0.74 & Reclassified. First ZTF-$r$ epoch is shock cooling. \\
\textbf{ZTF22aaetqzk} & 2022gzi & SN~IIn & 0.089 & 0.72 & \nodata \\
\textbf{ZTF23aamsetv$\ddagger$} & 2023kvy & TDE & 0.16 & 0.67 & Tagged on minimum required ZTF-$g,r$ obs. \\
\textbf{ZTF23abcinum$\dagger$} & 2023sds & SN~II & 0.065 & 0.67 & YSE target. In elliptical galaxy? \\
\textbf{ZTF23aajkisd} & 2023iex & SN~IIn & 0.029 & 0.66 & \nodata \\
\textbf{ZTF23aalgqsq} & 2023jdh & SN~IIn & 0.054 & 0.65 & Faint host. \\
\textbf{ZTF23aatdcey$\dagger$} & 2023nof & SN~IIn & 0.069 & 0.65 & With our spectrum, reclassify from Type II to IIn. \\
ZTF23aapgswu & 2023mcs & SN~Ia & 0.03 & 0.64 & First epochs are $\approx$+20~d after peak. \\
\textbf{ZTF23aatcsou$\dagger$} & 2023nwe & SN~IIn & 0.194 & 0.64 & Tagged anomalous 14~d before our spectrum. \\
\textbf{ZTF23aagpjyp} & 2023ggb & SN Ia-CSM & 0.08 & 0.59 & \nodata \\
\textbf{ZTF23aavtugd$\dagger$} & 2023omf & SN~IIn & 0.083 & 0.57 & Incorrect host. Match to SN~1996L. \\
\textbf{ZTF23abhafym} & 2023tsw & SN~Ia-91bg-like\footnote{We reclassify from Type Ia to Ia-91bg-like.} & 0.05 & 0.57 & Reclassified. Red, rapid fading favors 91bg-like. \\
\textbf{ZTF23aatabje$\dagger$} & 2023ocx & SN~Ia-CSM & 0.076 & 0.56 & Tagged anomalous 22~d before our spectrum. \\
ZTF23aawblmi$\ddagger$ & 2023otw & SN~II & 0.087 & 0.60 & Incorrect host.\footnote{Association is correct in current \texttt{GHOST} version.} \\
ZTF23aberpzw$\dagger$ & 2023swf & SN~II & 0.024 & 0.53 & Only 1 anomalous epoch. \\
\textbf{ZTF23aamsekn} & 2023khp & SN Ia-CSM & 0.09 & 0.52 & Incorrect host.\footnote{Association is correct in current \texttt{GHOST} version.} \\
ZTF23abayyjm$\dagger$ & 2023sed & SN~Ia & 0.14 & 0.50 & Only 1 anomalous epoch. Spiral host. \\
\hline
\hline
\multicolumn{6}{c}{Photometric} \\
\hline
\hline
\textbf{ZTF21abiggqx$^{\#}$} & 2021rjf & SN~IIn? & --- & 0.80 & Long-lived (800~d) CSM interaction? \\ 
\textbf{ZTF23aaveoxd} & 2023ofr & SN~IIn? & --- & 0.76 & Faint host. FLEET=82\% SN~II. \\
\textbf{ZTF23aaqbyzr} & 2023mic & SN~Ia-CSM? & --- & 0.74 & YSE target. Second peak likely CSM interaction. \\
ZTF23abhegfd & 2023tjc & SN~Ia? & --- & 0.66 & FLEET=59\% SN~I. SALT3 $c=-0.3$. \\
\textbf{ZTF23aaqqeek} & 2023mne & SLSN? & --- & 0.63 & Faint host. FLEET=82\% SLSN-II. \\
ZTF23abevrtm & 2023tim & SN~Ia? & --- & 0.63 & FLEET=76\% SN~I. SALT3 $c=-0.22$. \\
\textbf{ZTF23abbbypt} & 2023sap & SLSN? & --- & 0.61 & FLEET=74\% SLSN-II. \\
ZTF23abaurik & 2023she & SN~IIP? & --- & 0.60 & FLEET=75\% SN~II. \\
ZTF23abekzca & 2023tcq & SN~Ia? & --- & 0.59 & YSE target. FLEET=85\% SN~I. SALT3 $c=-0.15$. \\
\textbf{ZTF23aazfibd} & 2023puf & SN~II/IIn? & --- & 0.58 & Faint host. FLEET=90\% SN~II. \\
\textbf{ZTF23aasotjh} & 2023nie & SN~II? & --- & 0.57 & Peculiar rise. FLEET=80\% SN~II. \\
ZTF23abetluh & 2023tdy & SN~Ia? & --- & 0.57 & FLEET=46\% SN~I. SALT3 $c=-0.3$. \\
\textbf{ZTF23aaxbkgs} & 2023pdf & SN~II/IIn? & --- & 0.55 & Faint host. FLEET=41\% SN~II. \\
ZTF23abdpgvv & 2023sws & SN~Ia? & --- & 0.55 & Incorrect host. SALT3 $c=-0.27$. \\ 
\textbf{ZTF23aaufkak} & 2023nwk & SN~IIn? & --- & 0.54 & Faint host. FLEET=76\% SN~II.\\
\textbf{ZTF23abedgfr} & 2023syt & SN~Ib/c? & --- & 0.54 & Poor SALT3 fit. FLEET=52\% SN~I. \\
\textbf{ZTF23aaewyhm} & 2023gzn & SN~IIn? & --- & 0.50 & Behind Sun for $\approx$150~d. FLEET=56\% SN~II.\\
ZTF23abeujrk & 2023tnr & SN~Ia? & --- & 0.50 & FLEET=88\% SN~I. SALT3 $c=-0.29$. \\
\hline\\[-1.5ex]
\multicolumn{6}{c}{
\begin{minipage}{16cm}
NOTE: The bolded text designates a transient event that is likely anomalous, and we consider successfully tagged.\\
NOTE: FLEET does not distinguish between Type II and IIn.\\
$\dagger$Targets for which we got spectra and posted their (updated) spectroscopic classification to TNS.\\
$\ddagger$Spectroscopic follow-up in coordination with FLEET \citep{Gomez2023TDE} program.\\
$^{\#}$In the databank used for train/test split (see Table~\ref{tab:dataset_bank}).\\
*Filter tagged 1 known AGN (ZTF20acvfraq/ AGN~2020adpi) and 5 likely AGN (ZTF22abghche/ AT~2023tsa, ZTF23aaloouf, \\ ZTF22abplfmz/AT~2023gld, ZTF23aaxazht, ZTF23aaqxgan/ AT~2023tpk) which are not shown in this Table.
\end{minipage}}
\end{tabular}
\label{tab:real_time_any_phase}
\end{table*}

For the spectroscopic sample, we consider 17/21 (81\%) to be \emph{spectroscopic}, \emph{contextual}, or \emph{behavioral} anomalies (which aligns closely to the 86\% purity of our spectroscopic test set), marked by bold text. Note that the objects here are a mix of those that were already classified and active at the time of \emph{Filter} deployment (e.g., TDE~2022fpx), those which were tagged by our model as anomalous before a classification spectrum was acquired by others (e.g., SN~Ib~2023nlj), and those whose spectra were acquired by us---indicated by a dagger symbol (e.g., SN~Ia-CSM~2023ocx). We report new and/or updated classifications from our acquired spectra to TNS. \par

The distribution of objects is as follows: 6 SN~IIn, 3 SN~Ia-CSM, 1 M31 Nova, 2~TDE, 1 SLSN-II, 1 SN~Ib, 1~SN IIb, 1 SN-Ia-91bg-like, 3~SN II, and 2 SN~Ia. This is further evidence that our algorithm is capable at identifying an array of different spectroscopic anomalies with high purity in practice, with a preference at identifying long lived CSM-interacting events like SNe~IIn and SNe~Ia-CSM. Moreover, as we identified in the spectroscopic test set, our algorithm tags anomaly candidates at all common redshift ranges observable by ZTF: $z~\in~[0.029, 0.194]$. Likewise, these objects are varied in their apparent and intrinsic brightness, host galaxy type, morphology and apparent size, with no obvious difference compared to the test set. \par

Of our nine obtained spectra, we consider six to be anomalous. In fact, one was used to aid in the reclassification of a normal Type II to a IIn, as in the case of SN~2023nof \citep{Aleo2023nofTNS}. The three remaining are of a normal Type Ia and two Type II SNe, though two were only (marginally) above the anomaly score threshold for one epoch in its light curve evolution, and the other had an incorrect host association that may have played into the misclassification. Generally we find throughout this work that objects with higher anomaly scores and/or are above the $P(anom)=50\%$ anomaly threshold for more epochs are more likely to be anomalous, supported in part by the evidence from our purity as a function of prediction confidence as shown in Figure~\ref{fig:Panom_vs_PR}. \par

Three objects have likely incorrect host associations from \texttt{GHOST} (SN~2023omf, SN~2023khp, SN~2023otw), though oddly two of three are of anomalous classes. This could indicate that in some cases, truly anomalous phenomena as captured by the light curve evolution alone (or the pairing of the light curve evolution with an incorrect host that happens to be atypical for that light curve) is sufficient to be flagged anomalous. Although we make no formal analysis of misclassified hosts, \cite{Gagliano2021} estimates that the misassociation rate of \texttt{GHOST} is approximately 5\%. Continual improvements to the pipeline and addition of catalogs such as GLADE \citep{Dalya2018} since the original release likely will decrease the rate of misassociations. \par

Perhaps the most promising revelation from our real-time deployment beyond the high purity is that occasionally our algorithm flags a transient as anomalous \emph{well before ($\gtrsim$2 weeks) the first classification spectrum}. When we re-extract the light curve features for all epochs of all tagged transients and re-apply our model throughout its light curve evolution as a mock up for real-time deployment, we can investigate when these candidates \emph{would have been tagged} had our model been deployed before the transient was active. \par

Of our spectroscopic subset, two (TDE~2022fpx, SN~Ib 2023nlj) were tagged 39~d and 16~d before their respective classification spectra. With earlier anomaly classifications, we can more closely study the physics linked to its explosion \citep{2022Gagliano_CCA}, as well as more completely follow-up the most interesting events with spectrophotometric resources and create a fuller picture of the spectral energy distribution (SED) evolution \citep{Pierel2018, Vincenzi2019}. SED templates enable the construction of models and simulations that better capture genuine anomalies, supplementing existing idealized models focused on typical phenomena. Realistic simulations of anomalies can enhance machine learning algorithm training in classification and anomaly detection tasks---contrasting with the tendency of algorithms trained on idealized simulations to mischaracterize anomalies (see, e.g., \citealt{Muthukrishna2022, Aleo2023}). \par

For the photometric sample, unfortunately there is no way to confirm the anomalous nature of 18 anomaly candidates. However, we have proffered potential classifications and anomaly candidates for completeness. Potential classifications in Tables~\ref{tab:spec_test_set}, \ref{tab:real_time_any_phase},~\ref{tab:10k_full_phase},~\ref{tab:ysedr1_real_time_any_phase}, come from a combination of FLEET \citep{Gomez2020} predictions, the YSE~DR1 ParSNIP and SuperRAENN classifiers (\citealt{Aleo2023}, adapted from \cite{Boone2021} and \cite{Villar2020}, respectively), SN~Ia SALT3 fits \citep{Kenworthy2021}, and expert analysis. Although we cannot use the photometric subset to quantify our algorithm's results, we make an inference at their anomalous nature by highlighting in bold text those that are strong anomaly candidates (as we do with the spectroscopic subset). \par

Despite the lack of a classification spectrum, some objects stand out as likely anomaly candidates. For instance, AT~2021rjf, which exhibited no previous variability, shows signs of a normal SN~II light curve followed by a second bump likely powered by CSM interaction that has lasted 800~d and counting. AT~2023mic also appears to have prolonged CSM interaction. The rest are predominantly SN~II/IIn or SLSN candidates in faint host galaxies. \par

Our \texttt{LAISS\_RFC\_AD\_filter} \emph{Filter} continues to run on ANTARES. The full list of processed objects regardless of anomaly score can be found at \url{https://tinyurl.com/LAISSrfcADfilter}. \par

\subsection{Retrospective Characterization of the \texttt{iso\_forest\_anomaly\_detection} \emph{Filter}} \label{subsec:iso_forest_ad_filter}

Our anomaly detection model can also be used for retrospective characterization, similar to that of our spectroscopic test set (Table~\ref{tab:spec_test_set}). This enables us to potentially unearth new or overlooked anomalous objects worthy of study or attention. For this assessment, we run \laiss{} on 10,000 random loci tagged (out of over 25,000) by ANTARES' first anomaly detection \emph{Filter}, \texttt{iso\_forest\_anomaly\_detection}, and report objects whose maximum anomaly score is $\geq50\%$ at full light curve phase in Table~\ref{tab:10k_full_phase}. \par

The \texttt{iso\_forest\_anomaly\_detection} \emph{Filter} is simpler than \laiss{}. It aims to tag transient events like supernovae, cataclysmic variables, and weird or rare events/objects in the night sky. Originally deployed on 5 April 2021, it is an isolation forest algorithm \citep{Liu2012} trained on 1,000,000 random ANTARES loci using 1000 trees on 106 total light curve features\footnote{These 106 include all of the light curve features used in \laiss{} and more---for a comprehensive list see \url{https://antares.noirlab.edu/properties}.}, 53 for both ZTF-$g$ and ZTF-$r$ passbands. The only selection cuts used enforced that the loci does not reside in the galactic plane ($|b| \geq 10\degree$), and does not have its strongest period between 100 and 1000 days (e.g., to weed out the majority of Mira star contaminants). Thus, it uses no contextual host galaxy information, or selection cuts to reduce AGN, quasi-stellar objects (QSOs), or other variable stars. Approximately 1 out of 3 tagged objects are on TNS (reported by ANTARES or other teams), and the rest are contaminants. And the vast majority of tagged transients are normal SNe~Ia. \par

Despite the quantity of non-transients and those that are not successfully processed by the \texttt{LAISS\_RFC\_AD\_filter} \emph{Filter} (e.g., the locus is a star, the associated host is not found or does not have PS1-$grizy$ color information needed for our model, etc.) we do find spectroscopically-confirmed and photometric anomalies. \par

\texttt{LAISS} tags 10 objects, of which 6 fall into the spectroscopic anomaly category only: 5 SN~IIn and 1~SLSN-I. Because one was previously used in the training of our model (SN~2021qep), we exclude this object and say that 5 of 9 are anomalous, or 56\% purity. By happenstance, two of these objects (SN~2022vmg and SN~2023nwe) we had confirmed with spectra from previous follow-up campaigns. Of the four spectroscopic non-anomalies, two have likely incorrect host associations (SN~2023dgp, whose potential hosts are vastly different: either a small elliptical or large nearby spiral, and SN~2023eqx, which is between an irregular faint host or small elliptical), two have no rise information (SN~2023dgp as stated before, SN~2023cyx), and one has only $r$-band detections with $g$-band non-detections for the last 36~d of the light curve (SN~2023bfv; of which it was tagged anomalous on the last observation overall, meaning that the $g$-band light curve features are outdated by several epochs). \par

Of the photometric anomalies, our model tags 15 objects, of which we consider 9 to be likely anomalous. The strongest anomaly candidates are two missed likely TDEs (AT~2022zyh, AT~2023adr), notable for their strong blue color throughout the light curve duration and their location at the host galaxy nucleus \citep{vanVelzen2019}, and our discovery of the long-rising SN~II candidate AT~2023inr, which had a rising light curve of at least $\sim$75~d. \par

By leveraging \laiss{} and additional tests such as the \texttt{iso\_forest\_anomaly\_detection} \emph{Filter}, we can uncover a variety of anomalous objects in the night sky, some of which may have been previously overlooked. This underscores the significance of employing multiple anomaly detection techniques to maximize the scientific value we can extract from recovering known and likely anomalies, quantify common pitfalls for why objects were either missed or not targeted for follow-up, and limit missed opportunities for active study in the future. \par

\begin{table*}[ht]
\footnotesize
\centering
\caption{\textbf{Retrospective search:} \emph{Transient-only} loci with $max(P(anom))\geq50\%$ at \emph{full} light curve phase}
\begin{tabular}{lllccl}
\multicolumn{6}{c}{10,000 randomly selected tagged loci, \texttt{iso\_forest\_anomaly\_detection} \textit{Filter}, ordered by $P(anom)$.} \\
\hline
\hline
\multicolumn{6}{c}{Spectroscopic} \\
\hline
\hline
ZTF ID & IAU Name & Spec. Class & $z$ & $P(anom)$ & Remarks \\
\hline
\textbf{ZTF22abhwlnm} & 2022wed & SN~IIn & 0.114 & 0.95 & Tagged anomalous 118~d before BTDG's spectrum. \\
\textbf{ZTF22aadesjc} & 2022fnl & SN~IIn & 0.104 & 0.93 & Tagged anomalous 14~d before ZTF's spectrum. \\
\textbf{ZTF22abfdzrv} & 2022vmg & SLSN-I & 0.41 & 0.69 & Peak $M_{abs}\sim-22.4$~mag. \\ 
\textbf{ZTF21abgkfzh$^{\#}$} & 2021qep & SN~IIn & 0.086 & 0.67 & Faint host. Visible for $\sim$800~d. \\
\textbf{ZTF23aatcsou$\dagger$} & 2023nwe & SN~IIn & 0.194 & 0.64 & Tagged anomalous 14~d before our spectrum. \\
ZTF23aadjssg & 2023dgp & SN~Ia-91T-like & 0.045 & 0.64 & Likely incorrect host. No rise info. \\
ZTF18aaiwzie & 2023bfv & SN~Ia-91T-like\footnote{We reclassify from SN~Ia to SN~Ia-91T-like.} & 0.086 & 0.58 & Reclassified. No $g$-band obs for $\textgreater36$~d.  \\ 
ZTF23aaekebt & 2023eqx & SN~II & 0.02 & 0.58 & Incorrect host? \\
ZTF23aadbtou & 2023cyx & SN~Ia & 0.033 & 0.52 & No rise info. SALT3 $c=-0.09$. \\
\textbf{ZTF22abghrui} & 2022vwu & SN~IIn & 0.197 & 0.50 & Tagged anomalous 14~d before ZTF's spectrum. \\
\hline
\hline
\multicolumn{6}{c}{Photometric} \\
\hline
\hline
\textbf{ZTF21abiggqx$^{\#}$} & 2021rjf & SN~IIn? & --- & 0.80 & Long-lived (800~d) CSM interaction? \\ 
ZTF23aaefpfb & 2023fli & SN~II? & --- & 0.68 & Faint host, incorrect assoc. FLEET$\approx$28\% SN~II/SLSN-I. \\
\textbf{ZTF22aatwxrl} & 2022oym & SN~IIn? & --- & 0.64 & FLEET=51\% SN~II, 43\% SLSN-I. Visible for $\sim$375~d. \\
\textbf{ZTF23aaahnss} & 2023atr & SN~IIn? & --- & 0.64 & Faint host, incorrect assoc. No rise. FLEET=60\% SN~II. \\
\textbf{ZTF22absuavp} & 2022zyh & TDE? & --- & 0.61 & Blue. Faint host. Visible for $\sim$200~d. \\ 
\textbf{ZTF23aafgmaz} & 2023frg & SN~Ib/c? & --- & 0.60 & SALT3 $x_1=+3.00$. FLEET=84\% SN~I. \\
\textbf{ZTF23aaflptz} & 2023gbk & SN~Ia? & --- & 0.55 & No decline. Peak $M_{abs}\sim-18.3$~mag. Underluminous? \\
ZTF23aatcola & 2023noh & SN~Ia? & --- & 0.55 & Only 1 anomalous epoch. FLEET=53\% SN~I, 46\% SN~II. \\
\textbf{ZTF23aajestr} & 2023inr & SN~II? & --- & 0.54 & YSE Target. Long-rising ($\sim$75~d) SN~II candidate. \\
\textbf{ZTF23aahjdxa} & 2023gpp & SLSN/SN~IIn? & --- & 0.53 & SLSN/IIn? FLEET=46\% SLSN-II, 41\% SN~II. \\
ZTF22abrbohu & 2022ywi & SN~II? & --- & 0.53 & Incorrect host? FLEET=60\% SN~II. \\
ZTF23aarzzwu & 2023nfs & SN~Ia? & --- & 0.52 & No decline. Poor SALT3 fit ($x_1=+3.00, c=-0.3$). \\
\textbf{ZTF22abzajwl} & 2023adr & TDE? & --- & 0.51 & $max(P(anom))=0.84$. Blue, nuclear. FLEET=81\% TDE. \\
ZTF23aagxvad & 2023glx & SN~Ia? & --- & 0.50 & Barred spiral host galaxy. No $g$-band obs after peak. \\
ZTF23aaempzk & 2023fbj & SN~Ia? & --- & 0.50 & No rise. SALT3 $x_1=2.80, c=-0.28$. FLEET=46\% SN~I. \\
\hline\\[-1.5ex]
\multicolumn{6}{l}{
\begin{minipage}{16cm}
NOTE: The bolded text designates a transient event that is likely anomalous, and we consider successfully tagged.\\
$^{\#}$In the databank used for train/test split (see Table~\ref{tab:dataset_bank}). \\
*We encountered 8 AGN or non-transient activity (ZTF19abyfhvp, ZTF21ackmnbo, ZTF21achlwqg, ZTF22aaytzrb, \\ ZTF22aanvqhn, ZTF21abwyelp, ZTF23aaunnwa, ZTF23aaarpdm), which are not shown in this Table.\\
\end{minipage}}
\end{tabular}
\label{tab:10k_full_phase}
\end{table*}

\section{Retroactive Anomaly Detection with YSE~DR1} \label{sec:AD_ysedr1}

The premiere multi-band, multi-survey time-domain dataset is the Young Supernova Experiment Data Release~1 (YSE~DR1; \citealt{Aleo2023}), which includes the final photometry of 1975 transients observed by the Zwicky Transient Facility (ZTF, \citealt{Bellm2019}) in $gr$ and Pan-STARRS1 (PS1, \citealt{Chambers2016}) in $griz$ bands as conducted by the Young Supernova Experiment (YSE; \citealt{Jones2021}). YSE~DR1 stands out as the most extensive and consistent multi-band dataset of supernovae at low redshifts ever assembled. This dataset serves as an ideal testing ground: it features a diverse range of real objects (SN~Ia, SN~Ia-SC, SN~Ia-CSM, SN~Ia-91T–like, SN~Ia-91bg–like, SN~II, SN~IIn, SN~IIb, SN~Ib, SN~Ib-pec, SN~Ibn, SN~Ic, SN~Ic-BL, TDE, SLSN-II, SLSN-I, SNa~Iax, and rare SN impostors such as an LBV outburst and a luminous red nova), SNe observations spanning extensive time frames (2019~November~24 to 2021~December~20), a vast range of apparent magnitudes ($m~\in~[12, 22]$) and absolute magnitudes ($M_{abs}~\in~[-13.5, -22.5]$), and covers a redshift distribution up to approximately $z\approx0.5$. Moreover, because ZTF and PS1 share $gr$ photometry, we treat PS1-$gr$ and ZTF-$gr$ as equivalent by stacking their light curves, effecting reducing the overall cadence and adding depth information\footnote{While this approximation is generally accurate, minor discrepancies may arise due to variations in passband transmission profiles and disparities in photometric pipelines. Moreover, ZTF is not color-calibrated, it is calibrated to PS1 assuming that all ($g-r$) colors are zero. We do not account for the ZTF color correction coefficient. Therefore, the more rigorous approach involves considering passbands independently, which is left for future work. See \url{http://nesssi.cacr.caltech.edu/ZTF/Web/gettingto1.html} for more details.}. \par 

As an extension of the original work from \cite{Aleo2023}, we run \laiss{} to find the most anomalous objects in YSE~DR1. We do not use the redshift or latent embedding information from the ParSNIP photometric classifier in our model. Of the 1975 total YSE~DR1 dataset, 1153 ($\approx$58\%) are successfully processed through our model after selection cuts for requisite number of observations in the combined PS1+ZTF-$gr$ bands and associated hosts with PS1-$grizy$ photometry. Unlike Table~\ref{tab:spec_test_set} and Table~\ref{tab:10k_full_phase}, we consider the max anomaly score $max(P(anom))$ at \emph{any} light curve phase. Note that we use the the same RFC model that is trained on the full light curves, but the input YSE~DR1 light curve features are constantly re-calculated (the host features do not change) with the \texttt{light-curve} extractor and passed through model at each epoch, to mimic a real-time anomaly detection scenario. \par

There are 198/1153 ($\approx$17\%) transients that achieve $max(P(anom))\geq50\%$, but only 42/1153 ($\approx$3.6\%) that achieve $max(P(anom))\geq70\%$. For brevity, we show the results of these 42 objects in Table~\ref{tab:ysedr1_real_time_any_phase}, ordered by highest anomaly score. If two or more objects have the same score, we secondarily rank them by the number of observations for which the anomaly score of the object was $P(anom)\geq50\%$ (``Anom. Obs"). \par

\begin{table*}[ht]
\footnotesize
\centering
\caption{\textbf{Retrospective search:} The 42 most anomalous YSE~DR1 \citep{Aleo2023} events with $max(P(anom))\geq70\%$ at \emph{any} light curve phase.}
\begin{tabular}{llccccl}
\multicolumn{7}{c}{1153 processed, from 1975 total \texttt{YSE~DR1} objects, ordered by $max(P(anom))$.} \\
\hline
\hline
\multicolumn{7}{c}{Spectroscopic (Spec-$z$)} \\
\hline
\hline
IAU Name & Spec. Class & $z$ & Peak $M_{abs}$ & $max(P(anom))$ & Anom. Obs & Remarks \\
\hline
\textbf{2020xsy} & SLSN-II & 0.27 & $-22.3$ & 0.98 & 93 & Tagged anomalous 13~d before YSE's spectrum. \\
\textbf{2020tip} & SN~Ia-9T-like\footnote{We reclassify from SN~Ia to SN~Ia-91T-like.} & 0.095 & $-19.2$ & 0.90 & 14 & Spiral host? Lack of Si. Match to SN~2006oa. \\
2020qkx & SN~Ia & 0.127 & $-19.8$ & 0.89 & 51 & Luminous Ia?  \\
\textbf{2020opy} & TDE & 0.159 & $-20.5$ & 0.89 & 83 & Tagged 23~d before YSE's spectrum. \\
\textbf{2021nxq} & SLSN-I & 0.15 & $-20.6$ & 0.89 & 27 & Faint host, incorrect assoc. \\
2020qql & SN Ia & 0.076 & $-19.8$ & 0.88 & 35 & Luminous Ia? \\
\textbf{2021hrj} & SN~Ib & 0.022 & $-17.4$ & 0.88 & 79 & Tagged 7~d before YSE's spectrum (Fig.~\ref{fig:2021hrj}). \\
\textbf{2021bmv} & SN~IIn & 0.09 & $-19.3$ & 0.85 & 11 & Faint host.  \\
\textbf{2021aadc} & SLSN-II & 0.1953 & \textless$-20.8$ & 0.85 & 15 & Faint host. \\
\textbf{2020qmj} & SN~IIn & 0.022 & $-18.8$ & 0.82 & 51 & \nodata  \\
\textbf{2020kre} & SN~Ia-CSM & 0.136 & $-19.9$ & 0.82 & 70 & Tagged anomalous 37~d before YSE's spectrum. \\
\textbf{2021too} & SN~Ic-BL & 0.07 & $-19.4$ & 0.82 & 36 & \nodata \\
\textbf{2021pnp} & SN~IIb & 0.03 & $-17.0$ & 0.79 & 46 & \nodata \\
\textbf{2021gno} & SN~Ib-pec & 0.0062 & $-15.0$ & 0.78 & 29 & See \cite{Galan2022gno}. \\
\textbf{2020kbl} & SN~Ia & 0.079 & $-18.8$ & 0.77 & 22 & SALT3 $c=+0.3$. In flocculent spiral/merger galaxy. \\
\textbf{2021btn} & SN~II & 0.083 & $-19.6$ & 0.75 & 43 & Bright. No $g$- obs for $\sim$50~d. H$_{\alpha}$ origin unclear. \\
\textbf{2021udc} & SN~IIb & 0.035 & $-18.3$ & 0.75 & 8 & \nodata \\
2021bpq & SN~Ia & 0.1 & $-19.3$ & 0.75 & 5 & \nodata \\ 
2021ojn & SN~Ia & 0.08 & $-19.5$ & 0.74 & 18 & Faint host, misassoc. Early Ia-bump candidate. \\ 
\textbf{2020tan} & SN~IIn & 0.079 & $-19.0$ & 0.73 & 60 & Tagged 75~d before YSE's spectrum. \\
2020acun & SN~II & 0.0216 & $-17.3$ & 0.73 & 25 & Misclassified by ParSNIP, SuperRAENN (Ib/c). \\
\textbf{2020kvl} & SN~Ia-91T-like\footnote{We reclassify from SN~Ia to SN~Ia-91T-like.} & 0.12 & $-20.6$ & 0.72 & 8 & Possible SN~Ia-SC. \\
2020wfg & SN~Ia & 0.108 & $-19.4$ & 0.72 & 3 & \nodata \\ 
\textbf{2020ivg} & SN~IIb & 0.053 & $-17.3$ & 0.72 & 40 & Tagged 18~d before YSE's spectrum. \\
\textbf{2020kpz} & SN~II & 0.039 & \textless$-18.2$ & 0.71 & 65 & Pre-explosion data? Peculiar LC. Visible $\sim$400~d. \\
\textbf{2020acct} & SN~Ic & 0.035 & $-18.0$ & 0.70 & 32 & See Angus et al. (in prep). \\
2019wmr & SN~II & 0.038 & $-17.5$ & 0.70 & 2 & Misclassified by ParSNIP, SuperRAENN (Ib/c). \\
2021vwx & SN~Ia & 0.06 & \textless$-19.3$ & 0.70 & 9 & No rise or peak. In spiral. \\
\hline
\hline
\multicolumn{7}{c}{Spectroscopic (Host-$z$)} \\
\hline
\hline
2020rss & SN~Ia? & 0.1288 & $-19.1$ & 0.77 & 11 & \nodata \\ 
2021ofr & SN~Ia? & 0.0848 & $-19.2$ & 0.76 & 8 & \nodata \\ 
\textbf{2020kmj} & SN~II? & 0.0799 & $-18.4$ & 0.76 & 49 & Elliptical host galaxy. \\
\hline
\hline
\multicolumn{7}{c}{Photometric (Photo-$z$)} \\
\hline
\hline
2020hjv & SN~Ia? & 0.141 & $-19.9$ & 0.91 & 7 & Luminous Ia? \\ 
\textbf{2020acyu} & SN~IIn? & 0.254 & $-20.4$ & 0.90 & 33 & Faint host. \\ 
\textbf{2020jvi} & SN~IIn? & 0.192 & $-20.2$ & 0.84 & 9 & Faint host. \\ 
\textbf{2019vuz} & TDE? & 0.21 & $-20.8$ & 0.79 & 17 & Blue. Nuclear. FLEET=89\% TDE. \\
\textbf{2021dpa} & SLSN? & 0.152 & $-18.9$ & 0.73 & 20 & Faint host, misassoc. Peculiar long rise ($\sim$100~d). \\
\textbf{2020itp} & SN~Ia-CSM? & 0.16 & $-19.7$ & 0.73 & 44 & ZTF-$r$ declines slowly but ZTF-$g$ fades quickly. \\
\textbf{2021rmq} & SN~II? & 0.131 & $-19.1$ & 0.73 & 10 & Candidate member of long-rising SN~II class. \\
2020wwt & SN~Ia? & 0.157 & $-19.2$ & 0.72 & 14 & SALT3 $c=-0.21$. Spiral. Artifact in PS1-$r$ image. \\
2020iga & SN~Ia? & 0.13 & $-19.2$ & 0.71 & 7 & SALT3 $c=-0.20$. Nuclear. \\
2020sgy & SN~Ia? & 0.204 & $-20.6$ & 0.71 & 18 & Luminous Ia? Faint host. \\
2020vpn & SN~Ia? & 0.136 & $-19.4$ & 0.70 & 5 & \nodata \\
\hline\\[-1.5ex]
\multicolumn{7}{l}{
\begin{minipage}{16cm}
NOTE: The bolded text designates a transient event that is likely anomalous, and we consider successfully tagged.\\
\end{minipage}}
\end{tabular}
\label{tab:ysedr1_real_time_any_phase}
\end{table*}

When considering \emph{spectroscopic} anomalies alone, we find a purity of 15/29, or 52\%, aligning well with the results we find in Figure~\ref{fig:Panom_vs_PR}. If expanded to consider \emph{behavioral} and \emph{contextual} anomalies, we achieve a purity of 20/29, or 69\%, which slightly underperforms given what we expect from Figure~\ref{fig:Panom_vs_PR}. However, our anomaly detection model is not trained nor tested with PS1 photometry nor ZTF PSF-fit forced-photometry\footnote{\url{https://web.ipac.caltech.edu/staff/fmasci/ztf/forcedphot.pdf}} (as opposed to the ZTF Alert Stream) as in YSE~DR1, and thus we cannot expect our model to perform in exactly the same way. \par

The original work from \cite{Aleo2023} did not do an extensive study of anomalous objects. Broadly, they found transients with latent embeddings (from the YSE-ZTF trained ParSNIP classifier) outside the bulk of their member class distribution were more likely to be misclassified. They indicated that these objects likely deviated from the normal, more representative objects that share their spectroscopic class. Because the ParSNIP classifier was trained on simulations from idealized templates, objects whose real behavior differed from this normality was likely to be poorly characterized. Despite classifying SNe~Ia with high purity ($\geq90\%$), they found misclassified SNe~Ia tended to fall within two categories: 1) observing effects such as significant ($\sim$100~d) gaps, only observed well after peak (e.g., SN~2020zmi, SN~2021van, SN~2021vwx), and SN Ia which require a large correction for dust extinction (e.g., red SNe~Ia with SALT3 $c$~\textgreater~0.3; SN~2020pki, SN~2020zfn, SN~2021aamo); or 2) physical effects linked to rare phenomena (91T-like/91bg-like) properties (SN~2021bmu, SN~2021ctn), and long-lived CSM interaction (e.g., SN~Ia–CSM: SN~2020aekp, SN~2020kre, SN~2021uiq).  \par

If we investigate misclassified SNe~Ia due to observing effects, we find the following: the hosts of SN~2020pki and SN~2020zmi did not have the requisite $grizy$ photometry, and thus were not processed by our model; SN~2021van did not have enough $g$-band photometry; SN~2020zfn and SN~2021aamo were successfully processed by our model, but not tagged anomalous; SN~2021vwx was tagged anomalous, though likely due to having no rise nor peak information in the light curve, which is rare in our training set (see Section~\ref{subsec:ad_failures} for details). We have insufficient statistical evidence to determine how YSE~DR1 SNe affected by observing effects influences our \laiss{} anomaly detection classification. \par

We additionally investigate objects misclassifed by ParSNIP and SuperRAENN, as these are strong anomaly candidates. Note that both classifiers in \cite{Aleo2023} are only trained via simulations to classify into SN~Ia, SN~II, and SN~Ib/c (thus SLSN, TDE, and more will always be misclassified). We find 13 such objects flagged by our anomaly detection model: 
\begin{itemize}
    \item SN~2020xsy (SLSN-II): The most anomalous object from YSE~DR1 according to our model, in terms of $max(P(anom))$ at 98\% and number of epochs with an anomaly classification (93). Also at the furthest redshift, $z=0.27$. ParSNIP misclassified as SN~II; SuperRAENN misclassified as SN~Ia.
    \item TDE~2020opy (TDE): Tagged anomalous 23~d before the first classification spectrum from YSE. The second highest number of epochs with an anomaly classification (83). ParSNIP misclassified as SN~Ia; SuperRAENN misclassified as SN~Ibc.
    \item SN~2021nxq (SLSN-I): Even with an incorrect host association (likely due to the faint host), this object is tagged anomalous at peak light. ParSNIP misclassified as SN~Ib/c; SuperRAENN misclassified as SN~Ib/c. 
    \item SN~2021aadc (SLSN-II): Tagged anomalous 2~d before the first classification spectrum from BTDG, approximately two weeks before peak light. There is no decline information for this object. ParSNIP misclassified as SN~Ib/c; SuperRAENN misclassified as SN~Ib/c. Located in a faint host.
    \item SN~2020qmj (SN~IIn): From peak light onwards, there is an approximately constant red color $g-r\approx0.7$~mag in the light curve evolution. ParSNIP misclassified as SN~Ia; SuperRAENN misclassified as SN~Ia. Located in an edge-on spiral galaxy.
    \item SN~2020kre (SN~Ia-CSM): Our anomaly detection algorithm tagged this object as anomalous 21~d after the original SN~Ia-normal classification but 37~d \emph{before} YSE's SN~Ia-CSM reclassification spectrum. Located in a faint, blue host.
    \item SN~2021pnp (SN~IIb): From peak light onwards, there is an approximately constant red color $g-r\approx0.9$~mag in the light curve evolution, photometrically similar to SN~2020qmj. No strong evidence of shock breakout in the photometry. ParSNIP misclassified as SN~Ib/c; SuperRAENN misclassified as SN~Ib/c.
    \item SN~2021btn (SN~II). Unusually bright for a normal SN~II at $M_{abs}\sim-19.6$~mag. Lack of $g$-band observations for the first 50~d of the event. It is unclear if the H$_{\alpha}$ emission in the spectrum is from the host or the transient, and whether there are signs of P-cygni. A possible alternative is a SN~IIn classification. ParSNIP misclassified as SN~Ib/c; SuperRAENN misclassified as SN~Ib/c.
    \item SN~2021udc (SN~IIb): No strong evidence of shock breakout in the photometry. ParSNIP misclassified as SN~Ia; SuperRAENN misclassified as SN~Ia.
    \item SN~2020acun (SN~II): From peak light onwards, there is an approximately constant red color $g-r\approx0.9$~mag in the light curve evolution, photometrically similar to SN~2020qmj and SN~2021pnp. \texttt{Superfit} \citep{Howell2005} gives matches to SN~IIb, but the He~I 5875\AA~absorption is very low S/N if it is present at all, and hence we keep the SN~II classification. ParSNIP misclassified as SN~Ib/c; SuperRAENN misclassified as SN~Ib/c. 
    \item SN~2020kpz (SN~II): A peculiar SN with possible pre-explosion data, marked by a sudden shift from blue to red before the sharp rise to peak. ParSNIP misclassified as SN~Ib/c; SuperRAENN misclassified as SN~Ib/c. 
    \item SN~2020acct (SN~Ic*): The peculiar, distinctly double peaked SN separated by $\sim$60~d, which does not fall neatly into any existing SN classification. See Angus et al. in prep for details. ParSNIP misclassified as SN~Ia; SuperRAENN misclassified as SN~Ib/c.
    \item SN~2019wmr (SN~II): From peak light onwards, there is an approximately constant red color $g-r\approx0.9$~mag in the light curve evolution, photometrically similar to SN~2020qmj, SN~2021pnp, and SN~2020acun. ParSNIP misclassified as SN~Ib/c; SuperRAENN misclassified as SN~Ib/c. 
\end{itemize}

An interesting observation is that of the 28 most anomalous spectroscopic YSE~DR1 objects (Table~\ref{tab:ysedr1_real_time_any_phase}), 13 (or 46\%) are what we consider bonafide anomalies (bold text) \emph{and} are misclassified by both ParSNIP and SuperRAENN. When accounting for only spectroscopic classes for which ParSNIP and SuperRAENN were trained (SN~Ia, SN~II, SN~Ib/c; and thus excluding SLSN and TDE), we find this drops to only 9 (or 32\%). \par

\begin{figure}
    \centering
    \includegraphics[width=\columnwidth]{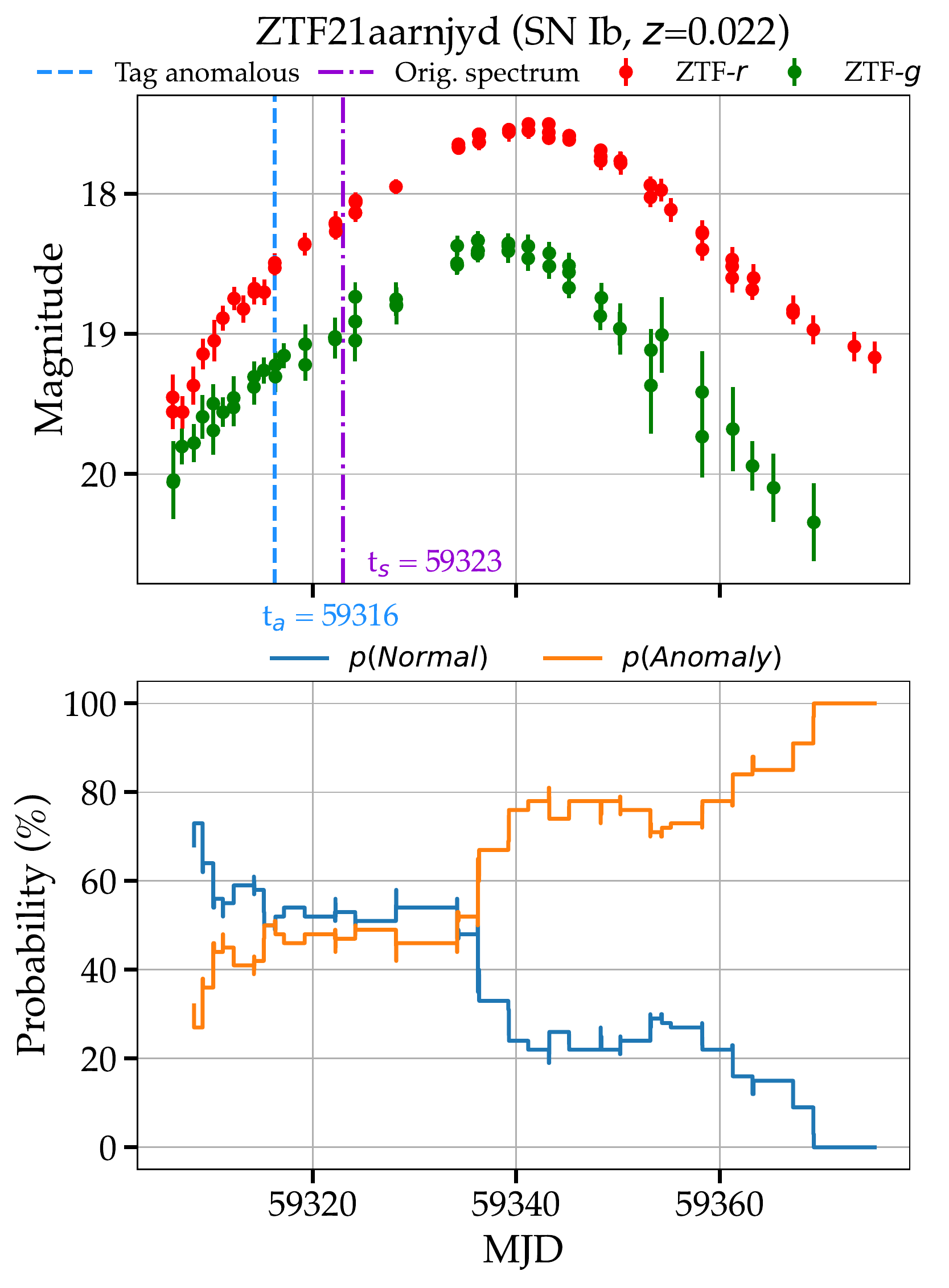}
    \caption{
    The Type~Ib SN~2021hrj/ZTF21aarnjyd was retrospectively tagged as anomalous ($P(anom)\geq50\%$) on MJD 59316 (2023-04-12) by our \texttt{LAISS} AD algorithm (dashed blue line), \emph{one week earlier} than the first classification spectra from YSE (2023-04-19; dot-dashed purple line) and ZTF (2023-04-20) on TNS. The YSE team originally requested spectra when this object passed their magnitude limited sample criteria. If our algorithm was in place when SN~2021hrj was active, we could in principle have obtained a spectrum one week earlier in the light curve evolution, providing valuable information in regards to the early-time supernova physics and local environment of this peculiar, long-rising event.
    } 
    \label{fig:2021hrj}
\end{figure}

Although we do not formally consider the phase at which the anomaly detection successfully tags an object as anomalous in our metrics, there is great value at flagging anomalies at or before peak, and as early-time as possible, as these times are when the physics of the progenitor explosion is most correlated with the photometry \citep{Gagliano2022}. Of the successfully tagged objects in Table~\ref{tab:ysedr1_real_time_any_phase}, our algorithm correctly identifies 6/20 (30\%) as anomalous \emph{at least 1 week earlier than the first classification spectrum}. In the best case, we identified SN~IIn 2020tan 75~d prior. We show the ``worst" case of only 7~d prior for SN~Ib 2021hrj in Figure~\ref{fig:2021hrj}. The YSE team originally requested spectra when SN~2021hrj passed their magnitude-limited sample criteria. If our anomaly detection algorithm were in place when it was active, we could, in principle, have obtained a spectrum one week earlier in the light curve evolution, providing valuable information in regards to earlier-time supernova physics and local environment of this peculiar, long-rising event. 

We additionally identify strong anomaly candidates for which there was no spectroscopic follow-up (Table~\ref{tab:ysedr1_real_time_any_phase}). We consider such photometric objects as missed opportunities. Although it is impossible to know the exact classifications, we have evidence that 6 of 11 (55\%) of the most anomalous tagged photometric objects from YSE~DR1 were likely bonafide anomalies: \par
\begin{itemize}
    \item AT~2020acyu (SN~IIn?): Visible for $\sim$175~d, and approximately 2\farcs42 separation from the center of faint host PS1 galaxy catalog ID 1237654653639656020. At an estimated redshift of $z=0.254$ from \texttt{Easy Photoz} \citep{Aleo2023}, we calculate a peak absolute magnitude of $M_{abs}\approx-20.4$~mag. Likewise, if we use the photo-$z$ estimate of $z=0.224$ from SDSS \citep{Csabai2003}, we estimate a peak absolute magnitude of $M_{abs}\approx-20.2$~mag. FLEET \citep{Gomez2020} predicts a Type II (with no distinction between Type II and Type IIn) with 68\% confidence, followed by SLSN-II at 24\%. ParSNIP predicts a Type II with 55.9\%. With the luminous but not super-luminous brightness, extended duration, and color evolution, we predict this to be a missed SN~IIn. Beyond the simple explanation of being overlooked, a possible reason for why there was no spectroscopic follow-up requested for this object is because this object achieved a peak apparent magnitude of only $\sim$20.0~mag, which would require long exposure times for such a faint source, and thus would be a low priority. 
    \item AT~2020jvi (SN~IIn?): Visible for $\sim$120~d, approximately 0\farcs11 separation from the center of faint, blue host PS1 galaxy catalog ID 1237662636371739674. At an estimated redshift of $z=0.192$ from \texttt{Easy Photoz} \citep{Aleo2023}, we derive a peak absolute magnitude of $M_{abs}\approx-20.2$~mag. Likewise, if we use the photo-$z$ estimate of $z=0.173$ from SDSS \citep{Csabai2003}, we calculate a peak absolute magnitude of $M_{abs}\approx-20.0$~mag. This object rose approximately 2~mag in 20~d, and is marginally red ($\sim$0.1~mag) throughout its duration. FLEET \citep{Gomez2020} predicts an SLSN-II at 58\% confidence. To add to the confusion, ParSNIP predicts a Type Ib/c at 69\% confidence (however, ParSNIP is not trained for SLSN classification). A SALT3 fit reveals a poor fit and a stretch parameter value of $x_1=+3.00$. Similarly as AT~2020acyu, with the luminous but not super-luminous brightness, extended duration, and color evolution, we cautiously predict this to be a missed SN~IIn. This object achieved a peak apparent magnitude of $\sim$19.4~mag in ZTF-$r$, which is within reasonable limits for follow up. 
    \item AT~2019vuz (TDE?): Arguably the most anomalous of the photometric subset due to the intrinsic rarity, we predict this object as a likely missed TDE, due to its duration of $\sim$75~d, blue color, at a nuclear location of 0\farcs11 separation from the center of its red, possible elliptical host PSO J072914.403+420437.05. At an estimated redshift of $z=0.21$ from \texttt{Easy Photoz} \citep{Aleo2023}, we estimate a peak absolute magnitude of $M_{abs}\approx-20.8$~mag. Using the estimated redshift of $z=0.248$ from PS1-STRM \citep{Beck2021}, we estimate a peak absolute magnitude of $M_{abs}\approx-21.2$~mag. FLEET \citep{Gomez2023TDE} predicts a TDE at 89\% confidence. This object achieved a peak apparent magnitude of $\sim$19.3~mag in PS1-$g$ and was brighter than 19.5~mag for $\sim$20~d, which is well within reasonable limits for follow up. 
    \item AT~2021dpa (SLSN?): Exhibits an unusually prolonged slow rise of $\sim$0.8 mag over $\sim$100~d (estimated $\sim125$~d~mag$^{-1}$) in an extremely faint, small, red host. Likely due to its faint nature, \texttt{GHOST} \citep{Gagliano2021} does not find the likely host and results in a misassociation (to a nearby, faint galaxy PS1 Object ID 1237664871896777642). At approximately 2\farcs2 offset from its apparent host PS1 object ID 149701403944978764, there are two disparate estimates for the photo-$z$. From \texttt{Easy Photoz} \citep{Aleo2023}, the estimated redshift of $z=0.152 \pm 0.096$ results in a peak absolute magnitude of $M_{abs}\approx-18.9$~mag. However, the YSE~DR1 estimate is likely low, due to the host being perceptible in the DESI Legacy Survey Imaging Surveys\footnote{\url{https://www.legacysurvey.org/}} but only marginally so in SDSS and PS1 stacked templates. The SDSS estimated redshift of $z=0.482 \pm 0.1461$ \citep{Csabai2003} results in a peak absolute magnitude of $M_{abs}\approx-21.9$~mag. FLEET \citep{Gomez2023SLSN} estimates a SLSN-II with 40\% confidence, followed by SLSN-I with 31\% confidence. When considering proprietary YSE photometry beyond the YSE~DR1 cutoff, this object is visible for $\sim$370~d. All of this evidence together points to a SLSN. A compelling reason for why there was no spectroscopic follow-up requested for this object is because this object achieved a peak apparent magnitude of only $\sim$20.2~mag before disappearing behind the Sun, only to reappear at a faint 21.7~mag at $\sim$250~d later. It is unlikely one could acquire spectroscopic follow-up for this object even if it was known to be anomalous.
    \item AT~2020itp (SN~Ia-CSM?): Visible for $\sim$80~d, in elliptical host PS1 ObjectID 147052397040217754. The ZTF-$r$ photometry declines slowly but the ZTF-$g$ fades quickly. There is a fair agreement in photo-$z$. From \texttt{Easy Photoz} \citep{Aleo2023}, the estimated redshift of $z=0.16 \pm 0.051$ results in a peak absolute magnitude of $M_{abs}\approx-19.7$~mag whereas the SDSS estimated redshift of $z=0.109 \pm 0.033$ \citep{Csabai2003} results in a peak absolute magnitude of $M_{abs}\approx-19.0$~mag; both are consistent with SN~Ia. Regardless, for both photo-$z$ estimates, if there existed any CSM interaction, the H$_{\alpha}$ profile would reside in only the ZTF-$r$ wavelength range\footnote{\url{http://svo2.cab.inta-csic.es/svo/theory/fps3/index.php?id=Palomar/ZTF.r\&\&mode=browse\&gname=Palomar\&gname2=ZTF\#filter}}, a possible driver for the slow ZTF-$r$ decline. Because this object is as intrinsically bright as an SN~Ia but with a longer duration and slower ZTF-$r$ decline which is consistent with possible CSM interaction, we posit this to be a missed SN~Ia-CSM. This object achieved a peak apparent magnitude of $\sim$19.4~mag in ZTF-$r$, which is within reasonable limits for follow-up.
    \item AT~2021rmq (long rising SN~II): The YSE~DR1 light curve shows a rise of $\sim$60~d, with evolution consistent with Type II, and is thus a candidate member of the rare long-rising SN~II class \citep{Sit2022}. Additional ZTF~DR data photometry\footnote{\url{https://www.ztf.caltech.edu/ztf-public-releases.html}} indicate that this rise may be longer, closer to $\sim$80~d. It is offset 0\farcs23 from its blue (likely spiral) host PS1 objectID 146701848390420314. The \texttt{Easy Photoz} \citep{Aleo2023} photo-$z$ (0.131) is in good agreement with that of \cite{Beck2021} (0.146), placing the peak absolute magnitude at around $M_{abs}\approx-19.1$~mag. However, if the photo-$z$ is accurate, this would place the absolute magnitude brighter than any of the 13 events found in \cite{Sit2022} at $-17.5$~mag. It is also possible the long-rise could be a result of CSM interaction \citep{Nyholm2020}, and this object could instead be an SN~IIn.
\end{itemize}

The remaining 5 photometric objects with high anomaly scores (AT~2020hjv, AT~2020wwt, AT~2020iga, AT~2020sgy, AT~2020vpn) are likely SN~Ia. If photo-$z$ estimates are correct, two are luminous and potentially of a rare subtype at $M_{peak}\sim-20$~mag (AT~2020hjv, AT~2020sgy), but we find no conclusive evidence. Of the others, two are blue when compared to the SALT3 \citep{Kenworthy2021} model at $c\sim-0.2$ (AT~2020wwt, AT~2020iga), and one (AT~2020vpn) appears to be unremarkable. \par

Lastly, there are three objects with host spectra but no transient spectra. AT~2020rss and AT~2021ofr appear to be normal SN~Ia based on evolution and absolute magnitude estimates using the spectroscopic host-$z$. The likely \emph{contextual} anomaly is AT~2020kmj, which exhibits an SN~IIP profile with typical SN~II peak absolute magnitude but resides in what appears to be an elliptical galaxy (PS1 ObjectID 116361881748977931). 

Overall, of the objects which were successfully processed and characterized by \laiss{} from YSE~DR1 (1153/1975, or 58\%), 7/27 (26\%) of the most anomalous objects ($P(anom)\geq70\%$ at any phase in the light curve) which we consider likely anomalies (bold text) did \emph{not} have transient spectra. Of those that did, 6/20 (30\%) were flagged by \laiss{} at least a week prior to the classification spectrum. \par

\section{Approximate Nearest Neighbor Similarity Search with \annoy{}} \label{sec:annoy}

In this current age of large time-domain surveys like ZTF \citep{Bellm2019} and the imminent LSST \citep{Ivezic2019}, the bottleneck of object discovery lies in the ability of automated algorithms to quickly isolate and prioritize detections of interest amid the flood of alert events. ZTF generates approximately 70~GB/night \citep{Patterson2019} and discovers of order $\sim$10000~SNe/year\footnote{For this estimate, we use the Search function from \url{www.wis-tns.org/}, and search for all transients reported to TNS from ZTF first light on 2017~November~1 to 2023~November~1 with ZTF listed as the discovery source, and find the per year average.}, and LSST is estimated to produce an alerts volume of 10x that at approximately 782 GB/night, observing an estimated of order $\sim$1 million SNe/year\footnote{\url{https://dmtn-102.lsst.io/DMTN-102.pdf}}. \par

With high present-day discovery rates and climbing, we cannot rely on computationally expensive operations for SNe discovery and characterization. In this section, we argue that a low-latency, sublinear time-complexity approximate nearest neighbors search can be an effective tool to search for similar SNe for applications of SN discovery, reclassfication, and more. \par

Previous studies have shown that data mining strategies involving a nearest neighbors search for retroactive SN discovery can be successful \citep{Aleo2022}. However, \cite{Aleo2022} used a brute-force kD-tree approach to search for neighbors in feature space, which at worst case scales linearly with the number of data points in runtime ($\mathcal{O}(n)$). This approach would be impractical for the real-time data streams of ZTF and LSST. We use their approach as inspiration, and instead opt for an approximate similarity search, which at worst case scales in logarithmic time with respect to the number of data points, $\mathcal{O}(log(n))$. \par 

For our ANN algorithm, we use the open-source package \annoy{}\footnote{\url{https://github.com/spotify/annoy}} \citep{Github:annoy}, the method developed by Spotify for song recommendations. Essentially, \annoy{} allows for an efficient ANN search in high-dimensional spaces, such as the large light curve and host galaxy feature dataset we use in this work\footnote{Note that in this work we do not use data imputation. Although in principle \annoy{} can run on data sets with imputed entries, it may introduce bias and impact the nearest neighbor search accuracy, particularly if many values are imputed.}. This process involves dividing the high-dimensional space into smaller subspaces using random hyperplanes. These subspaces are organized into a binary tree structure, where each node represents a subspace and each leaf node corresponds to a specific point in the original high-dimensional space. During a search operation, the algorithm begins at the root of the tree and recursively traverses the tree, selecting the branch that is closest to the query point. At each leaf node, the distance between the query point and the represented point is recorded. Simultaneously, a list of the $k$ closest points encountered thus far is maintained. To optimize the search process, recorded distances are utilized to prune branches unlikely to contain points closer than the current $k$th closest point. This iterative search is repeated for each query point. The \annoy{} library is designed to ensure efficient memory usage and high performance, enabling indexing of millions of high-dimensional vectors. \par 


In the \laiss{} pipeline (Figure~\ref{fig:pipeline}), we build an \annoy{} index of our databank---the reference dataset to which new, incoming, or manually chosen objects will be compared. However, we do so after scaling our databank with the \texttt{sklearn.preprocessing.StandardScaler()}\footnote{\url{https://scikit-learn.org/stable/modules/generated/sklearn.preprocessing.StandardScaler.html}} module and applying Principal Component Analysis (PCA, \citealt{Jolliffe2002}) to reduce the dimensionality of our data from 120~dimensions to 60~dimensions, retaining $\sim$98\% of the variance. This will partially speed up the ANN search without important information loss; moreover, having less than 100~dimensions is recommended by the code authors for the best performance\footnote{Though, in principle it can perform ``surprisingly well'' up to 1,000~dimensions; see \url{https://github.com/spotify/annoy\#summary-of-features}}. \par

\laiss{} can run the ANN search to calculate any $k$ neighbors (default is $k=8$) for any ZTF object for which its light curve and host galaxy association and features can be extracted, regardless if it exists in the databank or not; all that is needed is a ZTF Object ID of the the user's choosing. For example, when supplied a ZTF Object ID, \laiss{} queries ANTARES for the object of interest's ZTF photometry and manually extracts the light curve features in the same manner as the \texttt{lc\_feature\_extractor} \emph{Filter}, followed by running \texttt{GHOST} \citep{Gagliano2021} to retrieve the matched host galaxy and its features. The object will undergo the same scaling and PCA transformation as the data set bank, and then \annoy{} will find the ANNs in the 60~dimensional PC-space. As the ANNs are found, \laiss{} queries the ANTARES broker to retrieve the up-to-date ZTF photometry, IAU name, spectroscopic classification (if exists), and spectroscopic redshift (if exists) from TNS for each ANN. Plotting functions are then run to overlay the reference transient and its $k$ ANNs' light curves and host galaxy thumbnails. Currently, this ANN similarity search feature is only available in the Python module on Github\footnote{\url{https://github.com/patrickaleo/LAISS-local}}. Efforts to create the same functionality within an ANTARES \emph{Filter} is ongoing. This will remove the additional outside API requests for extracting light curve and host galaxy features and up-to-date TNS information. \par

Within our databank of 5472 transients, the default $k=8$ ANN search alone takes anywhere from 1 to 350~ms on a 2 GHz Quad-Core Intel Core i5. This results in $\leq$5 minutes of total processing time to find $k=8$ ANNs for all $\sim$1000 SN/night estimated with LSST. If an object needs to have its light curve and host features extracted from scratch, the entire process takes about 1 minute per transient (thus, $\sim$16 hours of total processing time for all $\sim$1000 SN/night), where the overwhelming majority time bottleneck stems from the \texttt{GHOST} host association---specifically time spend on catalog queries (e.g., PS1); such catalogs could be downloaded to disk to dramatically speed up associations. Utilizing faster host association methods such as DELIGHT ($\sim$60~ms/transient, \citealt{Forster2022}) and modifying \texttt{GHOST} to use internal disk-downloaded catalogs is a subject of future work. \par

\subsection{ANN Similarity Search Results on Light Curves and Host Galaxies} \label{subsec:sim_search_ztf}

The ability to quickly find analogs to an object of interest is crucial to many useful applications, including but not limited to: 
\begin{itemize}
    \item \textbf{Predicting Behavior}: Based on the known properties and evolution of analogs, one can predict the potential behavior of newly observed transients across their time evolution. Nominal events will likely evolve as expected in common galactic environments, and deviations from this behavior (or in the case where analogs are anomalous) could indicate rare or misunderstood phenomena; 
    \item \textbf{Calibrating Models}: By comparing observations of known analogs with theoretical predictions, experts can improve the accuracy of models describing the underlying physical processes of transient events, such as those used for simulations (e.g., models developed for \texttt{SNANA} \citep{Kessler2009} and their use in PLAsTiCC \cite{Hlozek2020}).
    \item \textbf{Studying Rare Events}: By associating a similarity score, experts can measure how similar or dissimilar an event is from the rest of the existing dataset. Objects strongly dissimilar to its nearest neighbors can indicate truely unusual phenomena, or out-of-distribution (OOD) behavior for the particular reference dataset.
    \item \textbf{Probing Host Galaxy Properties}: Analogous transients in different host galaxies allow experts to investigate the influence of host galaxy properties on the characteristics of the transient events. This information is valuable for understanding the connections between stellar populations, galactic environments, and transient phenomena.
    \item \textbf{Improving Classification}: As more analogs are discovered and characterized, classification systems can be refined, leading to more accurate and automated identification of new transients.
    \item \textbf{Reclassification}: A given transient of a known type whose analogs are classified differently can direct the expert's attention into reinspecting their spectra and classifications, possibly updating their classification. We show this to be true in Section~\ref{subsec:reclass_sne} and Table~\ref{tab:reclassification_SNe}, where we update classifications for 17 unique SNe from our databank, prompted by an ANN search. 
    \item \textbf{Optimizing Observational Strategies}: Observers can prioritize targets based on the likelihood of finding analogs or because they are analogs to a known rare or scientifically useful event, helping to maximize the efficiency of observing campaigns and the utilization of telescope time. This can be used in tandem with active learning strategies like RESSPECT \citep{Kennamer2020}.
    \item \textbf{Retroactive Discovery}: Analogs of known transients can also be used as a data mining method to find similar but previously undiscovered (i.e., not publicly reported) transients in large datasets, successfully demonstrated by \cite{Aleo2022}. In this work, we showcase 84 transients discovered from the ZTF Alert Stream in 2018-2021 using this method, as shown in Table~\ref{tab:report_to_TNS_SNe} and Section~\ref{subsec:previously_unreported}. Similarly, analogs can be used to suggest likely candidates of a certain class that were originally reported but have no follow-up spectra for classification, and thus represent possible missed opportunities (Section~\ref{subsec:previously_reported}).
\end{itemize}

\begin{figure*}
    \centering
    \includegraphics[width=1.131\columnwidth]{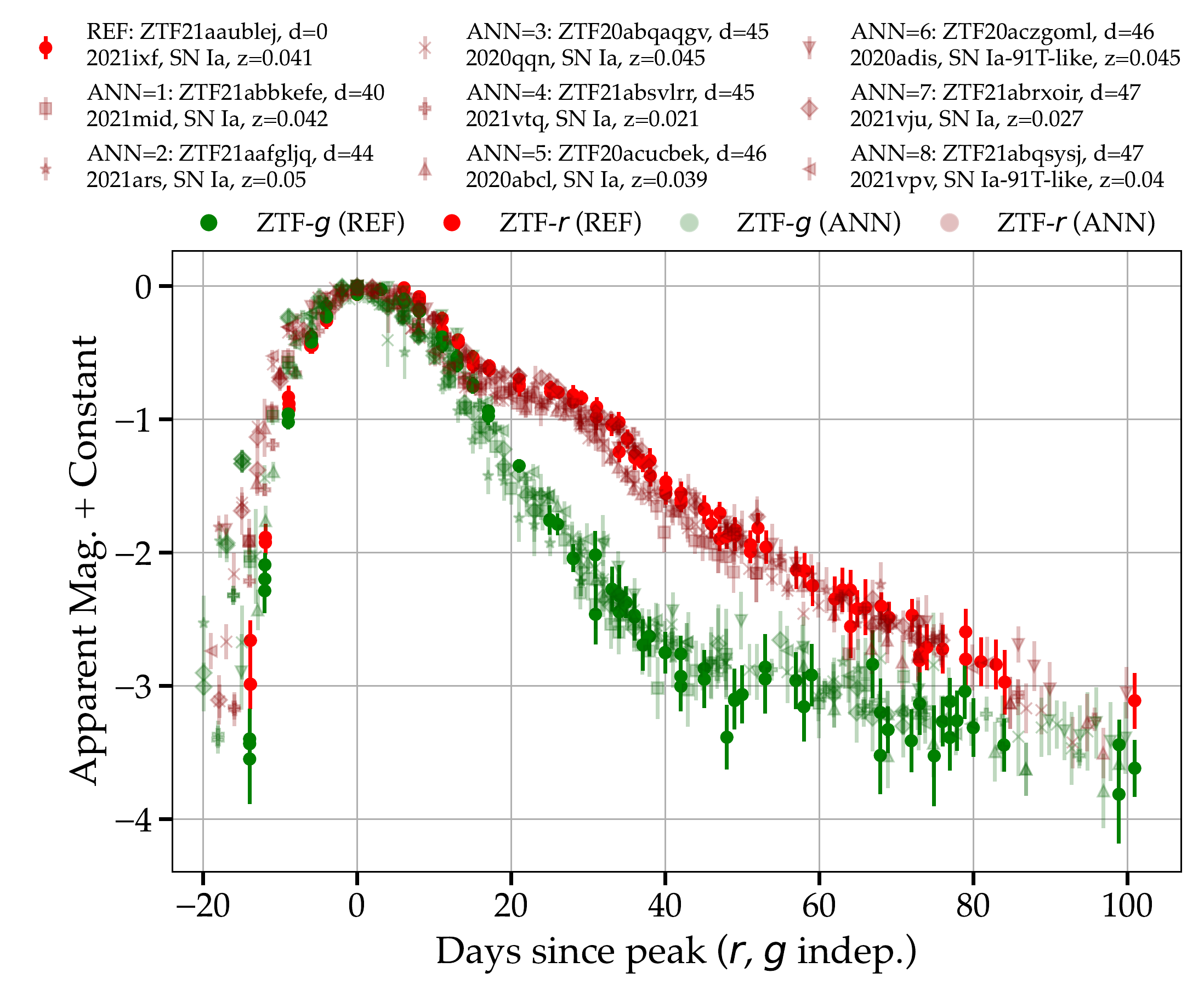}
    \includegraphics[width=0.859 \columnwidth]{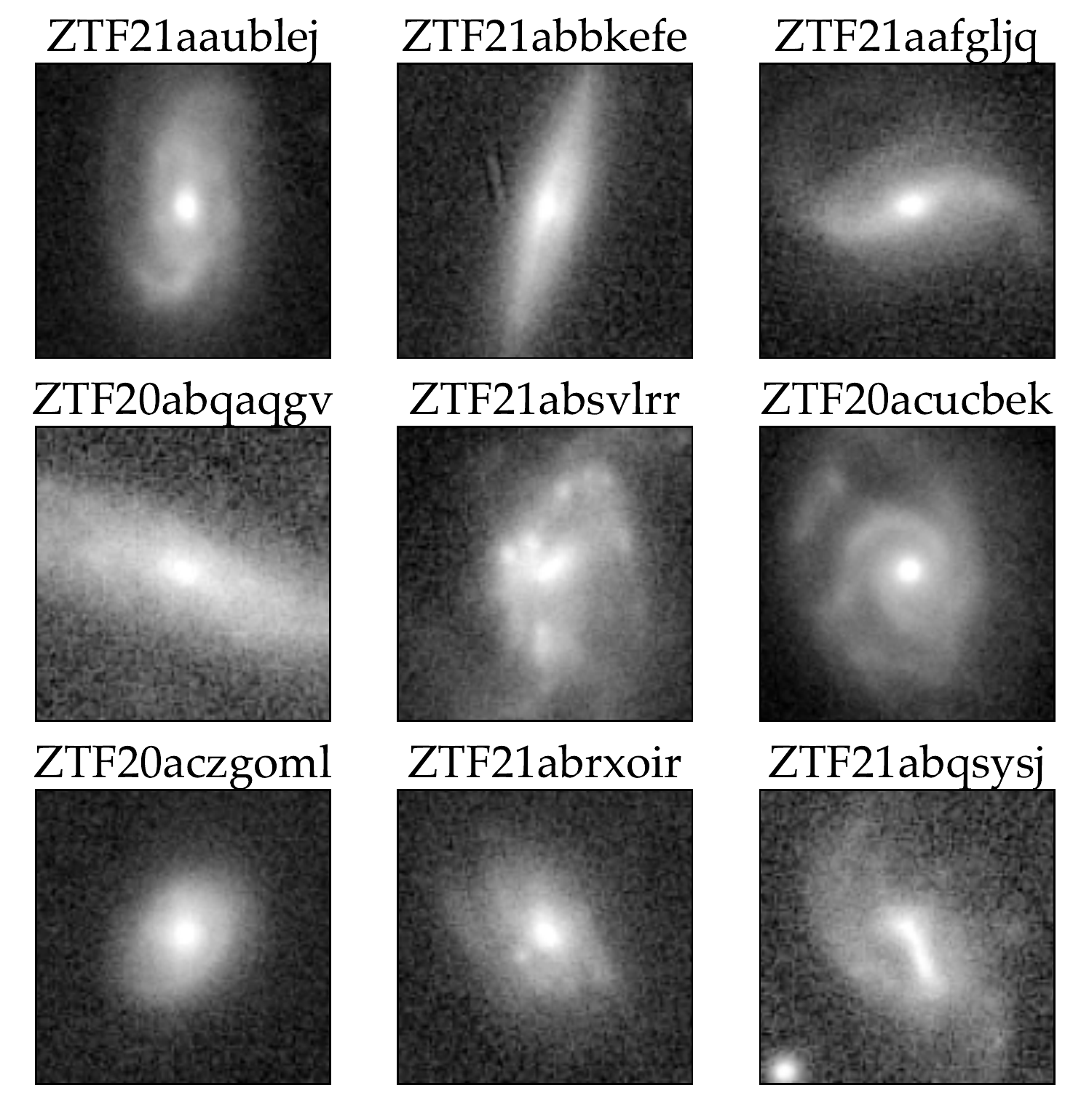}
    \caption{
    Stacked light curve (left panel) and grid of host galaxy thumbnails (right panel) of Type Ia SN~2021ixf (bolded) and its 8 approximate nearest neighbors within 120 d feature-space (faded). The similarity search in our database was completed in only 3~ms. Note that we do not include SN type or redshift information in the similarity search. Thus, the similar light curve evolution, SN types, redshift values, and host galaxy environments are as a result of the close proximity of SN~2021ixf and its neighbors in 120-dimensional feature-space.
    } 
    \label{fig:lc+host}
\end{figure*}

To our knowledge there is no popular and lightweight tool for finding transient analogs like SNe. To demonstrate such a utility, we present an ANN similarity search of the most commonly observed SNe (a Type~Ia) in Figure~\ref{fig:lc+host}. As a reminder, we only input the 60-dimensional PC feature-space (from the 120-dimensional light curve and associated host galaxy feature-space). We \emph{do not} input the photometric observations themselves, nor the redshift, SN type, galaxy type (spiral, elliptical), or thumbnail image (though this is a topic to be explored in future work). For visual aid, in the left panel we overlay the object of interest (ZTF21aaublej/SN~2021ixf) and its $k=8$ closest ANNs, shifting the light curves such that the peak brightness in the ZTF-$r$ band and ZTF-$g$ band is at 0~d since peak (both passbands fit independently). In the right panel, we show the thumbnail of the host galaxy of our object of interest in the top left, and the 8 ANNs continuing from left to right, top to bottom. This similarity search in our database was completed in 3~ms. \par

We find that the reference SN~2021ixf, in part because it is a well-sampled, full-phase, common SN~Ia at a nearby redshift, is indeed highly similar in light curve evolution and host galaxy environment as its 8 ANNs. All light curves increase in brightness by $\approx$3~mag in 16~d to 20~d to peak from their first observation, and have a prominent secondary ZTF-$r$ band bump at around 30~d post-peak, lasting about 100~d in total. Our reference object of class SN~Ia-normal at $z=0.041$ is matched to 6 SN~Ia-normal and 2 SN~Ia-91T-like spanning $z\in[0.021, 0.05]$, with a mean redshift of $\bar{z}=0.039$. The host galaxies are all nearby, face-on spiral galaxies except for one which is edge-on (SN~2021mid/ZTF21abbkefe). Another also exhibits evidence of a violent merging history (SN~2021vtq/ZTF21absvlrr). \par

\subsection{Reclassification of SNe}  \label{subsec:reclass_sne}

Here, we briefly test the ansatz that similar transient types broadly have similar light curve evolutions and host galaxy environments, investigating whether or not it proves fruitful to look for nearest neighbors of rare reference objects. Although there is no guarantee that nearest neighbors of an object are of the same type, it is a higher likelihood than random sampling. In this search, we find that in doing so we additionally recover transients of rare types that were initially misclassified.

\begin{figure*}
    \centering
    \includegraphics[width=1.131\columnwidth]{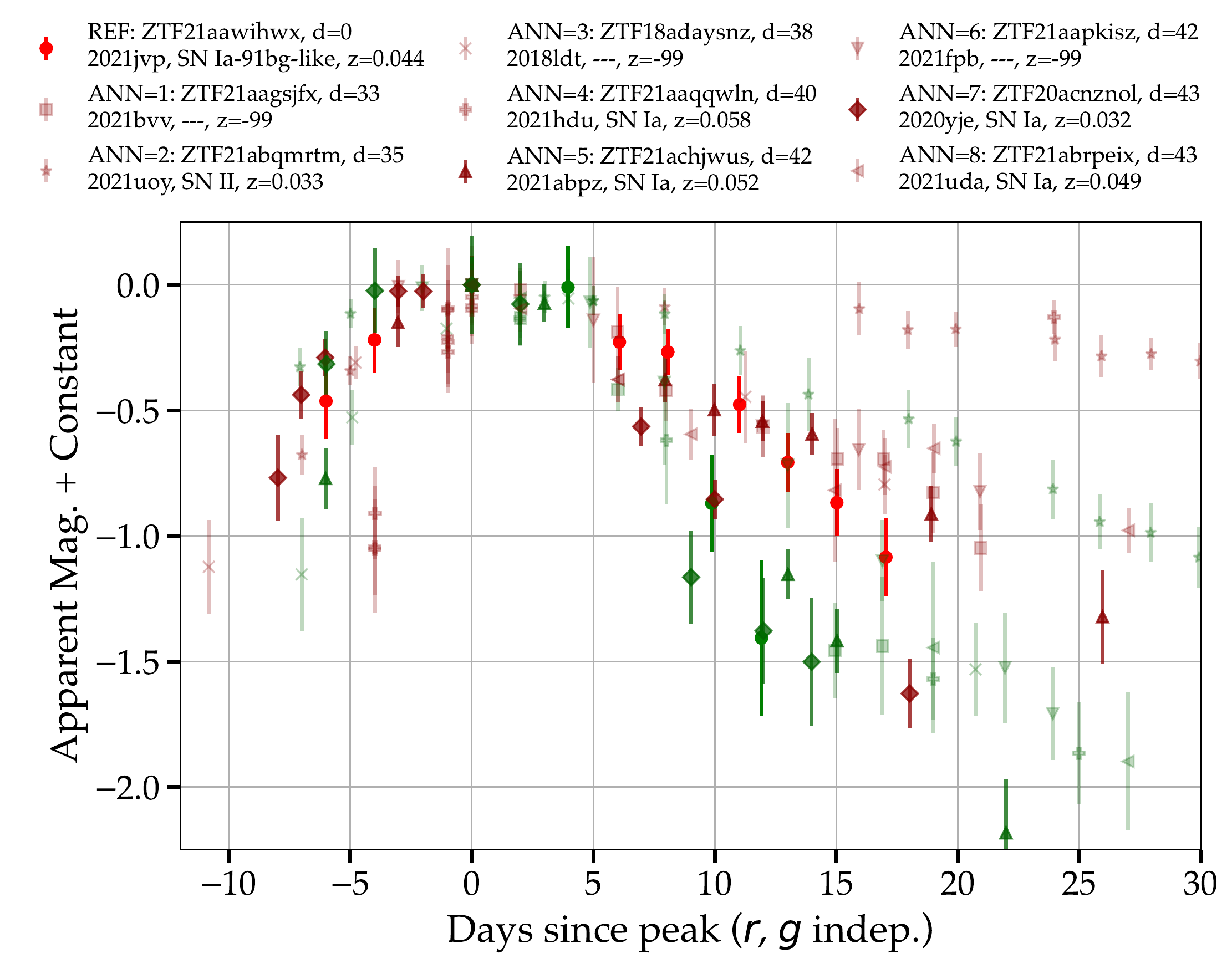}
    \includegraphics[width=0.859\columnwidth]{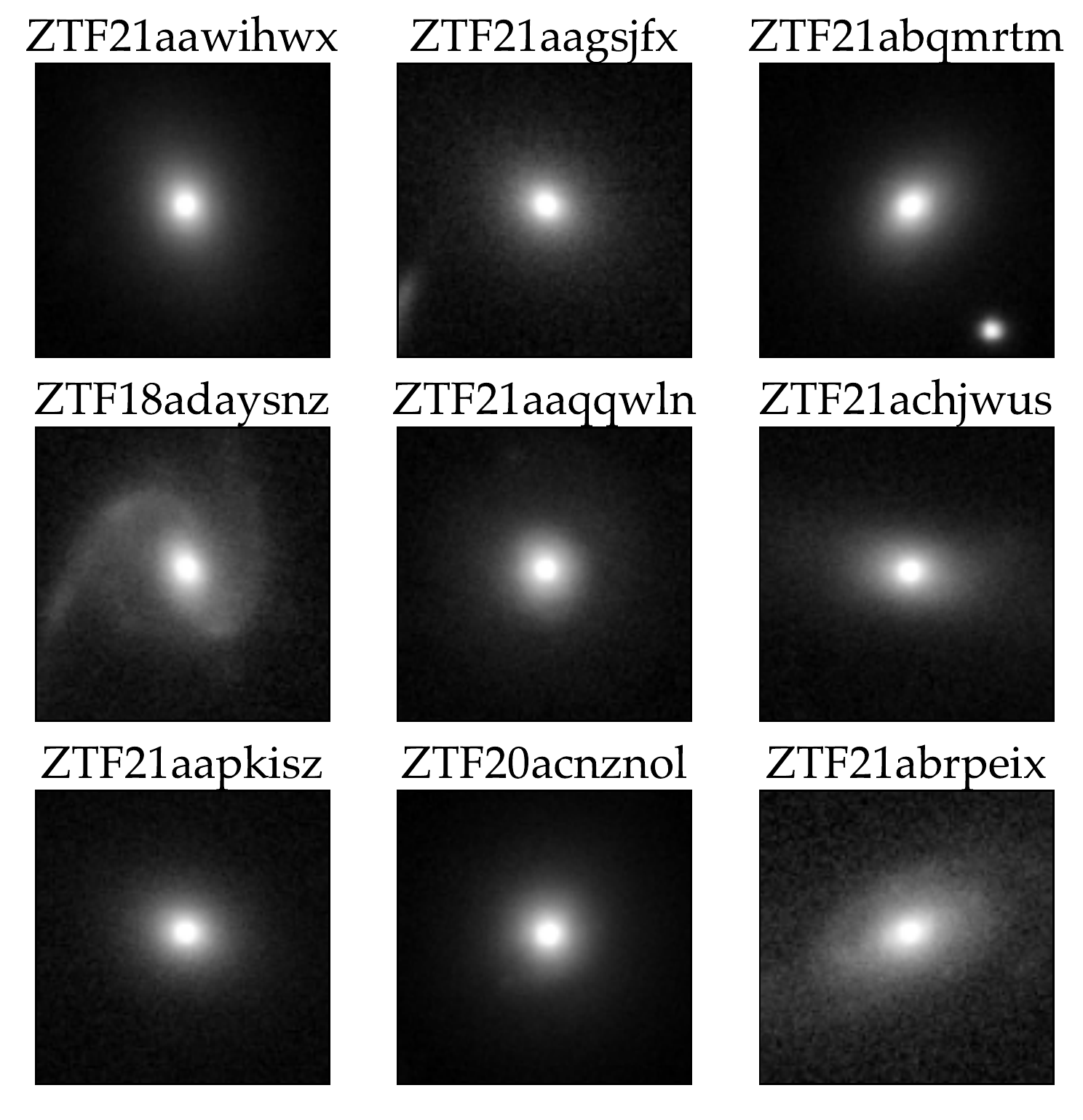}
    \caption{
    Same as Figure~\ref{fig:lc+host}, but for Type Ia-91bg-like SN~2021jvp/ZTF21aawihwx. In this case, ANN=5 (SN~2021abpz/ZTF21achjwus) and ANN=7 (SN~2020yje/ZTF20acnznol) are in fact SN~Ia-91bg-like SNe previously missclassified as SN~Ia-normal after re-examination of their TNS classification spectra, prompted by the results of this similarity search (see Table~\ref{tab:reclassification_SNe}). We bold these re-classified events (dark green and dark red for ZTF-$g$ and ZTF-$r$, respectively) for visual purposes. The similarity search in our database was completed in only $\sim$300~ms.
    } 
    \label{fig:lc_missed91bg}
\end{figure*}

In Figure~\ref{fig:lc_missed91bg}, we show the Type Ia-91bg-like SN~2021jvp/ZTF21aawihwx and its 8 ANNs, achieved in $\sim$300~ms. Initially, if we look at the original TNS classification labels only, we find 4 SN~Ia, 1 SN~II, and 3 with no classification spectrum (for which we make no definitive determination). Prompted by the nearest neighbor results, we manually re-inspect the classification spectra from TNS for these 5 objects using the Supernova Identification package (\texttt{SNID}, \citealt{Blondin2007}). We use the 5.0 version of \texttt{SNID} with additional template sets from the Berkeley Supernova Ia Program (BSNIP, \citealt{Silverman2012}), \cite{Modjaz2014, Liu2014, Liu2016, Modjaz2016, Gutirrez2017, Williamson2019}, totalling 6145 spectra from 811 templates. We use the \textit{forcez} argument for any object that has a known host-$z$. \par

\begin{table*}[ht]
\centering
\caption{\textbf{Reclassification of SNe} Updated classification of 17 unique SNe driven by ANN=8 nearest neighbor matches to all SNe classes using \texttt{SNID}}
\begin{tabular}{ccccccc}
\hline
\hline
ZTF ID & IAU Name & TNS Class. & New Class. & Reference SNe & ANN & Remarks \\
\hline
\hline
\multicolumn{7}{c}{\textit{Reference Class: SN~IIn (59 objects)}} \\
ZTF21abcjpnm & 2021njo & SN~II & SN~IIn & 2020abku & 1 & Second peak likely CSM interaction. \\
ZTF21aaizyqc & 2021ckb & SN~II & SLSN (He-rich) & 2021hur & 2 & Match to PTF10hgi \citep{Gal-Yam2019ARA}. \\
\hline
\multicolumn{7}{c}{\textit{Reference Class: SN~Ia-91T-like (37 objects)}} \\
ZTF21aagoliy & 2021cjc & SN~Ia & SN~Ia-91T-like & 2020zjv & 1 & Match to SN~1997br, SN~1991T. \\
ZTF21abjtqyq & 2021sis & SN~Ia & SN~Ia-91T-like & 2020acef & 1 & Match to SN~1999aa. \\
ZTF21abothvq & \textit{2021uib} & SN~Ia & SN~Ia-91T-like & " & 6 & Matches to SN~2001V, SN~1991T. \\
ZTF21abicgai & 2021sju & SN~Ia & SN~Ia-91T-like & 2020adis & 3 & Match to 1997br. \\
ZTF21abcxner & 2021nxh & SN~Ia & SN~Ia-91T-like & 2021qvg & 8 & Lack of Si. Match to SN~1991T. \\
\hline
\multicolumn{7}{c}{\textit{Reference Class: SN~Ib (25 objects)}} \\
ZTF21aaqwfqe & 2021hen & SN~I & SN~Ib & 2021gno & 2 & Match to iPTF13bvn. \\
ZTF21aabyifm & 2021qv & SN~Ib/c & SN~Ib & " & 5 & Matches to iPTF13bvn, SN~2009iz. \\
\hline
\multicolumn{7}{c}{\textit{Reference Class: SN~IIb (21 objects)}} \\
ZTF21abnvlnj & 2021tyf & SN~II & SN~IIb & 2021M & 8 & LC has shock-cooling peak. \\
\hline
\multicolumn{7}{c}{\textit{Reference Class: SN~Ic-BL (14 objects)}} \\
ZTF21aacufip & 2021vz & SN~Ic & SN~Ic-BL & 2021too & 4 & Match to SN~2007ce. Bad quality spectrum. \\
\hline
\multicolumn{7}{c}{\textit{Reference Class: SN~Ia-pec (10 objects)}} \\
ZTF20ackkejs & 2020xyd & SN~Ia & SN~Ia-91T-like & 2021cky & 8 & Lack of Si and bright ($M_{abs}=-19.65$). \\
ZTF21abothvq & \textit{2021uib} & SN~Ia & SN~Ia-91T-like & 2021ebb & 1 & Matches to SN~2001V, SN~1991T. \\
ZTF21abiawpf & 2021rce & SN~Ia & SN~Ia-91T-like & " & 5 & Matches to SN~2006cz, SN~2007S. \\
\hline
\multicolumn{7}{c}{\textit{Reference Class: SN Ia-91bg-like (6 objects)}} \\
ZTF20acnznol & \textit{2020yje} & SN~Ia & SN~Ia-91bg-like & 2021jvp & 7 & Matches to SN~2007ba, SN~2000cn. \\
" & \textit{"} & SN~Ia & SN~Ia-91bg-like & 2021wzb & 6 & " \\
ZTF21achjwus & \textit{2021abpz} & SN~Ia & SN~Ia-91bg-like & 2021jvp & 5 & Matches to SN~1986G, SN~1999bh. \\
" & \textit{"} & SN~Ia & SN~Ia-91bg-like & 2021wzb & 8 & " \\
ZTF21acfigoo & 2021aazj & SN~Ia & SN~Ia-91bg-like & 2021fnr & 2 & Matches to SN~1986G, SN~2007ax. \\
ZTF21abmwgow & 2021ttg & SN~Ia & SN~Ia-91bg-like & 2021wzb & 2 & Matches to SN~2008R, SN~2007ap. \\
\hline
\hline\\[-1.5ex]
\multicolumn{7}{c}{
\begin{minipage}{16cm}
NOTE: Italicised text designates transients that appear at least twice in the Table.\\
NOTE: The double quotation " marks a repeated entry from the above row.
\end{minipage}}
\end{tabular}
\label{tab:reclassification_SNe}
\end{table*}

We find SN~2021abpz (ANN=5) and SN~2020yje (ANN=7) are better explained by the SN~Ia-91bg-like classification---the same as reference SN~2021jvp. In SN~2021abpz, there is a visible titanium trough at 4200\AA~\citep{Hachinger2009, Heringer2017} and strong \texttt{SNID} matches to subluminous / Type Ia-91bg-like SN~1986G, SN~1999bh a few days after peak, in phase with when the spectrum was taken. SN~2021abpz has a peak absolute magnitude of $M\sim-18.5$~mag, consistent with subluminous Ia. Similarly, SN~2020yje has strong matches to Type Ia-91bg-like SN~2007ba and SN~2000cn a few days before peak, with a peak absolute magnitude of $M\sim-17.6$~mag, also consistent with subluminous Ia. Moreover, both SN~2021abpz and SN~2020yje are in elliptical hosts and offset from the center---common for Type Ia-91bg-like events \citep{Barkhudaryan2019}. Their light curves are red and fast declining. We additionally find both objects are among the first 8 ANNs of another SN~Ia-91bg-like object, SN~2021wzb. \par

We repeat this procedure and manually inspect the TNS classification spectra of all classified transients within the first 8 ANNs for the spectroscopic objects in our databank. Any updated classifications for classes in which at least one discovery was made is shown in Table~\ref{tab:reclassification_SNe}. In the majority of cases, the original classification labels are correct and we do not reclassify. Of our entire spectroscopic sample of 1620 objects, we find 17 unique objects that are better explained by a different classification label, or about 1\%. Although this percentage is small, of the 17 objects we reclassify, 8 (47\%) were reclassified from a non-anomalous class (which in this work are SN~Ia, SN~Ia-91T-like, SN~II, and SN~IIP) to an anomalous class. This brings the total anomalous spectroscopic sample from 228 objects to 236, an increase of 3.5\%. Because of the intrinsic rarity of anomalous objects, a 3.5\% increase in our overall spectroscopically anomalous sample from a directed manual re-evaluation of their spectra is a marked difference. \par

In our original spectroscopic sample of 1620 objects, only 6 were originally of this SN~Ia-91bg-like classification. After searching this same data set for the 8 ANNs of the 6 known SN~Ia-91bg-like objects, we find an additional 4 SN~Ia-91bg-like objects originally misclassified as SN~Ia-normal, an increase of 67\%. This nearly doubles our total to 10 SN~Ia-91bg-like objects. \par

\subsection{Missed Opportunity SNe - Previously Reported to TNS}  \label{subsec:previously_reported}

We can only reclassify SNe that have follow-up spectra. However, in principle we can repeat the same ANN=8 search of anomalous transients to look for potential anomalous candidates that were reported to TNS whilst active but have no classification spectrum. We call these ``missed opportunity" transients. Although these events have faded, we identify unclassified ZTF transients with a high likelihood of being anomalous. While the entire search is ongoing and its entirety is beyond the scope of this work, we do identify a few candidates for which there is a strong possibility of being a TDE, using the FLEET-TDE classifier from \cite{Gomez2023TDE}. \par

\begin{figure*}
    \centering
    \includegraphics[width=1.120\columnwidth]{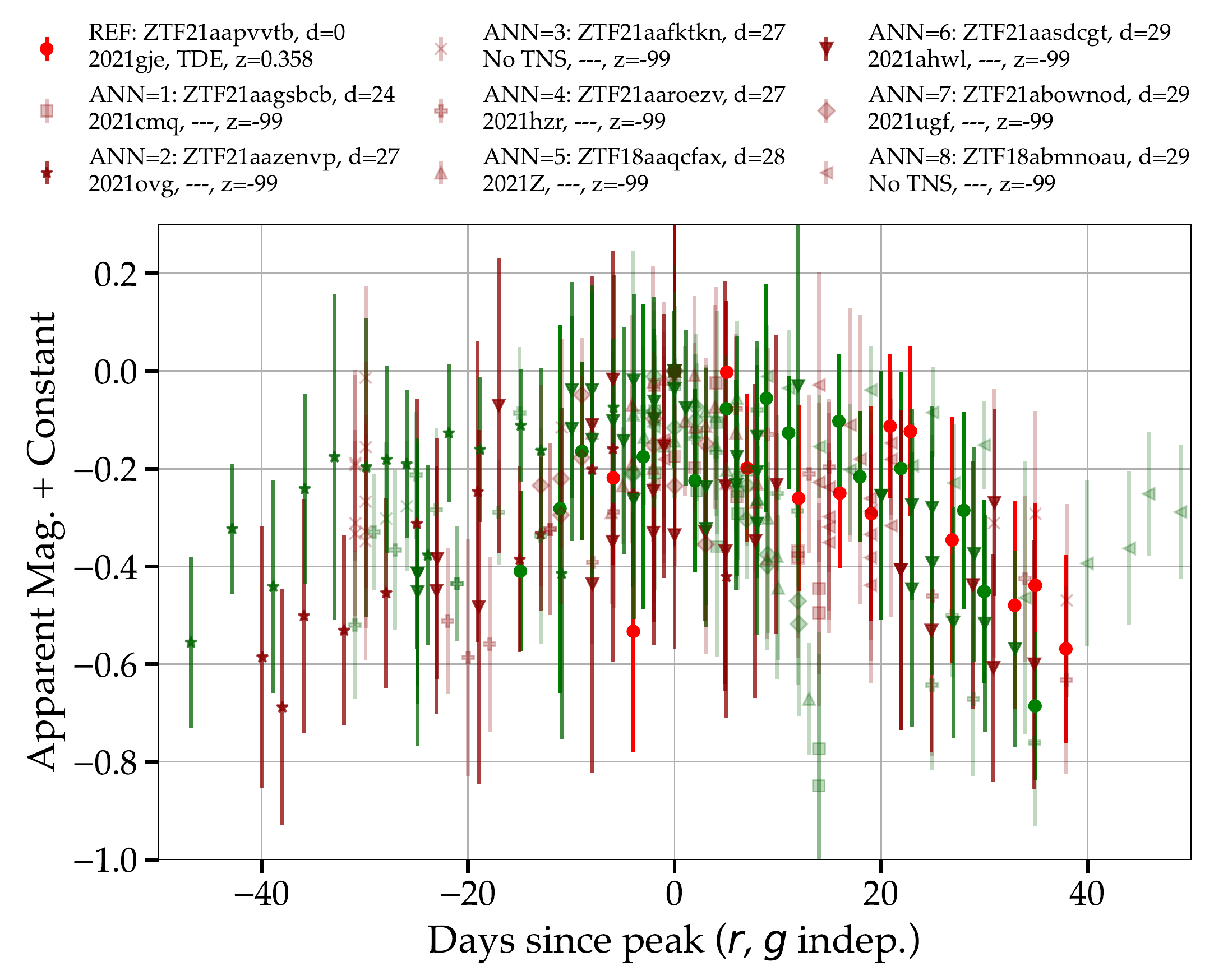}
    \includegraphics[width=0.870\columnwidth]{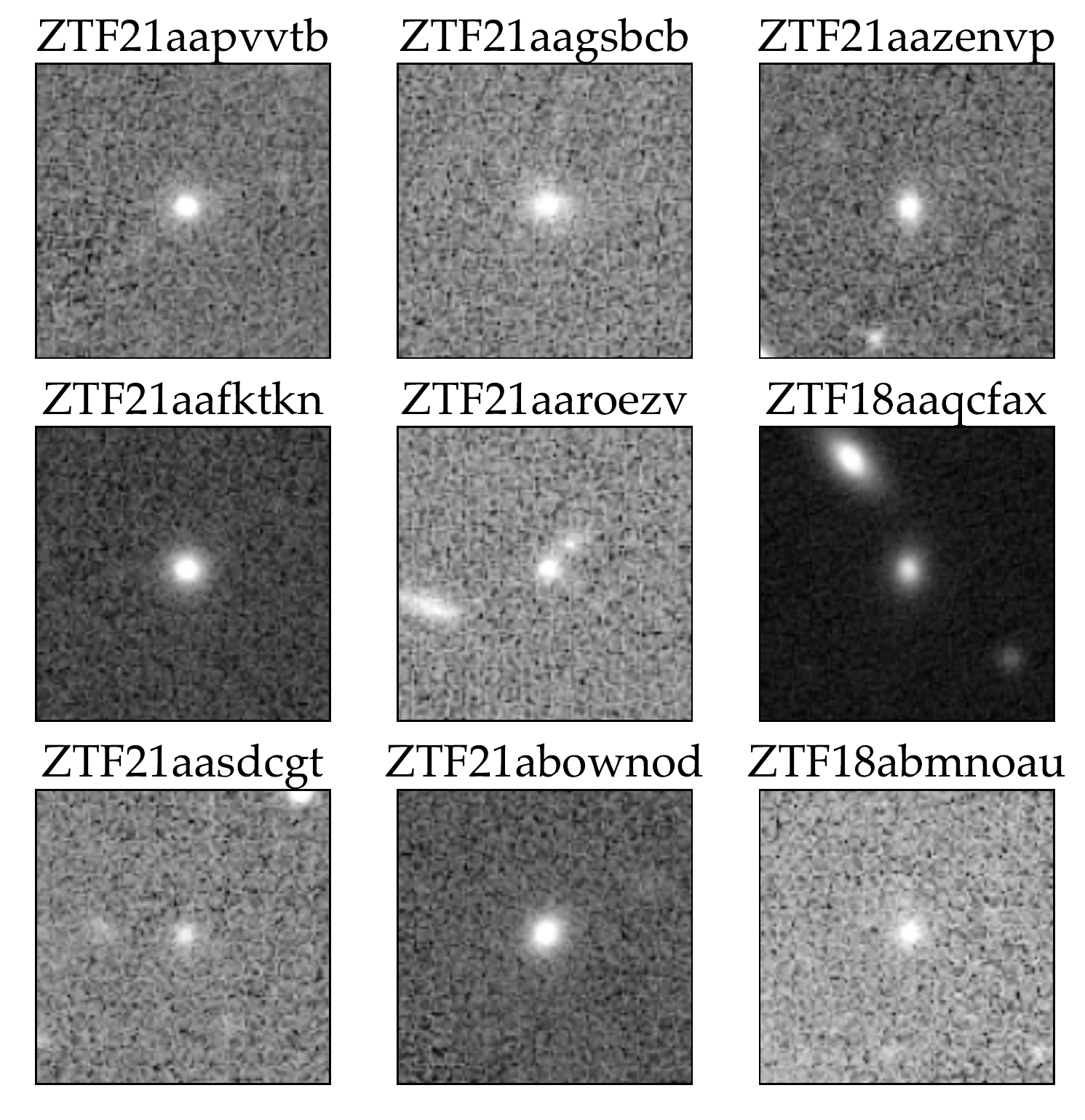}
    \caption{
    Same as Figures~\ref{fig:lc+host}, \ref{fig:lc_missed91bg} but for TDE~2021gje/ZTF21aapvvtb. In this case, ANN=2 (AT~2021ovg/ZTF21aazenvp) was previously reported to TNS but never previously identified as a TDE candidate until this work (FLEET=85\% TDE), and ANN=6 (AT~2021ahwl/ZTF21aasdcgt) is a missed TDE candidate (FLEET=70\% TDE; now reported to TNS from this work). We bold these new TDE candidates (dark green and dark red for ZTF-$g$ and ZTF-$r$, respectively) for visual purposes. The similarity search in our database was completed in only $\sim$230~ms.
    } 
    \label{fig:lc_missedTDE}
\end{figure*}

For a directed search, we examine the first 8 ANNs of each of the 20 spectroscopic TDEs from our databank to look for missed opportunity TDEs that have yet to be reported in the literature. For instance, we use as reference 2021gje, a TDE at $z=0.358$. We find the 8 ANNs in $\sim$230~ms, as shown in Figure~\ref{fig:lc_missedTDE}. Of those, 2 are AGNs which have survived all our selection cuts (ANN=3, ZTF21aafktkn; ANN=8, ZTF18abmnoau), 3 are likely SN~Ia (AT~2021cmq\footnote{This object likely has a host misassociation, but is correct in the currently running \texttt{GHOST} version.}, AT~2021Z, AT~2021ugf), 1 is an SLSN candidate\footnote{AT~2021hzr has $P(\text{SLSN-II})=68$\% according to FLEET, though, FLEET is not optimized for SLSN-II classifications.} (AT~2021hzr), and 2 are likely TDEs or AGN flares with no previous reporting as such in the literature (AT~2021ovg, AT~2021ahwl). Moreover, AT~2021ahwl was never reported to TNS until this work (and thus was not considered in \citealt{Gomez2023TDE}). \par

AT~2021ovg has a $P(\text{TDE})=85\%$ from FLEET from the late-time classifier, but there is no post-peak photometry, and thus the usage of the late-time classifier is not fully appropriate here. From the FLEET-TDE classifier for pre-peak, this object is assigned a $P(\text{AGN})=60\%$ and $P(\text{TDE})=32\%$. It is nuclear, blue, and there are no previous signs of variability according to ZTF~DR data. The ALeRCE light curve classifier \citep{SanchezSaez2021} assigns the highest probability as a SLSN at 34\% (there is no TDE option), with the second highest probability being an AGN (16\%). At an estimated redshift of $z=0.241$ from SDSS \citep{Csabai2003}, we estimate an absolute magnitude of at least $M_{abs}\approx-20.5$~mag, as there is no visible peak. AT~2021ovg was not reported in \cite{Gomez2023TDE}. \par

AT~2021ahwl is a moderately strong missed TDE candidate at $P(\text{TDE})=70\%$ from FLEET. It is nuclear, blue, visible for $\sim$60~d, and there are no previous signs of strong variability according to ZTF~DR data. At an estimated redshift of $z=0.364$ from SDSS \citep{Csabai2003}, we estimate a peak absolute magnitude of $M_{abs}\approx-21.8$~mag. Using the estimated redshift of $z=0.372$ from PS1-STRM \citep{Beck2021}, we estimate a peak absolute magnitude of $M_{abs}\approx-21.9$~mag. Both redshift estimates, if correct, are consistent with the reference $z$ at 0.358. \par

We repeat a likewise examination for the 20 spectroscopic TDEs from our databank, and find one strong TDE candidate (AT~2021agpi with $P(\text{TDE})=95\%$ from the FLEET late-time classifier), and three questionable ones (AT~2020hip, AT~2020yaf, AT~2021stx) that could be TDE or possibly AGN/ AGN flares, or enhanced accretion-driven AGN flares \citep{Trakhtenbrot2019}. Spectroscopic observations are warranted for confirmation, though this opportunity has passed. \par

Similarly, because FLEET is also optimized for SLSN-I classification and discovery, we perform another directed search and examine the first 8 ANNs of each of the 11 spectroscopic SLSN-I and 14 spectroscopic SLSN-II from our databank to look for missed opportunity SLSNe-I that have yet to be reported in the literature. We identify one moderately strong SLSN-I candidate (AT~2021fao with $P(\text{SLSN-I})=64\%$ from the FLEET late-time classifier) and one borderline candidate (AT~2021lnu with $P(\text{SLSN-I})=49\%$ from the FLEET late-time classifier\footnote{Although not formally discussed in the literature, this object does have an Astronote identifying it as a ZTF SLSN candidate \citep{Perley2021lnu}.}). \par

\subsection{Missed Opportunity SNe - Previously Unreported to TNS}  \label{subsec:previously_unreported}

\begin{figure*}
    \centering
    \includegraphics[width=1.131\columnwidth]{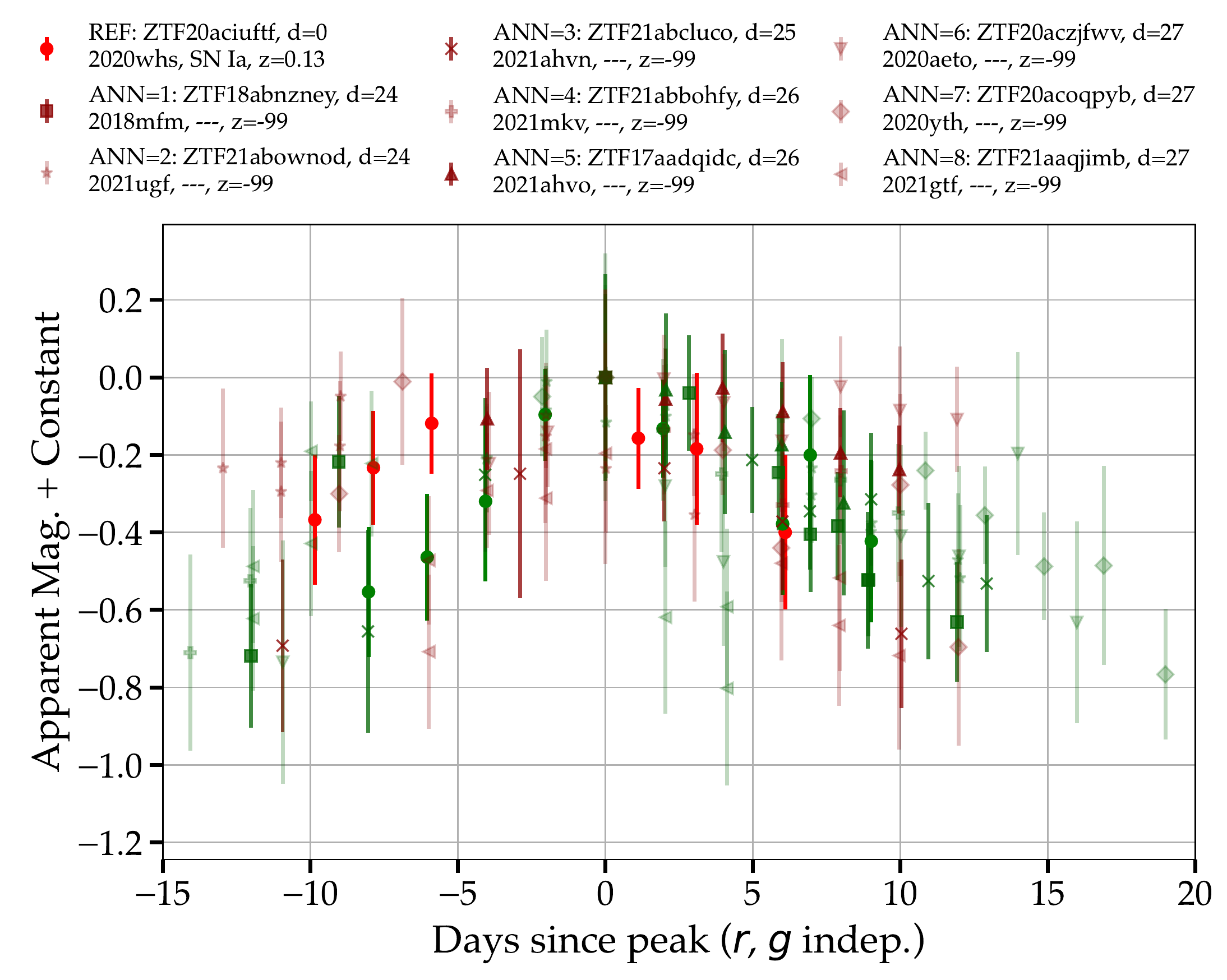}
    \includegraphics[width=0.859\columnwidth]{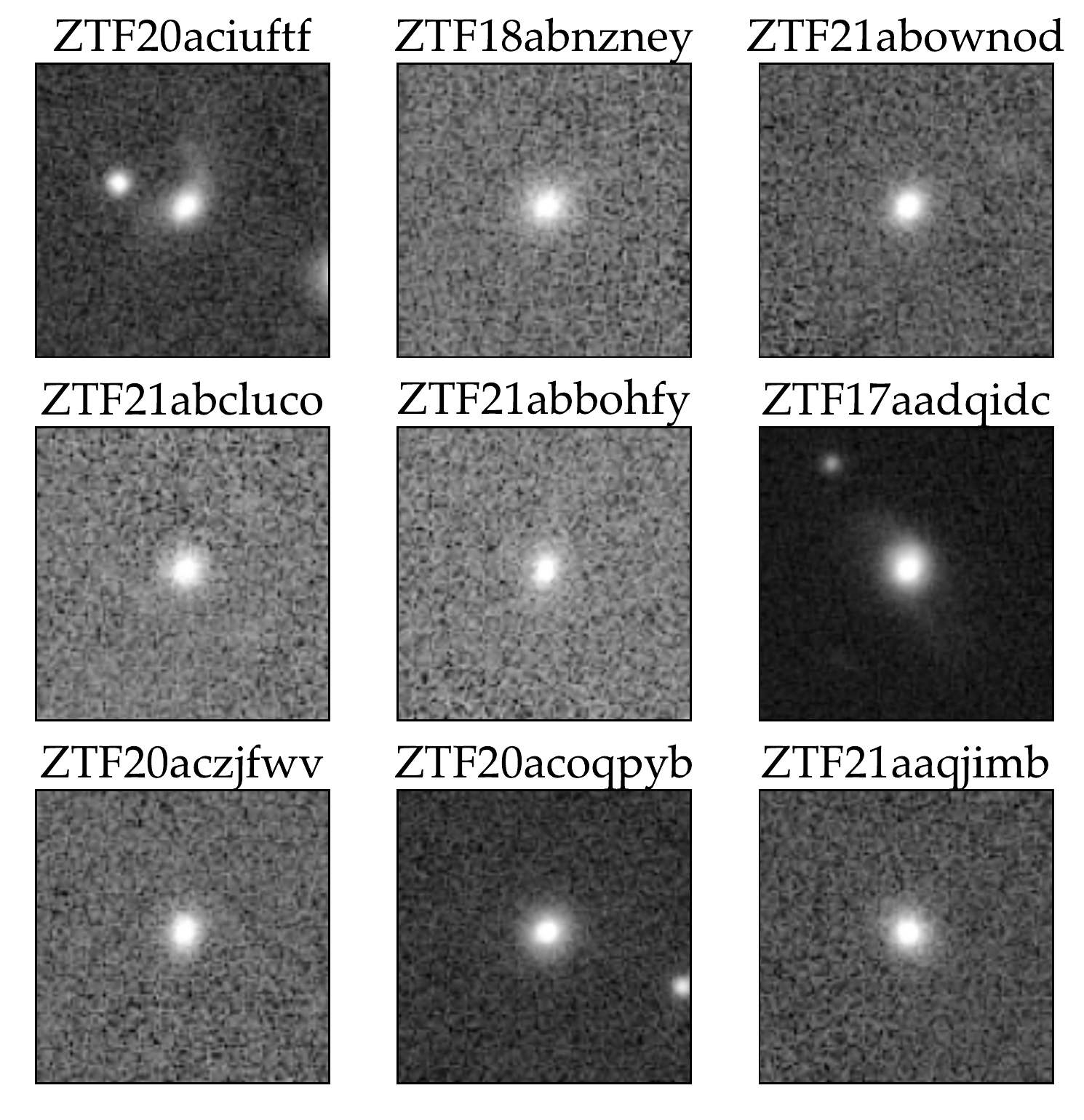}
    \caption{
    Same as Figures~\ref{fig:lc+host}, \ref{fig:lc_missed91bg}, \ref{fig:lc_missedTDE}, but for Type Ia SN~2020whs/ZTF20aciuftf. In this case, ANN=1 (AT~2018mfm/ZTF18abnzney), ANN=3 (AT~2021ahvn/ZTF21abcluco), and ANN=5 (AT~2021ahvo/ZTF17aadqidc) were missed SNe candidates (now reported to TNS from this work). We bold these SN discoveries (dark green and dark red for ZTF-$g$ and ZTF-$r$, respectively) for visual purposes. The similarity search in our database was completed in only $\sim$140~ms.
    } 
    \label{fig:lc_missingSNe}
\end{figure*}

Beyond identifying possible rare transients or anomaly candidates that were reported but have no spectroscopic follow-up, we can go the next step further and search for ANNs of known transients to find previously undiscovered or unreported transients (i.e., not reported to public catalogs like TNS). The idea here is that if a known transient has a specific feature distribution, (approximate) nearest neighbors with similar distributions are likely to also be transients, and perhaps a fraction are not reported. An illustration of this directed search is found in Figure~\ref{fig:lc_missingSNe}, whose search was completed in $\sim$140~ms.

The reference used is Type Ia SN~2020whs/ ZTF20aciuftf at $z=0.13$. From this ANN=8 search, 5 were already reported to TNS, and thus we manually investigated the remaining 3 with no submission. Upon expert analysis, we find that all 3 are missed SNe candidates: ANN=1 (AT~2018mfm/ ZTF18abnzney), ANN=3 (AT~2021ahvn/ ZTF21abcluco), and ANN=5 (AT~2021ahvo/ ZTF17aadqidc).

The 1st ANN, AT~2018mfm/ ZTF18abnzney is approximately 1\farcs38 from its host (WISEA J000044.86+152956.7). It peaks in ZTF-$g$ at an apparent magnitude of 19.8~mag with a rise of 11~d, compared to the 19.6~mag peak apparent magnitude of reference SN~2020whs with an 8~d rise. At the time of the transient, AT~2018mfm has a peak brightness about 1~mag brighter than the baseline from ZTF~DR5. Because of the short timescale and evolution, this is likely an SN~Ia. \par

The 3rd ANN, AT~2021ahvn/ ZTF21abcluco, is remarkably similar in evolution to reference SN~2020whs. Both have a first epoch in ZTF-$g$ at an apparent magnitude of 20.2~mag, rise to peak in 8~d to a peak apparent magnitude in ZTF-$g$ of 19.6~mag, and fade in 12-13~d. This candidate has an estimated photo-$z$ from from PS1-STRM \citep{Beck2021} of $z=0.168$, similar to the spectroscopic redshift of reference SN~2020whs ($z=0.13$). This candidate, at approximately 0\farcs61 from the center of host WISEA J115609.31+210137.8, has an estimated peak absolute magnitude consistent with that of an SN~Ia. \par
 
The 5th ANN, AT~2021ahvo/ ZTF17aadqidc, is at the center (0\farcs17) of a galaxy (WISEA J100313.80+283944.8) with a spectroscopic redshift of 0.087, placing the peak $M_{abs}\sim-18.6$~mag, well within the expected $M_{abs}$ range of an SN~I. Moreover, this object has 7 ZTF-$r$ and 5 ZTF-$g$ alert stream observations and an apparent peak magnitude of $m\sim19.4$~mag, which is about one magnitude brighter than the limiting magnitude of ZTF, making this a viable candidate for SN detection algorithms. A possible reason for why this was initially missed is due to its nuclear location, though this candidate's photometry is $\sim0.3$~mag above the baseline of from the ZTF~DR and is red $g-r\approx0.4$~mag (disfavoring an AGN flare). The ALeRCE light curve classifier \cite{SanchezSaez2021} predicts this candidate as an SN~Ib/c. \par

We likewise perform a directed search by calculating the first 8 ANNs of all spectroscopic transients in our databank, and investigate any non-reported ANNs for transient activity. In summary, we find that a fraction were never reported in public catalogs, totaling 84 transient candidate discoveries (not including likely AGN), all of which can be found in Table~\ref{tab:report_to_TNS_SNe}. \par

Finally, to be complete in recovering missed SNe from our databank, we perform an exhaustive search, visually inspecting each unique object not previously reported to TNS and not previously discovered through our ANN search. That search yielded 241 SNe candidates, as shown in Table~\ref{tab:remaining_no_TNS}. \par

In total, we report the discovery of 325 transients, all observed between 2018-2021 and absent from public catalogs ($\sim$1\% of all ZTF AT reports to TNS through 2021). \par



\begin{figure}
    \centering
    \includegraphics[width=\columnwidth]{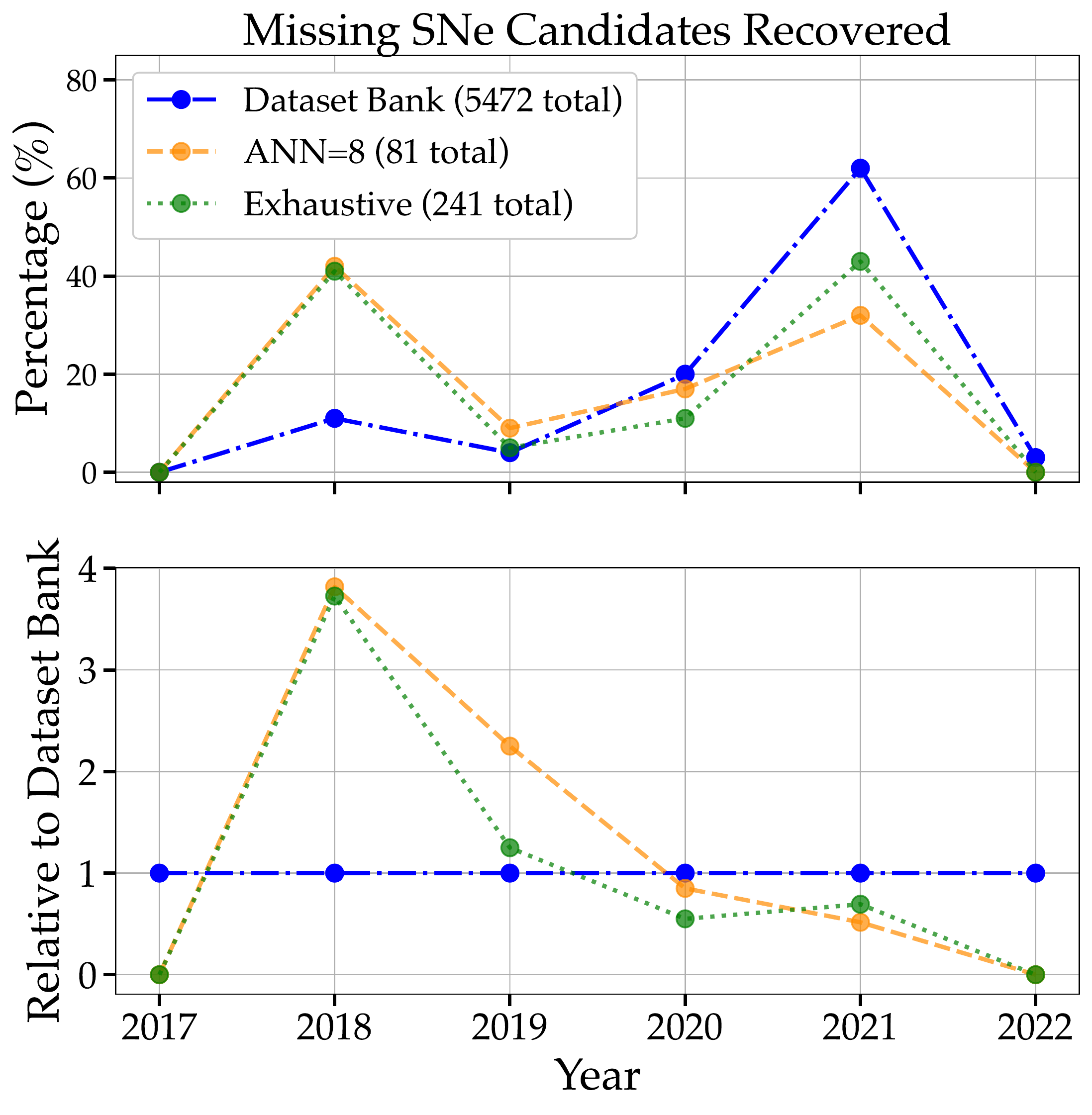}
    \caption{
    The unweighted percentage (top panel) and weighted (bottom panel) percentage relative to the fraction present in our original databank (blue) of the retroactive SNe candidate discoveries recovered by our ANN=8 search method (gold) and exhaustive method (green). From both methods combined, we recover nearly equal numbers from 2018 and 2021 (134 out of 325 (41\%) and 131 out of 325 (40\%), respectively). However, relative to the yearly fractions that comprise our databank, we find that the fraction of our total 2018 discoveries relative to all discoveries from this work (41\%) is nearly 3 times that of the fraction relative to that which comprise our databank (11\%), whereas our total 2021 discoveries (40\%) make up nearly two-thirds relative to our databank (62\%). Overall, the weighted fraction of discoveries is greatest in 2018, the first full year of the ZTF survey, and decreasing every year through 2022, implying an improvement in detection and reporting methods via increasing numbers of active broker teams in the ZTF Alert Stream.
    } 
    \label{fig:recovered_SNe}
\end{figure}

To better understand how our retroactive candidate SNe discoveries are distributed as a function of time, we plot in Figure~\ref{fig:recovered_SNe} the unweighted percentage (top panel) and weighted (bottom panel) percentage relative to the fraction present in our original databank (blue) identified by our ANN=8 search method (gold) and exhaustive method (green). The majority (134 out of 325, or 41\%) of our candidate discoveries are from 2018 (with contributions of 35 out of 84 (42\%) from our ANN=8 search method and 99 out of 241 (41\%) from our exhaustive search method). Contrary to our expectation, this is nearly equal to that of 2021, where we identify 131 out of 325 (also 40\%) total missed candidates. However, our databank contains an uneven distribution of reported candidates per year. Grouping the IAU names, approximately 0\% are from 2017, 11\% are from 2018, 4\% are from 2019, 20\% are from 2020, 62\% are from 2021, and 3\% are from 2022. Thus, when we account for this unequal weighting, our result of 40\% of our identified missed candidates from 2021 becomes unsurprising given that 62\% of our databank is from 2021; it follows that there would be relatively more missed transients (numbers wise) for that year. \par

Note the weighted fraction of objects (bottom panel) missed in 2018 and 2019 is greater than that relative to 2020 and 2021. In fact, the weighted fraction is greatest in 2018, the first full year of the ZTF survey, and decreasing every year through 2022, implying an improvement in detection and reporting methods via increasing numbers of active broker teams in the ZTF Alert Stream.\footnote{Only three objects in our databank have a ZTF ObjectID from 2017, so although it is included for completeness, it effectively bears no impact on this analysis.} For instance, the ALeRCE broker did not start real–time ML classification of the ZTF alert stream until early 2019 \citep{Forster2021}, and the ANTARES broker did not start reporting to TNS until 2021. Fink did not begin processing the ZTF public live-alert stream until November 2019 \citep{Moller2021}. Between additional teams reporting to TNS and the addition of object detection algorithms characterizing the ZTF alert stream in subsequent years \citep[e.g.,][]{Muthukrishna2019,Andreoni2021, Duev2021,vanRoestel2021,Coughlin2021,Forster2021,SanchezSaez2021,Carrasco-Davis2021,Leoni2022,Aleo2022,Reyes-Jainaga2023}, it is sensible that we discover progressively fewer SNe candidates relative to those present in our databank. We encourage more retroactive study of the ZTF survey, particularly in 2018 and 2019 as these data are likely rife with undiscovered SNe and interesting candidates. \par

\section{Discussion} \label{sec:discussion}

\subsection{Light curve extrapolation} \label{subsec:lc_extrapolation}

Astronomical transients are the observational counterpart of terminal events, and a study of their photometric evolution can offer insights to underlying astrophysical properties. This can be performed on an individual basis (e.g., \citealt{Gagliano2021}), or across a population (e.g., \citealt{Nyholm2020}). Due to cadence, weather, and other factors intrinsic to astronomical observations, our photometry is irregularly sampled, noisy, and at times has large gaps spanning several epochs. Thus, in order to obtain homogeneous data suited for feature extraction and training machine and deep learning models requires fast but accurate approximation. For such reasons, there are popular parametric light curve models in the literature \citep{Bazin2009, Villar2019, Russeil2023} and non-parameteric models such as Gaussian process \citep[GP,][]{Boone2019,Demianenko2022} both of which predict the temporal flux evolution. \par

Note that although we make no such rigorous attempt at a formal light curve fit model, we can approximate a possible light curve evolution through considering the additional photometry and host galaxy environment from analogs to ``reconstruct" a light curve. To our knowledge, this idea has not been demonstrated in the literature for light curve fits, but similar ideas of applying ``twin" supernovae in regards to spectra for SN Ia standardization \citep{Fakhouri2015, Boone2021TEII} have shown promise. \par

\begin{figure*}
    \centering
    \includegraphics[width=1.120\columnwidth]{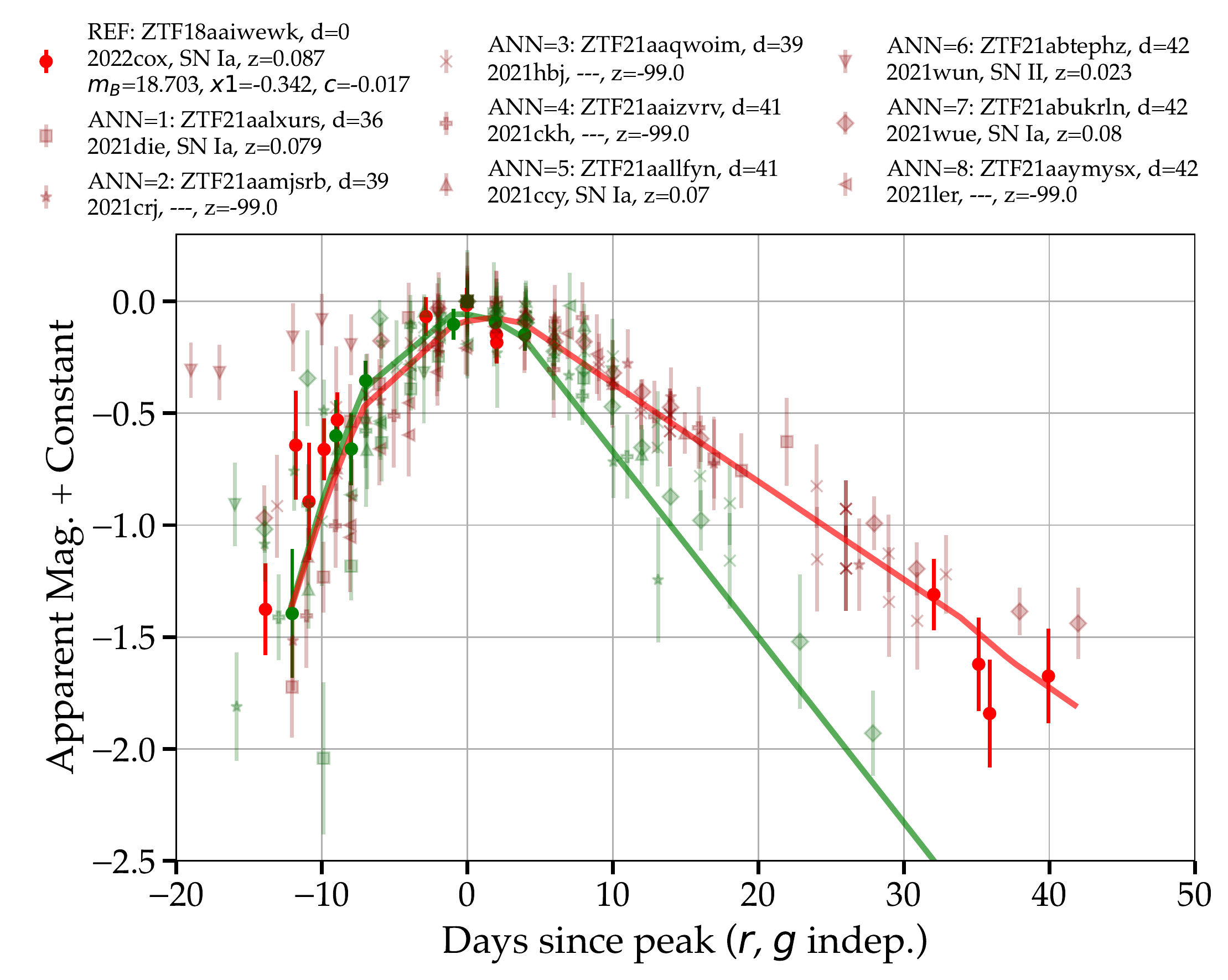}
    \includegraphics[width=0.870\columnwidth]{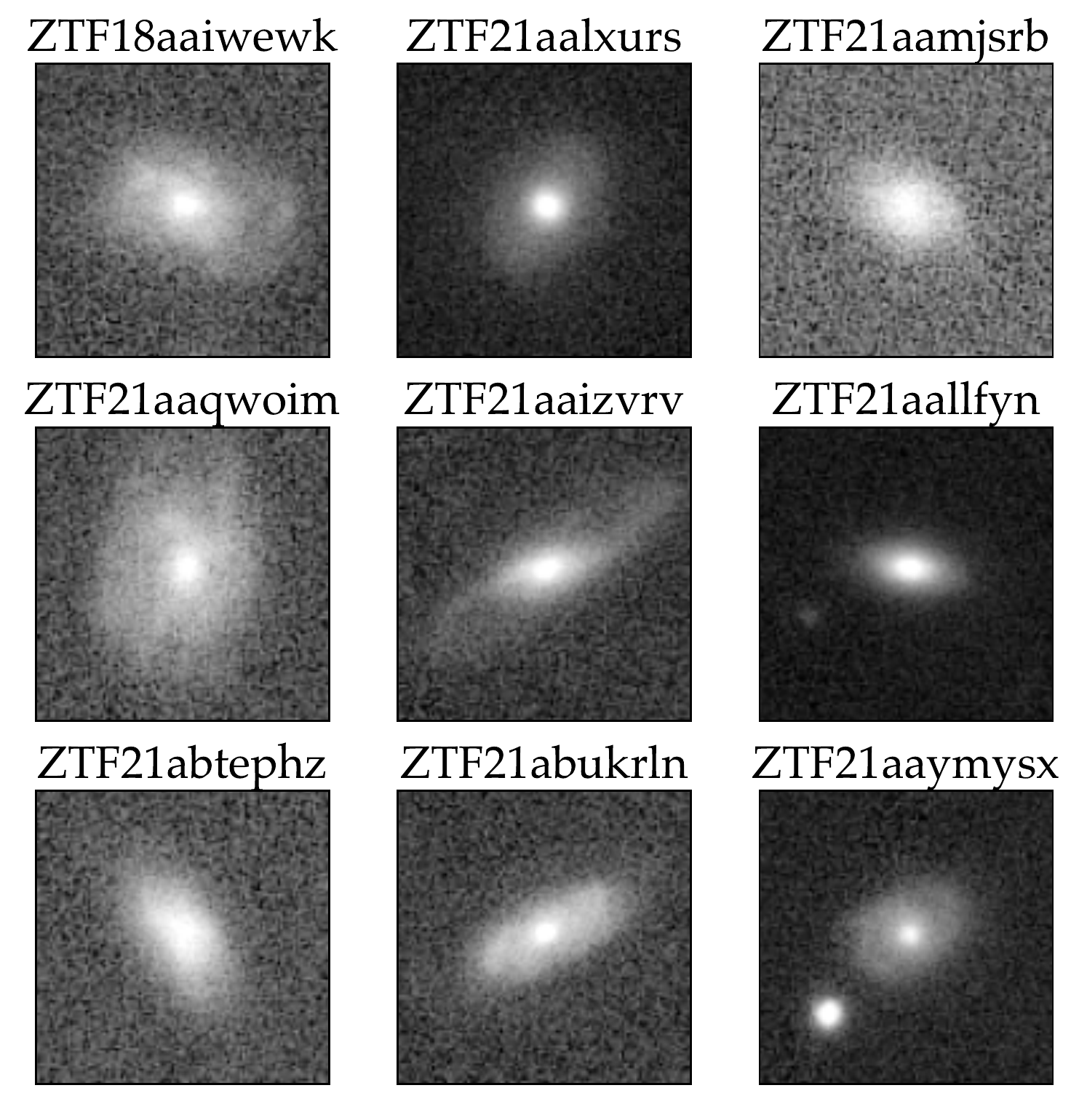}
    \caption{
    Same as Figures~\ref{fig:lc+host}, \ref{fig:lc_missed91bg},  \ref{fig:lc_missedTDE}, \ref{fig:lc_missingSNe}, but for Type Ia SN~2022cox/ZTF18aaiwewk, overplotted with the best-fit SALT3 model \citep{Kenworthy2021}. In this case, even though SN~2022cox is poorly sampled after peak with a 30~d gap, we can see fuller light curve evolution and accompanying host galaxies of analogs through the similarity search, providing useful insights as to the possible classification and redshift range  as well as the possible (unobserved) light curve evolution of our reference. The SALT3 fit provides additional context, and we observe that analog light curve epochs are positioned on or near this curve. The similarity search in our database was completed in only $\sim$2~ms. 
    } 
    \label{fig:lc_extrapolation}
\end{figure*}

Although we cannot recover photometry that was missed, we can place a soft constraint on the possible evolution by stacking light curves of ANNs, which broadly will be of a similar SN class, redshift, and host galaxy environment to the reference. Despite poor sampling of the light curve, or poor sampling at a critical phase, it is reasonable to investigate ANNs of these objects to gain insight of possible light curve evolutionary paths and the associated variance. We show an example of reference SN~Ia 2022cox/ZTF18aaiwewk in Figure~\ref{fig:lc_extrapolation}, which is poorly sampled post peak. \par 

Despite having a large 30~d gap immediately after peak, we know from spectra SN~2022cox is an SN~Ia at $z$=0.087. Four of its ANN matches have spectroscopic classifications; three SN~Ia with redshift ranging $z$=0.07 to $z$=0.08, and one poor SN~II match at $z$=0.023. The photometric evolution of the other 4 ANNs appear to also be of an SN~Ia nature. In total, only one appears to occur in an early-type galaxy. Stacking the light curves, the possible evolution post-peak (where the original gap is) is now well-sampled. In fact, if we overplot the best-fit SALT3 model \citep{Kenworthy2021}, we find that the ZTF-$r$ fit is well-traced by the majority of the analogs' $r$-band photometry post-peak as well as the late-time photometry of our reference. Moreover, despite there being no ZTF-$g$ photometry post-peak, the SALT3 fit is, too, well-traced by the analogs' $g$-band photometry. This method may provide viable supporting evidence for light curve constraints (or priors), particularly in cases where the SALT3 fit is poorly constrained.  \par

The attraction of this approach is that this provides critical potential photometry where there is little to none, which contrasts nicely with a popular light curve approximator: GP \citep{Boone2019,Demianenko2022,2022Gagliano_CCA}. GP regression \citep{Rasmussen2005}, while producing zero error at the learning points, incurs a large error in large gaps between subsequent epochs. A rigorous analytic comparison between the error of GP-fit light curves to those fit with analogs' photometry is left for future work. \par

Because our approach relies on the quality of the ANN search and consistency in light curve profiles, it is necessary for there to be sufficient numbers of the SN type at varying redshifts and host galaxies to have requisite matches for tracing possible light curve evolutions. Thus, for this work, it is unlikely this method will be effective for any SNe beyond normal SN Ia. \par

\subsection{Common failure modes of our anomaly detection model} \label{subsec:ad_failures}

Despite the success of our anomaly detection model and our hyperparameter tuning to create a high fidelity sample, there are still cases where our model tags a non-anomaly as anomalous. Visual inspection of such objects from Tables~\ref{tab:spec_test_set}, \ref{tab:real_time_any_phase},~\ref{tab:10k_full_phase},~\ref{tab:ysedr1_real_time_any_phase}, including their light curve evolution and host galaxy association and environment, reveals some common patterns. The majority of misclassified anomalies tend to:

\begin{enumerate}
    \item have an incorrect host association (SN~2021aff, SN~2021uzt, SN~2021ojn, AT~2022ywi, SN~2023dgp, SN~2023eqx, AT~2023fli, AT~2023sws). This often occurs where the likely true host is faint and small (correlated with high-$z$). Thus, normal light curve behavior in one host galaxy environment could be exceedingly peculiar if wrongly associated to another vastly different host galaxy. However, we note that there are many instances where a spectroscopic or behavioral anomaly is correctly tagged anomalous despite a misassociation (SN~2018lnb, AT~2021dpa, AT~2023atr, SN~2021axu, SN~2023khp, SN~2023omf, AT~2023otw), potentially indicating that the object's light curve behavior is peculiar enough alone to warrant being flagged anomalous. 
    \item be very blue (SALT3 $c\leq-0.2$) SN~Ia or SN~Ia candidates (AT~2020iga, AT~2020itp, AT~2020wwt, AT~2023tdy, AT~2023tim, AT~2023tjc, AT~2023tnr, AT~2023sws). This is likely a byproduct of \texttt{feature\_linear\_fit\_slope\_magn\_g}) being our model's most discriminating feature for anomaly detection, where strong blue color is correlated with TDEs and some pre-peak SNe~IIn light curves. Thus, it is unsurprising that SNe~Ia significantly bluer than the SALT3 model are flagged anomalous;
    \item be luminous (transient peak $M_{abs}\leq-19.8$, using spectroscopic or photometric redshift estimates) Ia or Ia candidates whose host galaxy is faint (apparent brightness) and/or small (apparent size) and/or blue (likely spiral, star-forming) according to the DESI Legacy Survey Imaging Surveys (AT~2020hjv, SN~2020qkx, SN~2020qql, AT~2020sgy, SN~2020wfg)\footnote{Note that SN~Ia-CSM 2020kre is an example of successfully tagged event that falls in this category.};
    \item have no rise information (SN~2023cyx, SN~2023dgp, AT~2023fbj, AT~2023huz) or have no observations until the light curve is well in decline (SN~2020tnq, SN~2021vwx, SN~2023mcs). The latter possibly masquerades as a transient that is far fainter than is typical or is at a lower $z$ than the host galaxy would imply, and would cause light curve features that are out-of-distribution (OOD) with respect to similar spectroscopic members due to the features' dependence on quantifying early (and intermediate) phases of light curve behavior. In other words, two of the exact same SN~Ia, where one is observed post peak and one that is observed for the entirety of the evolution will have some differing light curve features despite being the same event;
    \item be SN~II with constant color evolution, thus have light curves that appear more SN~Ib/c-like (SN~2019wmr, SN~2020acun). Both examples were in YSE DR1 and misclassified by both ParSNIP and SuperRAENN as SN~Ib/c, potentially indicating these events are in fact of the SNe~IIb subtypes (the most common such case, as cited in \citealt{Aleo2023});
    \item have a significant (\textgreater10~d) gap in one passband during a period of intense color evolution of the light curve (SN~2023bfv, AT~2023glx). Because ZTF-$g$ and ZTF-$r$ features are calculated independently, the passbands then no longer represent a similar phase in the evolution, but offset phases (where one lags behind), leading to behavior that is flagged as anomalous;
    \item or have some combination of the above.
\end{enumerate}

\subsection{Host-galaxy SED Modeling with \texttt{PROSPECTOR}} 
\label{subsec:fit-prospector}

As an additional analysis, we fit a spectral energy density model using stellar population synthesis to matched host galaxies for each real-time anomalous transient candidate (i.e., tagged by \texttt{LAISS\_RFC\_AD\_filter} \textit{Filter}) to understand the connection between transients and their host galaxy environments. We perform Bayesian statistical inference to report the star formation rate (SFR [$M_{\odot}$/Gyr]), the stellar mass of the galaxy ($\log(M)$), the specific star formation rate ($\log(\text{spSFR})$), the age of the stellar population (Age [Gyr]), and the V-band optical depth ($A_V$). \par

Our goal is to begin an investigation of the locations of tagged anomalies in host galaxy parameter feature-space (e.g., log(SFR) vs. $\log(M)$), relative to larger derived systematic samples across many normal and anomalous SNe classes from \cite{PalomarHostModelingSchulze2021} and \cite{Sharma2023}. We note that the sample from \cite{PalomarHostModelingSchulze2021} contains the entirety of core-collapse SNe from 2009 and 2017 by the Palomar Transient Factory --- 888 SNe of 12 distinct classes out to $z~\approx~1$ --- including host galaxy photometric properties spanning the far-ultraviolet (FUV) to the mid-infrared (MIR) with modeled host-galaxy spectral energy distributions (SEDs) to derive physical properties. Meanwhile, the sample from \cite{Sharma2023} added 12 BTS SNe~Ia-CSM hosts in comparison to BTS SNe~Ia collected by \cite{Irani2022}, with all such works overplotting hosts into SFR vs. stellar mass feature-space. By comparing the locations of derived SFR and stellar mass for our tagged anomalies to larger systematic samples, we can incrementally add to the literature of derived host galaxy parameters for rare SNe, as well as investigate the nature of \laiss{}' tagged spectroscopically ``Normal" SNe (e.g., SN~Ia, SN~II)---do they reside in sparse regions of host galaxy parameter space?  \par

We analyze the set of matched-host galaxies from Table~\ref{tab:real_time_any_phase} using \texttt{PROSPECTOR}\footnote{\url{https://github.com/bd-j/prospector}}, a framework integrating Bayesian statistical inference through nested (\texttt{DYNESTY}\footnote{\url{https://github.com/joshspeagle/dynesty}}) or distributed (\texttt{EMCEE}\footnote{\url{https://github.com/dfm/emcee}}) samplers with stellar population synthesis models  (\texttt{FSPS}\footnote{\url{https://github.com/cconroy20/fsps}}) \citep{EMCEEForemanMackey2013,Prospector2Leja2017,Prospector1Johnson2021}. We begin with the previously matched PanSTARRS host galaxies of anomaly candidate transients calculated via \texttt{GHOST} \citep{Gagliano2021} within the \texttt{LAISS\_RFC\_AD\_filter} \textit{Filter}. Using the matched host astrometry, we re-query MAST\footnote{\url{https://mast.stsci.edu/portal/Mashup/Clients/Mast/Portal.html}} for PanSTARRS photometry \citep{Chambers2016}, choosing the Kron magnitude \cite{Kron1980} from the Forced Mean table in each band. The Kron magnitude utilizes an adaptive size aperture, the size of which we also query. Using the Kron aperture in the PanSTARRS $r$ band, we query for the nearest fixed-size aperture and uncertainties in GALEX \citep{Gezari13}, 2MASS \citep{Skrutskie2006-2MASS}, AllWISE \citep{Jarrett2011}, and UKIDSS \citep{Lawrence2007-UKIRT} catalogs. Because the matched host galaxies via \texttt{GHOST} are inherently from the PanSTARRS catalog, we are guaranteed to have PanSTARRS photometry. \par

To properly model the SED to extract physical and phenomenological parameters of the host galaxies, we require a vast range of host galaxy photometry from the FUV to the MIR. Thus, we only include tagged anomalies that have at least one observation in GALEX (filters FUV, NUV spanning 1542–2274\AA), PanSTARRS (filters $g$, $r$, $i$, $z$, $y$ spanning 4776–9603\AA), either 2MASS (filters $J$, $H$, $K_s$ spanning 12350–21590\AA) or UKIDSS (filters $Z$, $Y$, $J$, $H$, $K$ spanning 8360-23800\AA), and AllWISE (filters $W1$, $W2$, $W3$, $W4$ spanning 33526–285500\AA). Moreover, we require a known redshift from spectroscopy, and the transient class (which is not an AGN) from at least one classification spectrum. This results in 5 matched hosts from the original sample of 45 tagged anomalies. \par

Our fitting routine follows that given in \citet{ProspectorAlphaLeja2017}; we first make a fit of our SPS model parameters using the Levenberg–Marquardt algorithm \citep{More1978} to initialize the MCMC chains to a sensible parameterization. We then use the \texttt{Dynesty} sampler to perform Bayesian parameter estimation.  Additional details on the SED fitting methods can be found in Appendix~\ref{subsec:APP_SED_Prospector}. \par 

Our reported galaxy properties are calculated from the parameter samples following Section~4.1 of \cite{GRBParameterDefinitionsNugent2020} and we report the parameter estimates in Table~\ref{tab:prospector}. It is important to note that while \texttt{PROSPECTOR} samples over the total solar masses formed, we estimate and report the stellar mass of the galaxy using the approximation given in \cite{Leja_2013}. \par



Of the five remaining objects after cuts, we only consider two bonafide anomalies, both of which are \emph{spectroscopic} anomalies. However, we will find that all five objects lie in sparse regions of SFR vs. stellar mass feature-space compared to \cite{PalomarHostModelingSchulze2021} and \cite{Sharma2023}, and could indicate that \laiss{}' tagged spectroscopically ``Normal" SNe have uncommon host galaxy parameters for their respective SNe classes despite small number statistics. \par

ZTF23aajkisd (SN~IIn) at a $\log(M)$ of $10.118^{+0.1125}_{-0.110}$~$M_{\odot}$ and $\log(\text{SFR})$ of $-1.469^{+0.167}_{-0.181}$~$M_{\odot}$~yr$^{-1}$, resulting in a $\log(\text{spSFR})$ of $-11.594^{+0.204}_{-0.209}$~yr$^{-1}$, lies beyond the outer 90\% contour of host galaxy populations in the mass–SFR plane as determined by a kernel density estimate of 111 SN~IIn in the \cite{PalomarHostModelingSchulze2021} sample (see their Figure~10). This indicates that SN~2023iex has an uncommonly low SFR for its relatively larger galaxy mass, resulting in a lower spSFR. However, in the mass-redshift plane (see their Figure~9), the host of SN~2023iex resides in the inner $\sim$20\% contour, indicating a common mass for its redshift (0.029).  \par

ZTF23aatdcey (SN~IIn) at a $\log(M)$ of $10.128^{+0.100}_{-0.096}$~$M_{\odot}$ and $\log(\text{SFR})$ of $-1.591^{+0.198}_{-0.238}$~$M_{\odot}$~yr$^{-1}$, resulting in a $\log(\text{spSFR})$ of $-11.719^{+0.241}_{-0.272}$~yr$^{-1}$, follows the same pattern as SN~2023iex (ZTF23aajkisd). The host of SN~2023nof (ZTF23aatdcey) also lies beyond the outer 90\% contour of host galaxy populations in the mass–SFR plane from the \cite{PalomarHostModelingSchulze2021} sample. Its host galaxy also resides in a sparsely populated, low region of spSFR, where the SFR is low considering the relatively larger mass of the galaxy. In the mass-redshift plane, it resides in the inner $\sim$20\% contour, indicating a common mass for its redshift (0.069).  \par

ZTF23aberpzw (SN~II), at a $\log(M)$ of $10.204^{+0.071}_{-0.083}$~$M_{\odot}$ and $\log(\text{SFR})$ of $-1.949^{+0.175}_{-1.87}$~$M_{\odot}$~yr$^{-1}$, resulting in a $\log(\text{spSFR})$ of $-12.154^{+0.232}_{-0.216}$~yr$^{-1}$, lies beyond the outer 90\% contour of host galaxy populations in the mass–SFR plane as determined by a kernel density estimate of 498 SN~II in the \cite{PalomarHostModelingSchulze2021} sample (see their Figure~10). Moreover, it lies in what \cite{PalomarHostModelingSchulze2021} found to be in the lower extreme of SFR---up to $\lesssim$2\% of regular CC~SNe (SN~Ibc, SN~IIb, SN~II, and SN~IIn) whose hosts exhibit SFR~$\lesssim$~0.01~$M_{\odot}$~yr$^{-1}$ (a $\log(\text{SFR})$ of $-2$~$M_{\odot}$~yr$^{-1}$) and spSFR between $10^{-14}$~yr$^{-1}$ and $10^{-11}$~yr$^{-1}$, often with sizeable uncertainties. Thus, we consider this object to be a member of the lower extreme SFR hosts. In the mass-redshift plane the host of SN~2023swf resides in the inner $\sim$10\% contour, indicating an extremely common mass at its redshift (0.024). Despite the light curve and spectra indicating a normal SN~II, there is some evidence this object may be anomalous when considering the extremely low spSFR, and thus may be considered a \emph{contextual} anomaly. \par 

Overall, all core-collapse SNe (ZTF23aajkisd, ZTF23aatdcey, ZTF23aberpzw) reside in a sparsely populated region of uncommonly low spSFR, where their galaxies have low SFR for their relatively larger size (all above the median and mode $\log(M)$ values as shown in \cite{PalomarHostModelingSchulze2021}, their Table~5). \par

ZTF23aapgswu (SN~Ia) at a $\log(M)$ of $8.367^{+0.077}_{-0.095}$~$M_{\odot}$, is a peculiar case. Its host is a relatively compact, strongly blue and likely high star-forming galaxy. One such consequence is a heavily negatively skewed SFR distribution, with one peak and long tail, resulting in a (more negative) median value of $\log(\text{SFR})$ of $-4.389^{+0.570}_{-59.850}$~$M_{\odot}$~yr$^{-1}$ and a $\log(\text{spSFR})$ of $-12.751^{+0.600}_{-59.850}$~yr$^{-1}$. However, because of the heavy negative skew that biases the median, we also note a marginally more realistic (but still unphysical) mode value of $\log(\text{SFR})$ of $-3.998$~$M_{\odot}$~yr$^{-1}$ and a $\log(\text{spSFR})$ of $-12.424$~yr$^{-1}$. Such an unphysically low SFR value can likely be traced to our choice of SFH function: it has been suggested that parameterized SFH are not flexible enough to deal with the actual complexities inherent in galaxy systems \citep{Leja_2019_nonparametricSFH}. As a further point of caution, no SN~Ia from \cite{Irani2022} of BTS SNe~Ia have such a low $\log(\text{SFR})$, and thus we make no strong claims as to the validity of the fit. More investigation of the host galaxy and derived host parameters of SN~2023mcs is encouraged, especially when considering it is host to a SN~Ia. \par

ZTF23abayyjm (SN~Ia) at a $\log(M)$ of $10.815^{+0.110}_{-0.044}$~$M_{\odot}$ (the largest in our sample of five objects) and $\log(\text{SFR})$ of $0.147^{+0.263}_{-0.842}$~$M_{\odot}$~yr$^{-1}$, resulting in a $\log(\text{spSFR})$ of $-10.705^{+0.234}_{-0.761}$~yr$^{-1}$, is the only event that resides in a host galaxy with above Milky Way metallicity ([Z/H] = 0.0; see \citet{Choi_2016_MIST}). When compared to BTS SNe~Ia from \cite{Irani2022} and \cite{Sharma2023}, it is at the upper regime of massive, highly star-forming and high metallicity host galaxies beyond the red and blue contours corresponding to Galaxy Zoo ellipticals and spirals (see Figure~14 of \citet{Irani2022}). However, the spSFR is average after normalizing SFR by the very large galaxy mass, and is among the highest mass hosts compared to the \cite{Irani2022} and \cite{Sharma2023} samples. \par

The goal of this exercise was to lay the groundwork of building a downstream \emph{Filter} to fit SPS host galaxy models. In this pursuit, we have discerned valuable insights. Moreover, we recognize the need for additional technologies and software infrastructure that needs to be developed before this process can be scaled to \emph{Rubin}-era data streams. \par

First, it was critical to use matched-aperture size photometry to get sensible results. Even so, we chose to only use pre-calculated magnitudes from reported fixed-size apertures, when in principle the apertures should be matched exactly across each band. Additional work would be necessary to create a robust pipeline that works on galaxies at various distances. Our fits could likely be improved by utilizing non-parameteric SFH functions. Additionally, MCMC sampling is computationally expensive. Each object's sampling takes approximately 2~hr on a shared compute node equipped with an AMD EPYC 7502 CPU. To rectify this, other works have suggested speeding up SPS models with neural network emulators \citep{Kwon_2023_PROVemulator}, which are reported to achieve a 100x speed-up; or with approximate posterior distributions of galaxy parameters efficiently via Simulation-Based inference \citep[SBI;][]{Cranmer2020} and amortized Neural Posterior Estimation \citep[NPE;][]{Papamakarios2016} techniques, which can achieve an amortized calculation of galaxy parameter posteriors in less than 1s \citep[see][and references therein]{Hahn2022, Khullar2022, Wang2023}; or with a physics-informed variational autoencoder \citep{Gagliano2023rapidgalinf}. These efforts are promising and ongoing, but beyond the scope of this work. \par

\begin{table*}[ht]
\footnotesize
\centering
\caption{\textbf{Host Galaxy Parameter Estimation with Prospector.} We fit a spectral model using photometry from matched host galaxies to each transient in Table~\ref{tab:real_time_any_phase} that pass strict selection cuts. We report the transient redshift, the log star formation rate ($\log(\text{SFR})$ [$M_{\odot}$]), the total stellar mass of the galaxy ($\log(M)$~[$M_{\odot}$~yr$^{-1}$]), the log specific star formation rate ($\log(\text{sSFR})$~[yr$^{-1}$]), the age of the stellar population (Age [Gyr]) and the V-band optical depth ($A_V$). We also report the value of $\chi^2$/n.o.f. (number of filters). For all distributions, we report the median values with upper and lower bounds equivalent to the $84^{\text{th}}$ and $16^{\text{th}}$ quantiles, respectively, unless otherwise noted (italicised text; see below).}
\begin{tabular}{llllllllll}
\hline
\hline
ZTF ID & Spec. Class & Redshift & $\log(M)$ & Age [Gyr] & $A_V$ & $\log(\text{SFR})$ & $\log(\text{spSFR})$ & $\chi^2/n.o.f$ \\
\hline
\textbf{ZTF23aatdcey} & SN IIn & $0.069$ & $10.128^{+0.100}_{-0.096}$ & $4.406^{+1.952}_{-1.331}$ & $1.636^{+0.267}_{-0.278}$ & $-1.591^{+0.198}_{-0.238}$ & $-11.719^{+0.241}_{-0.272}$ & 39.451/14 & \\ 
\textbf{ZTF23aajkisd} & SN IIn & $0.029$ & $10.118^{+0.125}_{-0.110}$ & $3.120^{+1.959}_{-1.031}$ & $2.097^{+0.191}_{-0.184}$ & $-1.469^{+0.167}_{-0.181}$ & $-11.594^{+0.204}_{-0.209}$ & 23.446/14 & \\ 
ZTF23aapgswu & SN Ia & $0.030$ & $8.367^{+0.077}_{-0.095}$ & $10.401^{+2.368}_{-2.913}$ & $0.117^{+0.189}_{-0.087}$ & \textit{--4.389}$^{+0.570}_{-59.729}$ & \textit{--12.751}$^{+0.600}_{-59.850}$ & 28.920/10 & \\ 
ZTF23aberpzw & SN II & $0.024$ & $10.204^{+0.071}_{-0.083}$ & $9.418^{+2.611}_{-2.387}$ & $1.221^{+0.194}_{-0.188}$ & $-1.949^{+0.175}_{-0.187}$ & $-12.154^{+0.232}_{-0.216}$ & 19.036/14 & \\ 
ZTF23abayyjm & SN Ia & $0.140$ & $10.815^{+0.110}_{-0.044}$ & $1.471^{+1.006}_{-0.461}$ & $3.275^{+0.184}_{-0.170}$ & $0.147^{+0.263}_{-0.824}$ & $-10.705^{+0.234}_{-0.761}$ & 28.451/11 & \\ 
\hline
\hline\\[-1.5ex]
\multicolumn{9}{c}{
\begin{minipage}{16cm}
The italicised text indicates distributions that have unreliable median values (displayed as default in the table) due to their heavy negative skew. For these, the mode is better representative. For ZTF23aapgswu, the mode of $\log(\text{SFR})$ is $-3.998$~$M_{\odot}$~yr$^{-1}$ and the mode of $\log(\text{spSFR})$ is $-12.424$~yr$^{-1}$. 
\end{minipage}}
\end{tabular}
\label{tab:prospector}
\end{table*}

\section{Conclusion} \label{sec:conclusion}

In this work, we present \laiss{} (Lightcurve Anomaly Identification and Similarity Search), a pipeline for real-time anomaly detection and approximate similarity searches of astronomical transients within large volumetric data streams. We debut our anomaly detection model as a \emph{Filter} on the ANTARES broker to process the nightly ZTF Alert Stream. Our model is based on a RFC architecture using extracted light curve and contextual host galaxy features without the need for redshift information, designed to classify several types of transients as anomalies (``Anomaly") with high purity: 
\begin{enumerate}
    \item Spectroscopic anomalies (i.e., designate all SNe other than normal Type Ia, Type Ia-91T-like, and normal Type II, IIP as anomalous); 
    \item Contextual anomalies (e.g., a SN in an atypical galactic environment);
    \item Behavioral anomalies (e.g., a SN re-brightening due to CSM-interaction).
\end{enumerate}
We run our anomaly detection model on the active ZTF alert stream for real-time discovery, as well as legacy subsets of the ZTF alert stream and the Young Supernova Experiment Data Release 1 for retroactive discovery. Moreover, we construct a low-latency approximate similarity search model within our derived light curve and host galaxy feature-space to find transient analogs of similar light curve evolution and host galaxy environments. \par

Our conclusions and key takeaways for the \laiss{} anomaly detection and approximate similarity search pipeline are as follows:
\begin{itemize}
    \item The most important \emph{light curve features} for distinguishing anomalies from other transients according to our RFC model using both impurity and permutation importances are the light curve slope in a least squares fit of the linear stochastic model with Gaussian noise described by observation errors \{$\delta_{i}$\} (\texttt{feature\_linear\_fit\_slope\_magn\_g}) and its error in $r$-band \\(\texttt{feature\_linear\_fit\_slope\_sigma\_magn\_r}), and the unbiased Anderson–Darling normality test statistic for $g$-band flux \\ (\texttt{feature\_anderson\_darling\_normal\_flux\_g}). 
    \item The most important \emph{host galaxy features} for distinguishing anomalies from other transients according to our RFC model using both impurity and permutation importances are radial offset and color-derived features, namely the aperture magnitude $i-z$ color difference (\texttt{i-z}) and the normalized host separation \texttt{dist/DLR}. Notably, the high importance of these contextual host galaxy features are in agreement with that of FLEET \citep{Gomez2020, Gomez2023SLSN, Gomez2023TDE} and \texttt{GHOST} \citep{Gagliano2021}, despite the nature of their different anomaly detection and SNe classification tasks.
    \item Using the spectroscopic label alone from TNS (before vetting) as indicator of an anomalous transient, we can recover anomalies with a purity of $\approx$50\% using the full light curve classifier for events with $P(anom)~\textgreater~0.5$. If we assume anomalies (defined in this work as everything not Type Ia-normal, 91T-like, II-normal, and IIP) comprise about 10\% of the SNe population in a magnitude-limited survey like ZTF, this is a factor of $\sim$5 improvement compared to random selection. The corresponding completeness for this threshold is 29\%. However, if we additionally consider \emph{updated spectroscopic classifications}, \emph{contextual anomalies}, and \emph{behavioral anomalies} (such as SNe in an atypical environment or peculiar light curve or spectral behavior) as a result of expert vetting, we can recover anomalies with an \emph{observed purity} of $\approx$90\% using the full light curve classifier for events with $P(anom)~\textgreater~0.5$ at a completeness threshold of 38\%. This is a factor of $\sim$8 improvement compared to random selection.
    \item When using any SNe as input into our approximate similarity search, we can retrieve 8 ANNs in our 5472 object, 60-dimensional principal component-space in $\mathcal{O}(100)$~ms. If needed, we can re-extract light curve features in in $\sim$1~s and host galaxy association and features in $\sim$1~min. We show that an ANN solution to finding SNe analogs with pre-computed features can scale to \textit{Rubin} data streams.  
    \item We demonstrate an approximate similarity search for finding SNe analogs is useful in many applications, such as but not limited to: finding similar SNe evolution and host galaxy environment, finding missed SNe in legacy datasets, providing possible light curve evolution of poorly sampled SNe, and prompting some reclassification of SNe (often to a rarer subclass of the SNe type).
    \item From our approximate similarity search, we report 17 unique reclassified SNe and 84 previously undiscovered SNe candidates found by an ANN=8 search of our databank spectroscopic sample. From a final exhaustive search of our databank, we report the remaining 241 SNe candidates to TNS. In total, we report 325 discoveries, all from 2018-2021 and absent from public catalogs ($\sim$1\% of all ZTF AT reports to TNS through 2021\footnote{\url{https://tinyurl.com/ZTF-AT-TNS-reports-thru2021}}).
    \item Despite small number statistics, there is some evidence to suggest that objects tagged anomalous by \laiss{} have host galaxy properties such as mass, star formation rate, and specific star formation rate, that reside in sparse regions of latent space of equivalent SN type when compared to larger samples in the literature. 
\end{itemize}

\subsection{List of Anomalies}
\label{subsec:list_of_anomalies}

We report the following new \emph{spectroscopic} and \emph{behavioral} anomalies, either flagged in real-time or retrospectively by our anomaly detection model. They are confirmed via follow-up observations or from a retroactive reclassification of existing spectra prompted by approximate similarity search. Additionally, we report likely anomalous transient candidates based on their light curve evolution, host galaxy environment, and results from photometric classifiers such as FLEET \citep{Gomez2020}. We newly identify a myriad of peculiar and rare transients, including:
\begin{itemize}
    \item \textbf{SLSN (11 total; 2 spectroscopic; 9 photometric)}: We identify 11 new candidate members: 2 spectroscopic (ZTF21aaizyqc / SN~2021ckb, ZTF22abfdzrv / SN~2022vmg); 9 photometric (ZTF20acbiwfi / AT~2020afew, ZTF20aczhbki / AT~2020afex, ZTF21aajrgkw / AT~2021dpa, ZTF21aaualdi / AT~2021ahwa, ZTF21abasbpe / AT~2021ahwh, ZTF20acyroio / AT~2021ahwp, ZTF23aahjdxa / AT~2023gpp, ZTF23aaqqeek / AT~2023mne, ZTF23aawblmi / AT~2023otw).
    \item \textbf{TDE (10 total; 1 spectroscopic, 9 photometric)}: We identify 9 new candidate members: 1 spectroscopic (ZTF23aamsetv / SN~2023kvy); 8 photometric (ZTF18abtjrbt / AT~2018mfz, ZTF18acvwkvc / AT~2018mkd, PS19gzf / AT~2019vuz, ZTF20acpgnmw / AT~2020afev, ZTF20acpzjsk / AT~2020affx, ZTF21aazenvp / AT~2021ovg, ZTF21aasdcgt / AT~2021ahwl, ZTF22absuavp / AT~2022zyh, ZTF22abzajwl / AT~2023adr).
    \item \textbf{Long-rising ($\textgreater$~40~d) SN~II (6 total; 2 spectroscopic, 4 photometric)}: We identify 6 new candidate members: 2 spectroscopic (ZTF20abqlmwn/ SN~2020rmk, ZTF21aaqyifh/ SN~2021hqe); 4 photometric (ZTF18abiitmq/ AT~2018mhh, ZTF18abkmfaj/ AT~2018mhj, ZTF21abasjcd/ AT~2021rmq, ZTF23aajestr / AT~2023inr).
    \item \textbf{SN~Ic-BL (1 total; 1 spectroscopic, 0 photometric)}: We identify 1 new candidate member; 1 spectroscopic (ZTF21aacufip / SN~2021vz); 0 photometric.
    \item \textbf{SN~Ib (2 total; 2 spectroscopic, 0 photometric)}: We identify 2 new candidate members: 2 spectroscopic (ZTF21aaqwfqe / SN~2021hen, ZTF21aabyifm / SN~2021qv); 0 photometric.
    \item \textbf{SN~Ib/c (6 total; 0 spectroscopic, 6 photometric)}: We identify 6 new candidate members: 0 spectroscopic; 6 photometric (ZTF18abwhsnx / AT~2018mgd, ZTF18aajgowk / AT~2018mgw, ZTF20acuyxki / AT~2020afes, ZTF21aaiapis / AT~2021ahwg, ZTF23aafgmaz / AT~2023frg, ZTF23abedgfr / AT~2023syt).
    \item \textbf{SN~IIb (4 total; 3 spectroscopic, 1 photometric)}: We identify 4 new candidate members: 3 spectroscopic (ZTF21aajvukh/ SN~2021cpi, ZTF21abnvlnj / SN~2021tyf, ZTF23aarktow/ SN~2023myo); 1 photometric (ZTF18acvgmpx / AT~2018mkc).
    \item \textbf{SN~IIn (19 total; 4 spectroscopic, 15 photometric)}: We identify 19 new candidate members: 4 spectroscopic (ZTF21abcjpnm / SN~2021njo, ZTF23aatdcey / SN~2023nof, ZTF23aatcsou / SN~2023nwe, ZTF23aavtugd / SN~2023omf); 15 photometric (ZTF18ablqjws / AT~2018mga, PS20czv / AT~2020jvi, PS20mgo / AT~2020acyu, ZTF20aclghmy / AT~2020affa, ZTF20acbptqx / 2020affg, ZTF20acngnvf / AT~2020affn, ZTF20acrssoi / AT~2020affz, ZTF20actkqax / AT~2020afgb, ZTF20acxyrkk / AT~2020afge, ZTF21abiggqx / AT~2021rjf, ZTF22aatwxrl / AT~2022oym, ZTF23aaahnss / AT~2023atr,  ZTF23aaveoxd / AT~2023ofr, ZTF23aaufkak / AT~2023nwk,  ZTF23aaewyhm / AT~2023gzn).
    \item \textbf{SN~Ia-CSM (3 total; 2 spectroscopic, 1 photometric)}: We identify 3 new candidate members: 2 spectroscopic (ZTF20acpbboa / SN~2020ywo, ZTF23aatabje / SN~2023ocx); 1 photometric (ZTF23aaqbyzr / SN~2023mic).
    \item \textbf{SN~Ia-91bg-like (10 total; 7 spectroscopic, 3 photometric)}: We identify 10 new candidate members: 7 spectroscopic (ZTF20acnznol / SN~2020yje, ZTF21abmwgow / SN~2021ttg, ZTF21acfigoo / SN~2021aazj, ZTF21achjwus / SN~2021abpz, ZTF21acjgafq / SN~2021ackd, ZTF21acmnpqa / SN~2021adpx, ZTF23abhafym / SN~2023tsw); 3 photometric (ZTF19aavoqbe / AT~2019aatq, ZTF21aaiahsu / AT~2021ahxg, ZTF23aaflptz / AT~2023gbk).
    \item \textbf{Unknown blue transients (2 total; 2 photometric)}: We identify 2 blue transients of unknown type: 0 spectroscopic; 2 photometric (ZTF21aabyoxk / AT~2021ahwu, ZTF21aawzmne / AT~2021ahyj).    
\end{itemize}

Lastly, we aggregate the \emph{contextual} anomalies:
\begin{itemize}
    \item \textbf{SN candidates in rare ring host galaxy (2)}: ZTF18acvgmpx / AT~2018mkc, ZTF21aakjxhg / AT~2021ahxn.
    \item \textbf{SN~Ia in flocculent spiral/merger (1)}: ZTF20aazpwen / SN~2020kbl.
    \item \textbf{SN~II or SN~II candidate in possible elliptical host galaxy (3)}: ZTF19abljudj / AT~2019aats, ZTF20aawlmfu / AT~2020kmj, ZTF23abcinum / SN~2023sds.
\end{itemize}


\textit{Facilities:} ZTF \citep{ZTF_image}, ADS, TNS, NED \citep{https://doi.org/10.26132/ned1}, ATel, ANTARES, ALeRCE

\textit{Software:} \texttt{ANNOY} \citep{Github:annoy}, \texttt{antares-client} \citep{Gitlab:antares-client}, \texttt{Astropy} \citep{astropy:2013, astropy:2018}, \texttt{FLEET} \citep{Gomez2020}, \texttt{GHOST} \citep{Gagliano2021}, \texttt{Matplotlib} \citep{hunter2007matplotlib}, \texttt{numpy} \citep{walt2011_numpy}, \texttt{Pandas} \citep{reback2020_pandas}, \texttt{Scikit-Learn} \citep{scikit-learn}, \texttt{YSE-PZ} \citep{Coulter2022_YSEPZ, Coulter2023YSE-PZ}

\section{Acknowledgments} 
\label{sec:acknowledgments}

We thank Anya Nugent, Nell Byler, Maggie Verrico, Nicholas Earl, Yuxin Dong, Charlie Conroy, and Ben Johnson for helpful correspondence regarding host-galaxy SED Modeling.

Author contributions are listed below. \\
P.~D.~Aleo as project lead and YSE collaboration meeting co-lead contributed to ideation, development, and deployment of \texttt{LAISS}; statistical and data analysis; analysis of classified objects; lead writing and lead editing; figures. \\
A.~Engel contributed to writing text, making figures, and performing analysis using \texttt{Prospector}.\\
G.~Narayan as YSE Executive Committee member contributed to oversight; editing; helpful discussions.\\
C.~R.~Angus contributed to observing and reducing NOT data used for classification; analysis of classified objects; and sniffing and/or sorting and/or flagging transients.\\
K.~Malanchev contributed to ANTARES operations (software); \emph{Filter} management; draft review.\\
K.~Auchettl contributed to WiFeS observations to confirm candidates; to draft review.\\
V.~F.~Baldassare as YSE Executive Committee member contributed to helpful discussions.\\
A.~Berres contributed to the creation of the Slack-bot.\\
T.~J.~L.~de~Boer contributed to PS1 operations.\\
B.~M.~Boyd contributed as a YSE junior review panelist.\\
K.~C.~Chambers contributed to PS1 operations.\\
K.~W.~Davis contributed to observing and reducing Lick-3m data used for classification.\\ 
N.~Esquivel contributed to ANTARES operations (software); ANTARES operations (hardware).\\
D.~Farias contributed as a YSE junior review panelist.\\
R.~J.~Foley as YSE Executive Committee member contributed to helpful discussions, managed the YSE survey, and was PI of spectroscopic observing programs.\\ 
A.~T.~Gagliano contributed to draft review; helpful discussions and to sniffing and/or sorting and/or flagging transients.\\ 
C.~Gall contributed to sniffing and/or sorting and/or flagging transients.
H.~Gao contributed to PS1 operations.\\
S.~Gomez contributed to observing and reducing SOAR data used for classification in co-ordination with the FLEET program; to helpful discussions.\\
M.~Grayling contributed as lead YSE junior review panelist.\\
D.~O.~Jones contributed to helpful discussions and to PS1 operations.\\
C.-C.~Lin contributed to PS1 operations.\\
E.~A.~Magnier contributed to PS1 operations.\\
K.~S.~Mandel as YSE Executive Committee contact contributed to oversight and helpful discussions.\\
T.~Matheson contributed to ANTARES operations (software); discussions; draft review.\\
S.~I.~Raimundo contributed to sniffing and/or sorting and/or flagging transients.\\
V.~G.~Shah contributed as a YSE junior review panelist.\\
M.~D.~Soraisam contributed to ANTARES operations (software); discussions; draft review.\\
K.~M.~de~Soto contributed as a YSE junior review panelist.\\
S.~Vicencio contributed to ANTARES operations (software); \emph{Filter} management.\\
V.~A.~Villar as YSE Executive Committee member contributed to helpful discussions.\\
R.~J.~Wainscoat contributed to PS1 operations.\\


P.D.A.\ and G.N.'s contributions to this work were directly supported by NSF AST--2206195, and we gratefully acknowledge this funding.
P.D.A.\ has also been supported by the Illinois Survey Science Graduate Fellowship from the Center for AstroPhysical Surveys (CAPS)\footnote{\url{https://caps.ncsa.illinois.edu/}} at the National Center for Supercomputing Applications (NCSA).
A.W.E.\ was partially supported by the Open Call Initiative, under the Laboratory Directed Research and Development (LDRD) Program at Pacific Northwest National Laboratory (PNNL). PNNL is a multi-program national laboratory operated for the U.S.\ Department of Energy (DOE) by Battelle Memorial Institute under Contract No.\ DE-AC05-76RLO 1830.
G.N.\ is also supported by NSF CAREER grant AST--2239364, funded in-part by a grant by Charles Simonyi, and NSF OAC--2311355, DOE support through the Department of Physics at the University of Illinois, Urbana-Champaign (\# 13771275), and support from the HST Guest Observer Program through HST-GO-16764 and HST-GO-17128 (PI: R. Foley).
Support was provided by Schmidt Sciences, LLC. for KM.
V.F.B.'s contributions to this work were supported by NSF AST-2206165.
B.M.B.\ is supported by the Cambridge Centre for Doctoral Training in Data-Intensive Science funded by the UK Science and Technology Facilities Council (STFC).
The UCSC team is supported in part by NASA grant NNG17PX03C, NSF grants AST--1815935 and AST--2307710, the Gordon \& Betty Moore Foundation, the Heising-Simons Foundation, and by a fellowship from the David and Lucile Packard Foundation to R.J.F.
A.T.G.\ is supported by the National Science Foundation under Cooperative Agreement PHY-2019786 (The NSF AI Institute for Artificial Intelligence and Fundamental Interactions, http://iaifi.org/).
C.G.\ is supported by a VILLUM FONDEN Young Investigator Grant (project number 25501).
M.G.\ and K.S.M.\ are supported by the European Union’s Horizon 2020 research and innovation programme under ERC Grant Agreement No.\ 101002652 and Marie Sk\l{}odowska-Curie Grant Agreement No.\ 873089.
K.M.dS.\ acknowledges support by the NSF through grant AST--2108676. K.dS.\ thanks the LSST-DA Data Science Fellowship Program, which is funded by LSST-DA, the Brinson Foundation, and the Moore Foundation; her participation in the program has benefited this work. 
V.S.\ acknowledges the support of the LSST Corporation's 2021 Enabling Science award for undergraduates.
V.A.V.\ acknowledges support by the NSF through grant AST--2108676.

The Young Supernova Experiment (YSE) and its research infrastructure is supported by the European Research Council under the European Union's Horizon 2020 research and innovation programme (ERC Grant Agreement 101002652, PI K.\ Mandel), the Heising-Simons Foundation (2018-0913, PI R.\ Foley; 2018-0911, PI R.\ Margutti), NASA (NNG17PX03C, PI R.\ Foley), NSF (AST--1720756, AST--1815935, AST--2307710, PI R.\ Foley; AST-1909796, AST-1944985, PI R.\ Margutti), the David \& Lucille Packard Foundation (PI R.\ Foley), VILLUM FONDEN (project 16599, PI J.\ Hjorth), and the Center for AstroPhysical Surveys (CAPS) at the National Center for Supercomputing Applications (NCSA) and the University of Illinois Urbana-Champaign.

The ANTARES project has been supported by the National Science Foundation through a cooperative agreement with the Association of Universities for Research in Astronomy (AURA) for the operation of NOIRLab, through an NSF INSPIRE grant to the University of Arizona (CISE AST-1344024, PI: R. Snodgrass), and through a grant from the Heising-Simons Foundation. \par

YSE-PZ was developed by the UC Santa Cruz Transients Team with support from The UCSC team is supported in part by NASA grants 
NNG17PX03C, 80NSSC18K0303, 80NSSC19K0113, 80NSSC19K1386, 80NSSC20K0953, 80NSSC21K2076, 80NSSC22K1513, 80NSSC22K1518, and 80NSSC23K0301; NSF grants AST--1720756, AST--1815935, and AST--1911206; grants associated with {\it Hubble Space Telescope} programs DD--14925, DD--15600, GO--15876, GO--16238, SNAP--16239, GO--16690, SNAP--16691, and GO--17128; the Gordon \& Betty Moore Foundation; the Heising-Simons Foundation; fellowships from the Alfred P.\ Sloan Foundation and the David and Lucile Packard Foundation to R.J.F.; Gordon and Betty Moore Foundation postdoctoral fellowships and a NASA Einstein fellowship, as administered through the NASA Hubble Fellowship program and grant HST-HF2-51462.001, to D.O.J.; and a National Science Foundation Graduate Research Fellowship, administered through grant No.\ DGE-1339067, to D.A.C.

Pan-STARRS is a project of the Institute for Astronomy of the University of Hawaii, and is supported by the NASA SSO Near Earth Observation Program under grants 80NSSC18K0971, NNX14AM74G, NNX12AR65G, NNX13AQ47G, NNX08AR22G, 80NSSC21K1572 and by the State of Hawaii.  The Pan-STARRS1 Surveys (PS1) and the PS1 public science archive have been made possible through contributions by the Institute for Astronomy, the University of Hawaii, the Pan-STARRS Project Office, the Max-Planck Society and its participating institutes, the Max Planck Institute for Astronomy, Heidelberg and the Max Planck Institute for Extraterrestrial Physics, Garching, The Johns Hopkins University, Durham University, the University of Edinburgh, the Queen's University Belfast, the Harvard-Smithsonian Center for Astrophysics, the Las Cumbres Observatory Global Telescope Network Incorporated, the National Central University of Taiwan, STScI, NASA under grant NNX08AR22G issued through the Planetary Science Division of the NASA Science Mission Directorate, NSF grant AST-1238877, the University of Maryland, Eotvos Lorand University (ELTE), the Los Alamos National Laboratory, and the Gordon and Betty Moore Foundation.

Parts of this work are based on observations obtained with the Samuel Oschin Telescope 48-inch and the 60-inch Telescope at the Palomar Observatory as part of the Zwicky Transient Facility project. ZTF is supported by the National Science Foundation under Grants No.\ AST--1440341 and AST--2034437 and a collaboration including current partners Caltech, IPAC, the Weizmann Institute of Science, the Oskar Klein Center at Stockholm University, the University of Maryland, Deutsches Elektronen-Synchrotron and Humboldt University, the TANGO Consortium of Taiwan, the University of Wisconsin at Milwaukee, Trinity College Dublin, Lawrence Livermore National Laboratories, IN2P3, University of Warwick, Ruhr University Bochum, Northwestern University and former partners the University of Washington, Los Alamos National Laboratories, and Lawrence Berkeley National Laboratories. Operations are conducted by COO, IPAC, and UW.
The ZTF forced-photometry service was funded under the Heising-Simons Foundation grant \#12540303 (PI: Graham). 

Parts of this research were supported by the Australian Research Council Discovery Early Career Researcher Award (DECRA) through project number DE230101069. 

A major upgrade of the Kast spectrograph on the Shane 3~m telescope at Lick Observatory was made possible through generous gifts from the Heising-Simons Foundation as well as William and Marina Kast. Research at Lick Observatory is partially supported by a generous gift from Google. 

This research has made use of the VizieR catalogue access tool, CDS, Strasbourg, France (DOI : 10.26093/cds/vizier). The original description of the VizieR service was published in 2000, A\&AS 143, 23


\bibliography{references}{}
\bibliographystyle{aasjournal}

\phantomsection
\appendix
\label{sec:appendix}

\section{Light Curve Features}
\label{appx:lc_features}

\counterwithin{figure}{section}
\counterwithin{table}{section}
\renewcommand{\thefigure}{A.\arabic{figure}}
\setcounter{figure}{0}
\renewcommand{\thetable}{A.\arabic{table}} \setcounter{table}{0}

Our light curve features are extracted with the \texttt{lc\_feature\_extractor} filter in ANTARES using the \texttt{light-curve} package. We extract the same features for $r$ and $g$ band, comprising 62 total features (31 $r$ band, 31 $g$ band). A brief description of each feature is as follows:

\begin{itemize}
    \item \texttt{feature\_amplitude\_magn}: Half amplitude of magnitude \citep{Malanchev2021}.
    \item \texttt{feature\_anderson\_darling\_normal\_magn}: Unbiased Anderson–Darling normality test statistic for magnitude \citep{Malanchev2021}.
    \item \texttt{feature\_beyond\_1\_std\_magn}: Fraction of observations beyond $n=1 \sigma_{m}$ from the mean magnitude $\langle m \rangle$ \citep{D'Isanto2016}.
    \item \texttt{feature\_beyond\_2\_std\_magn}: Fraction of observations beyond $n=2 \sigma_{m}$ from the mean magnitude $\langle m \rangle$ \citep{D'Isanto2016}.
    \item \texttt{feature\_cusum\_magn}: A range of cumulative sums dependent on the number of observations, mean magnitude, and magnitude standard deviation \citep{Kim2014}.
    \item \texttt{feature\_inter\_percentile\_range\_2\_magn}: Inter-percentile range for $p=0.02$, where $p$ is the pth quantile of the magnitude distribution \citep{Malanchev2021}.
    \item \texttt{feature\_inter\_percentile\_range\_10\_magn}: Inter-percentile range for $p=0.10$, where $p$ is the pth quantile of the magnitude distribution. A special case of the interpercentile range known as the interdecile range \citep{Malanchev2021}.
    \item \texttt{feature\_inter\_percentile\_range\_25\_magn}: Inter-percentile range for $p=0.25$, where $p$ is the pth quantile of the magnitude distribution. A special case of the interpercetile range known as the interquartile range \citep{Malanchev2021}.
    \item \texttt{feature\_kurtosis\_magn}: Excess kurtosis of magnitude \citep{Malanchev2021}.
    \item \texttt{feature\_linear\_fit\_slope\_magn}: The slope of the light curve in the least squares fit of the linear stochastic model with Gaussian noise described by observation errors \{$\delta_{i}$\} \citep{Malanchev2021}.
    \item \texttt{feature\_linear\_fit\_slope\_sigma\_magn}: The error of the slope of the light curve in the least squares fit of the linear stochastic model with Gaussian noise described by observation errors \{$\delta_{i}$\} \citep{Malanchev2021}.
    \item \texttt{feature\_magnitude\_percentage\_ratio\_40\_5\_magn}: The magnitude 40 to 5 ratio, written in terms of the magnitude distribution quantile function $Q$. \citep{D'Isanto2016}.
    \item \texttt{feature\_magnitude\_percentage\_ratio\_20\_5\_magn}: The magnitude 20 to 5 ratio, written in terms of the magnitude distribution quantile function $Q$. \citep{D'Isanto2016}.
    \item \texttt{feature\_mean\_magn}: The non-weighted mean magnitude.
    \item \texttt{feature\_median\_absolute\_deviation\_magn}: The median of the absolute value of the difference between magnitude and its median \citep{D'Isanto2016}.
    \item \texttt{feature\_percent\_amplitude\_magn}: The maximum deviation of magnitude from its median \citep{D'Isanto2016}.
    \item \texttt{feature\_median\_buffer\_range\_percentage\_10\_magn}: The fraction of observations inside Median($m$) $\pm 10 \times$ (max($m$)-min($m$))/2 interval \citep{D'Isanto2016}.
    \item \texttt{feature\_median\_buffer\_range\_percentage\_20\_magn}: The fraction of observations inside Median($m$) $\pm 20 \times$ (max($m$)-min($m$))/2 interval \citep{D'Isanto2016}.
    \item \texttt{feature\_percent\_difference\_magnitude\_percentile\_5\_magn}: Ratio of $p$=5th inter-percentile range to the median \citep{Malanchev2021}.
    \item \texttt{feature\_percent\_difference\_magnitude\_percentile\_10\_magn}: Ratio of $p$=10th inter-percentile range to the median \citep{Malanchev2021}.
    \item \texttt{feature\_skew\_magn}: Skewness of magnitude, $G_{1}$ \citep{Malanchev2021}.
    \item \texttt{feature\_standard\_deviation\_magn}: 	Standard deviation of magnitude, $\sigma_{m}$ \citep{Malanchev2021}.
    \item \texttt{feature\_stetson\_k\_magn}: Stetson $K$ coefficient described light curve shape of magnitude \citep{Stetson1996}.
    \item \texttt{feature\_weighted\_mean\_magn}: Weighted mean magnitude \citep{Malanchev2021}.
    \item \texttt{feature\_anderson\_darling\_normal\_flux}: Unbiased Anderson–Darling normality test statistic for flux \citep{Malanchev2021}.
    \item \texttt{feature\_cusum\_flux}: A range of cumulative sums dependent on the number of observations, mean flux, and flux standard deviation \citep{Kim2014}.
    \item \texttt{feature\_excess\_variance\_flux}: Measure of the flux variability amplitude \citep{Sanchez2017}.
    \item \texttt{feature\_kurtosis\_flux}: Excess kurtosis of flux \citep{Malanchev2021}.
    \item \texttt{feature\_mean\_variance\_flux}: Standard deviation of flux to mean flux ratio \citep{Malanchev2021}.
    \item \texttt{feature\_skew\_flux}: Skewness of flux \citep{Malanchev2021}.
    \item \texttt{feature\_stetson\_k\_flux}: Stetson $K$ coefficient described light curve shape of flux \citep{Stetson1996}.
\end{itemize}

The full documentation, including equations, can be found here: \url{https://docs.rs/light-curve-feature/0.2.2/light_curve_feature/features/index.html}.

\section{Host Galaxy Features}
\label{appx:host_gal_features}

\counterwithin{figure}{section}
\counterwithin{table}{section}
\renewcommand{\thefigure}{B.\arabic{figure}}
\setcounter{figure}{0}
\renewcommand{\thetable}{B.\arabic{table}} \setcounter{table}{0}

Our host galaxy features and a brief description are as follows\footnote{Those from PS1 can be found here: \url{https://outerspace.stsci.edu/display/PANSTARRS/PS1+Database+object+and+detection+tables}}:

\begin{itemize}
    \item \texttt{gmomentXX}: Second moment $M_{xx}$ for $g$ filter stack detection. 
    \item \texttt{gmomentXY}: Second moment $M_{xy}$ for $g$ filter stack detection. 
    \item \texttt{gmomentYY}: Second moment $M_{yy}$ for $g$ filter stack detection. 
    \item \texttt{gmomentR1}: First radial moment for $g$ filter stack detection.
    \item \texttt{gmomentRH}: Half radial moment ($r^{0.5}$ weighting) for $g$ filter stack detection.
    \item \texttt{gPSFFlux}: PSF flux from $g$ filter stack detection.
    \item \texttt{gApFlux}: Aperture flux from $g$ filter stack detection.
    \item \texttt{gKronFlux}: \cite{Kron1980} flux from $g$ filter stack detection.
    \item \texttt{gKronRad}: \cite{Kron1980} radius from $g$ filter stack detection.
    \item \texttt{gExtNSigma}: An extendedness measure for the $g$ filter stack detection based on the deviation between PSF and \cite{Kron1980} magnitudes, normalized by the PSF magnitude uncertainty.
    \item \texttt{rmomentXX}: Second moment $M_{xx}$ for $r$ filter stack detection. 
    \item \texttt{rmomentXY}: Second moment $M_{xy}$ for $r$ filter stack detection. 
    \item \texttt{rmomentYY}: Second moment $M_{yy}$ for $r$ filter stack detection. 
    \item \texttt{rmomentR1}: First radial moment for $r$ filter stack detection.
    \item \texttt{rmomentRH}: Half radial moment ($r^{0.5}$ weighting) for $r$ filter stack detection.
    \item \texttt{rPSFFlux}: PSF flux from $r$ filter stack detection.
    \item \texttt{rApFlux}: Aperture flux from $r$ filter stack detection.
    \item \texttt{rKronFlux}: \cite{Kron1980} flux from $r$ filter stack detection.
    \item \texttt{rKronRad}: \cite{Kron1980} radius from $r$ filter stack detection.
    \item \texttt{rExtNSigma}: An extendedness measure for the $r$ filter stack detection based on the deviation between PSF and \cite{Kron1980} magnitudes, normalized by the PSF magnitude uncertainty.
    \item \texttt{imomentXX}: Second moment $M_{xx}$ for $i$ filter stack detection. 
    \item \texttt{imomentXY}: Second moment $M_{xy}$ for $i$ filter stack detection. 
    \item \texttt{imomentYY}: Second moment $M_{yy}$ for $i$ filter stack detection. 
    \item \texttt{imomentR1}: First radial moment for $i$ filter stack detection.
    \item \texttt{imomentRH}: Half radial moment ($r^{0.5}$ weighting) for $i$ filter stack detection.
    \item \texttt{iPSFFlux}: PSF flux from $i$ filter stack detection.
    \item \texttt{iApFlux}: Aperture flux from $i$ filter stack detection.
    \item \texttt{iKronFlux}: \cite{Kron1980} flux from $i$ filter stack detection.
    \item \texttt{iKronRad}: \cite{Kron1980} radius from $i$ filter stack detection.
    \item \texttt{iExtNSigma}: An extendedness measure for the $i$ filter stack detection based on the deviation between PSF and \cite{Kron1980} magnitudes, normalized by the PSF magnitude uncertainty.
    \item \texttt{zmomentXX}: Second moment $M_{xx}$ for $z$ filter stack detection. 
    \item \texttt{zmomentXY}: Second moment $M_{xy}$ for $z$ filter stack detection. 
    \item \texttt{zmomentYY}: Second moment $M_{yy}$ for $z$ filter stack detection. 
    \item \texttt{zmomentR1}: First radial moment for $z$ filter stack detection.
    \item \texttt{zmomentRH}: Half radial moment ($r^{0.5}$ weighting) for $z$ filter stack detection.
    \item \texttt{zPSFFlux}: PSF flux from $z$ filter stack detection.
    \item \texttt{zApFlux}: Aperture flux from $z$ filter stack detection.
    \item \texttt{zKronFlux}: \cite{Kron1980} flux from $z$ filter stack detection.
    \item \texttt{zKronRad}: \cite{Kron1980} radius from $z$ filter stack detection.
    \item \texttt{zExtNSigma}: An extendedness measure for the $z$ filter stack detection based on the deviation between PSF and \cite{Kron1980} magnitudes, normalized by the PSF magnitude uncertainty.
    \item \texttt{ymomentXX}: Second moment $M_{xx}$ for $y$ filter stack detection. 
    \item \texttt{ymomentXY}: Second moment $M_{xy}$ for $y$ filter stack detection. 
    \item \texttt{ymomentYY}: Second moment $M_{yy}$ for $y$ filter stack detection. 
    \item \texttt{ymomentR1}: First radial moment for $y$ filter stack detection.
    \item \texttt{ymomentRH}: Half radial moment ($r^{0.5}$ weighting) for $y$ filter stack detection.
    \item \texttt{yPSFFlux}: PSF flux from $y$ filter stack detection.
    \item \texttt{yApFlux}: Aperture flux from $y$ filter stack detection.
    \item \texttt{yKronFlux}: \cite{Kron1980} flux from $y$ filter stack detection.
    \item \texttt{yKronRad}: \cite{Kron1980} radius from $y$ filter stack detection.
    \item \texttt{yExtNSigma}: An extendedness measure for the $y$ filter stack detection based on the deviation between PSF and \cite{Kron1980} magnitudes, normalized by the PSF magnitude uncertainty.
    \item \texttt{i-z}: Aperture magnitude for the $i$ filter stack detection minus aperture magnitude for the $z$ filter stack detection. (iApMag - zApMag).
    \item \texttt{gApMag\_gKronMag}: Aperture magnitude minus the \cite{Kron1980} magnitude for the $g$ filter stack detection.
    \item \texttt{rApMag\_rKronMag}: Aperture magnitude minus the \cite{Kron1980} magnitude for the $r$ filter stack detection.
    \item \texttt{iApMag\_iKronMag}: Aperture magnitude minus the \cite{Kron1980} magnitude for the $i$ filter stack detection.
    \item \texttt{zApMag\_zKronMag}: Aperture magnitude minus the \cite{Kron1980} magnitude for the $z$ filter stack detection.
    \item \texttt{yApMag\_yKronMag}: Aperture magnitude minus the \cite{Kron1980} magnitude for the $y$ filter stack detection.
    \item \texttt{4DCD}: A 4-dimensional color distance in $g-r$, $r-i$, $i-z$, and $z-y$ from the PS1 stellar locus, the path traced by stars in color-color space \citep{Tonry2012}.
    \item \texttt{dist/DLR}: Transient-host separation (arcsec) normalized by the directional light radius.  
\end{itemize}
 

\section{Feature Correlations}
\label{subsec:APP_feat_corr}

\counterwithin{figure}{section}
\counterwithin{table}{section}
\renewcommand{\thefigure}{C.\arabic{figure}}
\setcounter{figure}{0}
\renewcommand{\thetable}{C.\arabic{table}} \setcounter{table}{0}

With a large 120 dimensional feature-space, we are likely to have dependent or correlated features. In this subsection, we argue that this is an intentional choice that proffers quantitative benefits rather than suffers from poor design. \par

\begin{figure*}
    \centering
    \includegraphics[scale=0.72]{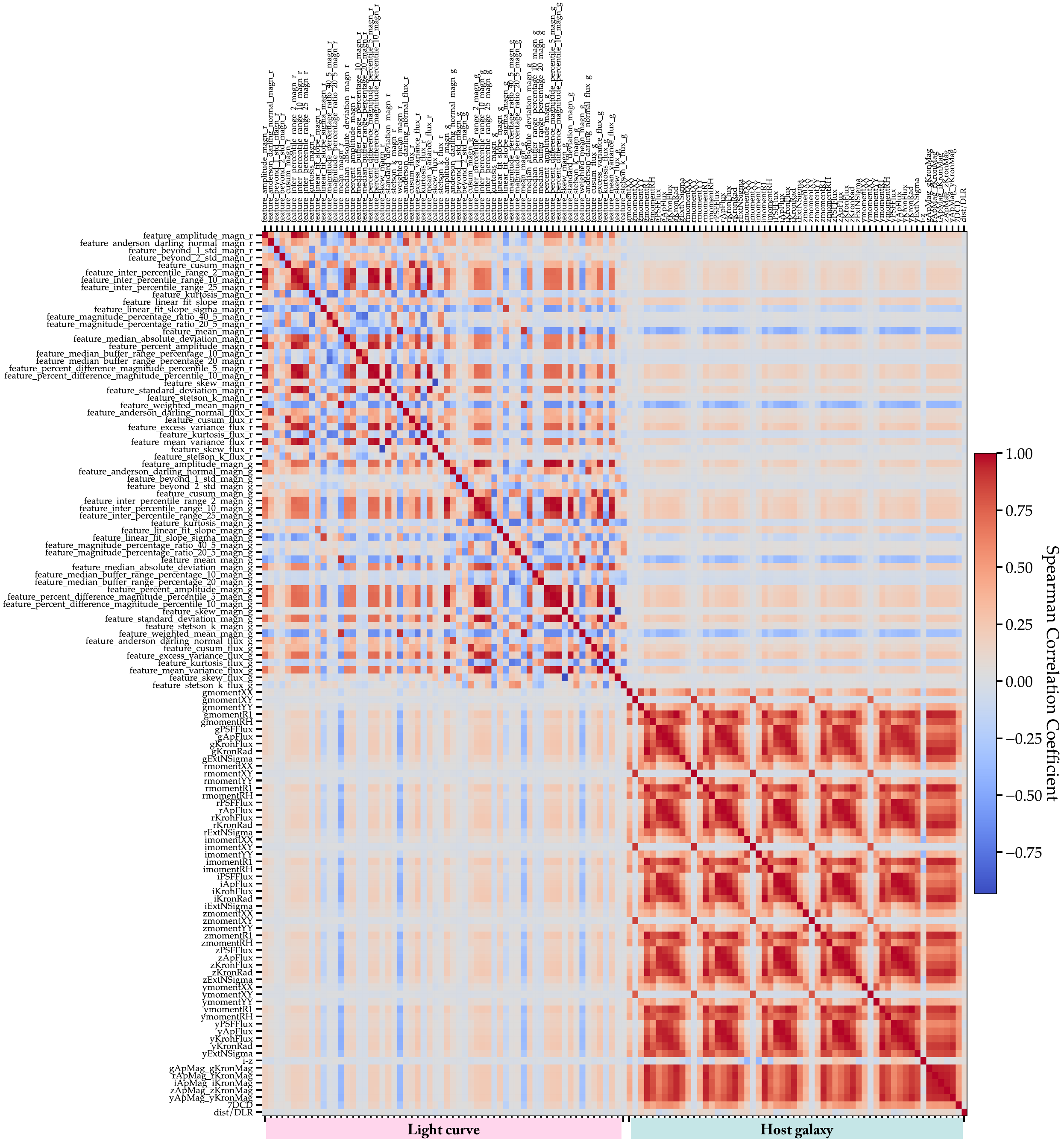}
    \caption{
    The Spearman rank correlation matrix for the \laiss{} databank of extracted ZTF-$g$, ZTF-$r$ light curve and PS1-$grizy$ host galaxy features (see~Appendix~\ref{appx:host_gal_features}), with red corresponding to positively correlated features and blue corresponding to negatively correlated features. Broadly, the light curve features that measure amplitude or amplitude-adjacent properties are correlated, more strongly intra-passband versus inter-passband. There exists strong positive correlations between brightness and radial moments of host galaxies in each band, forming a repeating block structure. From a high-level, the correlation between light curve features to host galaxy features is poor. Overall, these features are chosen to maximize color and capture related but different information of the supernova light curve and host galaxy, and to reduce the impact of dependent or irrelevant features.
    } 
    \label{fig:120d_corr}
\end{figure*}

We present a Spearman rank-correlation matrix of our training set in Figure~\ref{fig:120d_corr}. Note that we choose Spearman's rank correlation in favor of Pearson correlation because the former can quantifiably describe linear and non-linear relationships (those which can be described with a monotonic function), whereas the latter describes only linear relationships. \par

For light curve features, we use the same set of 31 for both ZTF-$r$, ZTF-$g$ bands. Broadly, the features that measure amplitude or amplitude-adjacent properties are correlated across intra-passband and inter-passband. However, those that are correlated intra-passband are measuring different degrees of amplitude or amplitude variation during a supernova's evolution (e.g., those measuring the 2nd, 10th, and 25th quantile of the magnitude distribution). Meanwhile, those that are inter-passband correlated provide utility because of the color information they capture. Moreover, inter-passband correlation vectors are less correlated than their intra-passband counterparts. \par

For host galaxy features, it is not only known that some features are highly correlated with each other; in fact, it has been leveraged in its use in empirical relationships such as the Fundamental Plane for ellipticals \citep{Dressler1987}, the color-magnitude relation \citep{Bell2004}, and more recently in cosmological analyses with SNe~Ia in regard to host-galaxy stellar mass \citep{Popovic2021, Kelsey2023}. See, e.g., \cite{Sullivan2010, Kelly2010, BroutScolnic2021, Grayling2024} for additional discussion regarding the relationship between SN and galaxy properties. \par 

We find that with the exception of the second moment \texttt{\{g,r,i,z,y\}momentXY}, all of our intra-passband host features are positively correlated, but to varying degrees. However, the same features inter-passband tend to exhibit strong correlations (e.g., \texttt{\{g,r,i,z,y\}momentR1}). Easier to see is the appearance of blocks within the full correlation matrix, which are the strong positive correlations between brightness and radial moments of host galaxies in each band. These findings are largely consistent with those found in \cite{Gagliano2021} (see further details in their Section~4).

From a broader viewpoint, the correlation between light curve and host galaxy features is poor at best, with the largest positive Spearman correlation of $\rho\sim+0.30$ existing between the slope of the light curve in the least squares fit of the linear stochastic model \texttt{feature\_linear\_fit\_slope\_magn\_{g,r}} and nearly all host galaxy features (with the exception of the second moment \texttt{\{g,r,i,z,y\}momentXY}, aperture magnitude $i-z$ color difference \texttt{i-z}, and normalized directional light radius \texttt{dist/DLR}). The most negative Spearman correlation of $\rho\sim-0.30$ relates the majority of host galaxy features to the non-weighted mean magnitude \texttt{feature\_mean\_magn\_\{g,r\}} and the weighted mean magnitude \texttt{feature\_weighted\_mean\_magn\_\{g,r\}}. Perhaps the most interesting feature is the normalized directional light radius \texttt{dist/DLR}, which is not correlated with any other feature except for itself, and is the most discriminating host galaxy feature used in our RFC model (see Figure~\ref{fig:RFC_feat_importance}).

Typically, the large number of features used in this work could fall prey to the \emph{curse of dimensionality} and would prohibit a brute-force ($\mathcal{O}(n)$ time) search across SNe. We circumvent this by 1) using a RFC for our AD model, which inherently performs feature selection when building each decision tree in the forest (where a random subset of \texttt{max\_features} features at each split is used\footnote{This process effectively reduces the dimensionality because not all features are used for each tree, helping to avoid overfitting, and is robust to more irrelevant features.}), and 2) using dimensionality reduction via PCA for our ANN similarity search with \texttt{ANNOY} (which itself is natively $\mathcal{O}(log(n))$ time) to efficiently search the entire feature space. \par

\section{Methods for \texttt{PROSPECTOR} Host-galaxy SED Modeling}
\label{subsec:APP_SED_Prospector}

\counterwithin{figure}{section}
\counterwithin{table}{section}
\renewcommand{\thefigure}{D.\arabic{figure}}
\setcounter{figure}{0}
\renewcommand{\thetable}{D.\arabic{table}} \setcounter{table}{0}

We correct photometry by subtracting an extinction correction using the corrected SFD map \citep{CSFDDustmapChiang2023}, provided through the \texttt{DUSTMAP}\footnote{\url{https://github.com/gregreen/dustmaps}} package. To make these corrections we assume an average $R_{v}$ = 3.1. We set a lower bound on the photometric error in any filter as 15\% to allow for systematic error--- which we note is a larger bound than used in \cite{Leja_2019}. We felt it was necessary to increase the lower bound of uncertainty given our inexact matched aperture sizes. 

The final vector of photometric observations and uncertainties are used to fit our stellar population synthesis model. Prospector calls the \texttt{PYTHON-FSPS}\footnote{\url{https://github.com/dfm/python-fsps}} library to perform the fits \citep{Conroy_2009_FSPS,Conroy_2010_FSPS}. The library uses the MIST isochrones to build the models \citep{Choi_2016_MIST,Dotter_2016_MIST}. We use a Chabrier initial mass function \citep{Chabrier2003} with a delayed-tau stellar formation history ($t$ $\times$ $\exp{-t/\tau}$). We set the redshift to the value of the known transient's redshift. We model dust attenuation using the Milky Way Dust extinction law \citep{MWExtinction}, and additionally attenuate young stars with an extra static dust contribution of 0.5 (for additional details see \citealt{Conroy_2009_FSPS}).
We model the contribution from Nebular Emission but fix the gas phase metallicity and the gas ionization parameters \citep{BylerNebularEmission2017}. We use the \cite{Gallazzi2005} mass-metallicity relationship as a prior on the stellar metallicity, following its use in \cite{RecentAnyaNugent23}. 
The final model is a function of 5 free parameters. Our priors are described in Table~\ref{tab:prospectorpriors}.


\begin{table*}[ht]
\centering
\caption{\textbf{\texttt{PROSPECTOR} Parameter Priors.} We describe our model, initial parameter estimates, and priors below. We broadly follow the suggestions of model complexity following the procedure of \cite{RecentAnyaNugent23}. Below, $\mathcal{U}(a,b)$ is the Top Hat distribution on the interval $(a,b)$. $\delta$ means that the parameter is always fixed.}
\begin{tabular}{lll}
\hline
\hline
Parameter & Prior distribution & Notes\\
\hline
mass & $\log{\mathcal{U}(1e6,1e13)}$ & \texttt{PROSPECTOR} samples masses formed; we report stellar masses \\
logzsol & $\mathcal{U}(-2,0.4)$ & Limited by availability of MIST Isochrones\footnote{The MIST models developed for ancient, metal-poor populations from -4.0$~\leq$[Z/H]$\textless$~-2.0 from \cite{Choi_2016_MIST} are not included in \texttt{FSPS}.} \citep{Choi_2016_MIST} \\
dust2 & $\mathcal{U}(0.0,4.0)$ & --- \\
tage & $\mathcal{U}(0.0,13.8)$ & Age of stellar population  \\
tau & $\log{\mathcal{U}(0.01,10)}$ & --- \\
\hline
IMF & $\delta(1)$ & Chabrier IMF  \\
Dust Type &  $\delta(1)$ &  Milky Way Extinction Law \\
SFH & $\delta(4)$ & Delayed-Tau  \\
\hline
\hline
\end{tabular}
\label{tab:prospectorpriors}
\end{table*}


\section{Spectra}
\label{subsec:APP_Spectra}

\counterwithin{figure}{section}
\counterwithin{table}{section}
\renewcommand{\thefigure}{E.\arabic{figure}}
\setcounter{figure}{0}
\renewcommand{\thetable}{E.\arabic{table}} \setcounter{table}{0}

Here we describe our observations and data reduction for the 9 spectra whose classifications are used in Table~\ref{tab:real_time_any_phase}. 

TDE~2023kvy was discovered on 2023~June~16 23.59 by \cite{ATLAS2023kvy} and classified as a TDE on 2023~September~9.97 \citep{FLEET2023kvy} as part of the FLEET program \citep{Gomez2020} from a SOAR / Goodman spectrum acquired on 2023~August~27. \texttt{LAISS} correctly tagged TDE~2023kvy as an anomaly on 2023~August~26. This spectrum was acquired in coordination with FLEET, whose program identified this object independently. \par

SN~2023sds was discovered on 2023~September~11.38 by \cite{ALeRCE2023sds} and classified by us as an SN~II on 2024~March~24.10 \citep{Aleo2023sds} from a Lick-3m / KAST spectrum acquired by us on 2023~October~22.19. \texttt{LAISS} correctly tagged SN~2023sds as an anomaly on 2023~September~24. \par

SN~2023nof was discovered on 2023~July~23.34 by \cite{ALeRCE2023nof} and originally classified as an SN~II on 2023~July~27.34 by \cite{DEbass2023nof} from an ANU-2.3m / WiFeS spectrum acquired on 2023~July~26. Later, we reclassified as an SN~IIn on 2023~October~06.77 from an ANU-2.3m / WiFeS spectrum acquired by us on 2023~September~29 \citep{Aleo2023nofTNS}. \texttt{LAISS} correctly tagged SN~2023nof as an anomaly on 2023~September~24. \par

SN~2023nwe was discovered on 2023~July~24.43 by \cite{PS12023nwe} and classified by us as an SN~IIn on 2023~September~11.69 \citep{Davis2023nwe} from a Lick-3m / KAST spectrum acquired by us on 2023~September~06.30. \texttt{LAISS} correctly tagged SN~2023nwe as an anomaly on 2023~August~28. \par

SN~2023omf was discovered on 2023~July~31.35 by \cite{ZTF2023omf} and classified by us as an SN~IIn on 2024~March~24.15 \citep{Aleo2023omf} from a Lick-3m / KAST spectrum acquired by us on 2023~October~22.11. \texttt{LAISS} correctly tagged SN~2023nwe as an anomaly on 2023~October~10. \par

SN~2023ocx was discovered on 2023~July~20.22 by \cite{ZTF2023ocx} and classified as an SN~Ia-CSM by us on 2023~October~02.52 \citep{Angus2023ocx} from a NOT / ALFOSC spectrum acquired by us on 2023~September~19. \texttt{LAISS} correctly tagged SN~2023ocx as an anomaly on 2023~August~28. \par

SN~2023otw was discovered on 2023~July~29.46 by \cite{ZTF2023otw} and classified as an SN~II on 2023~November~07.56 \citep{FLEET2023otw} as part of the FLEET program \citep{Gomez2020} from a SOAR / Goodman spectrum acquired on 2023~October~11. This spectrum was acquired in coordination with FLEET, whose program identified this object independently. \texttt{LAISS} incorrectly tagged SN~2023otw as an anomaly on 2023~September~20. \par

SN~2023swf was discovered on 2023~September~16.25 by \cite{ZTF2023swf} and classified as an SN~II on 2023~October~04.72 \citep{Ayala2023swf} from a ESO-NTT / EFOSC2-NTT (ePESSTO+) spectrum acquired on 2023~October~04.21. We retained the original SN~II classification from a ANU-2.3m / WiFeS spectrum acquired by us on 2023~October~11. \texttt{LAISS} incorrectly tagged SN~2023swf as an anomaly on 2023~October~09. \par

SN~2023sed was discovered on 2023~September~08.35 by \cite{ALeRCE2023sed} and classified by us as an SN~Ia on 2023~October~02.55 \citep{Angus2023sed} from a NOT / ALFOSC spectrum acquired by us on 2023~September~28. \texttt{LAISS} incorrectly tagged SN~2023sed as an anomaly on 2023~September~24. \par

To reduce the Kast data, we used the {\tt UCSC Spectral Pipeline}\footnote{\url{https://github.com/msiebert1/UCSC\_spectral\_pipeline}} \citep{Siebert20}, a custom data-reduction pipeline based on procedures outlined by \citet{Foley03}, \citet{Silverman2012}, and references therein.  The two-dimensional (2D) spectra were bias-corrected, flat-field corrected, adjusted for varying gains across different chips and amplifiers, and trimmed.  One-dimensional spectra were extracted using the optimal algorithm \citep{Horne86}.  The spectra were wavelength-calibrated using internal comparison-lamp spectra with linear shifts applied by cross-correlating the observed night-sky lines in each spectrum to a master night-sky spectrum.  Flux calibration and telluric correction were performed using standard stars at a similar airmass to that of the science exposures.  We combine the sides by scaling one spectrum to match the flux of the other in the overlap region and use their error spectra to correctly weight the spectra when combining.  More details of this process are discussed elsewhere \citep{Foley03, Silverman2012, Siebert20, Davis22}.

Data obtained with ALFOSC, GOODMAN, and WiFeS were reduced using standard techniques, which included correction for bias, overscan, and flat-field. Spectra of comparison lamps and standard stars acquired during the same night and with the same instrumental setting have been used for the wavelength and flux calibrations, respectively. We employed standard \textsc{IRAF} commands to extract all spectra.

A table summarizing all 9 spectroscopic follow-up observations follows in Table~\ref{tab:observations}.

\begin{table}[htbp]
    \centering
    \caption{Spectroscopic Observation Details}
    \begin{tabular}{cccccc}
    \hline
    \hline
    \textbf{IAU Name} & \textbf{Obs Date (UT)} & \textbf{MJD} & \textbf{Estimated Phase (Days)} & \textbf{Telescope} & \textbf{Instrument} \\
    \hline
    2023kvy & 2023-08-27 & 60183 & +65 & SOAR & GOODMAN \\
    2023sds & 2023-10-22 & 60239 & +34 & Lick-3m & KAST \\
    2023nof & 2023-09-29 & 60196 & +43 & ANU-2.3m & WiFeS \\
    2023nwe & 2023-09-06 & 60193 & +14 & Lick-3m & KAST \\
    2023omf & 2023-10-22 & 60239 & +52 & Lick-3m & KAST \\
    2023ocx & 2023-09-19 & 60206 & +33 & NOT & ALFOSC \\
    2023otw & 2023-10-11 & 60228 & +65 & SOAR & GOODMAN \\
    2023swf & 2023-10-11 & 60228 & +7 & ANU-2.3m & WiFeS \\
    2023sed & 2023-09-28 & 60215 & +4 & NOT & ALFOSC \\
    \hline
    \hline\\[-1.5ex]
    \multicolumn{6}{c}{
    \begin{minipage}{16cm}
    Log of spectroscopic observations for TDE~2023kvy, SN~2023sds, SN~2023nof, SN~2023nwe, SN~2023omf, SN~2023ocx, SN~2023otw, SN~2023swf, SN~2023sed.\\
    \end{minipage}}
    \end{tabular}
    \label{tab:observations}
\end{table}


\section{Tables}
\label{subsec:APP_add_tables}

\counterwithin{figure}{section}
\counterwithin{table}{section}
\renewcommand{\thefigure}{F.\arabic{figure}}
\setcounter{figure}{0}
\renewcommand{\thetable}{F.\arabic{table}} \setcounter{table}{0}

In Table~\ref{tab:report_to_TNS_SNe}, we report the following previously undiscovered SNe to TNS from ANN=8 nearest neighbor matches to all SNe classes. Bolded text indicates an object we consider a likely anomaly. Italicised text indicates an object that appeared more than once from the ANN=8 search (e.g., ZTF18abmofew/2018mfo was uniquely found 7 times during the ANN=8 search). These objects are listed for completeness for each match shown. Only classes for which at least one discovery was made is shown.

In Table~\ref{tab:remaining_no_TNS}, we report the following previously undiscovered SNe to TNS from an exhaustive search of our databank. We do not include objects discovered previously through our ANN search (see Table~\ref{tab:report_to_TNS_SNe}). 

\newpage




\hspace*{-10mm}
\begin{minipage}{\textwidth}
\footnotesize
\centering \captionof{table}{\textbf{84 Previously Undiscovered/Reported SNe Found by ANN=8 Search}}
\vspace*{-1mm}
\label{tab:report_to_TNS_SNe}
\begin{tabular}{cccccc}

\toprule
ZTF ID & IAU Name & Possible Class. & Reference SNe & ANN & Remarks \\
\hline 
\hline
\multicolumn{6}{c}{\textit{Reference Class: SN~Ia (1098 objects)}} \\
ZTF18achdfqm & 2018mfl & SN~Ia? & 2020vnl & 4 & Host-$z$=0.095. Peak $M_{abs}\sim-18.8$~mag. \\
ZTF18abnzney & \textit{2018mfm} & SN~Ia? & 2020whs & 1 & \nodata \\
ZTF21abcluco & \textit{2021ahvn} & SN~Ia? & " & 3 & \nodata \\
ZTF17aadqidc & 2021ahvo & SN~I? & " & 5 & Host-$z$=0.087. Peak $M_{abs}\sim-18.6$~mag. \\
ZTF18acpegrg & 2018mfn & SN~Ia? & 2020wts & 2 & Host-$z$=0.132. Peak $M_{abs}\sim-19.6$~mag. \\
ZTF18abmofew & \textit{2018mfo} & SN~Ia? & 2020xyh & 2 & Well sampled. \\
ZTF19aaaamwp & 2018mfp & SN~Ia? & 2020xit & 5 & Host-$z$=0.124. Peak $M_{abs}\sim-19.1$~mag. \\
ZTF21aamssts & \textit{2021ahvp} & SN~Ia? & 2020zcw & 5 & \nodata \\
ZTF21abvjcnb & 2021ahvq & SN~I? & " & 7 & \nodata \\
ZTF18acbvtjm & \textit{2018mfq} & SN? & 2020zbr & 1 & \nodata \\
ZTF20aconebc & 2020afeq & SN? & " & 5 & \nodata \\
ZTF21aamssts & \textit{2021ahvp} & SN~Ia? & 2020abrg & 1 & \nodata \\
ZTF18abrxvpd & 2018mfr & SN~Ia? & " & 4 & \nodata \\
ZTF20adadbsm & \textit{2021ahvr} & SN~I? & 2020acmi & 5 & \nodata \\
ZTF20adadbsm & \textit{2021ahvr} & SN~I? & 2020acvz & 4 & \nodata \\
ZTF21aalgilf & 2021ahvs & SN~Ia? & 2021ab & 8 & \nodata \\
ZTF21aarpnxt & 2021ahvt & SN~Ia? & 2021Y & 1 & \nodata \\
ZTF18aczerlj & \textit{2018mfs} & SN~II/IIP? & 2021by & 2 & Well sampled. Host-$z$=0.066. Peak $M_{abs}\sim-17.7$~mag. \\
ZTF18acbzvzm & \textit{2018mft} & SN~Ia? & 2021vt & 1 & Host-$z$=0.106. Peak $M_{abs}\sim-19.5$~mag. \\
ZTF21aamssts & \textit{2021ahvp} & SN~Ia? & 2021apk & 8 & \nodata \\
ZTF18acbwdym & \textit{2018mfu} & SN~Ia? & 2021arr & 2 & Host-$z$=0.091. Peak $M_{abs}\textgreater-19.0$~mag. \\
\textbf{ZTF18acbwgmi} & 2018mfv & SN~Ib/c? & 2021buy & 5 & Tagged by AD model. \\
ZTF19abxsehw & 2019aata & SN~I? & 2021eij & 3 & \nodata \\
ZTF20acvdqsy & 2020afer & SN~Ia? & " & 7 & \nodata \\
ZTF18abokvkt & \textit{2018mfw} & SN~Ia? & 2021gfi & 3 & Faint host. \\
ZTF19aaviczu & 2019aatb & SN~Ia? & 2021iok & 2 & Host-$z$=0.156. $M_{abs}\sim-19.6$~mag. \\
ZTF18abtgmuw & 2018mfx & SN~Ia? & " & 8 & Faint host. \\
ZTF18abnvnqb & 2018mfy & SN~Ia? & 2021lea & 4 & Well sampled. Faint host. \\
ZTF21aalimvt & 2021ahvu & SN? & 2021mhm & 2 & No decline. \\
ZTF21aadruss & 2021ahvv & SN~Ia? & 2021mla & 4 & \nodata \\
ZTF21aaxswjy & 2021ahvw & SN~Ia? & 2021nya & 5 & Faint host. \\
ZTF21aajtfas & \textit{2021ahvx} & SN~Ia? & 2021oia & 1 & Well sampled. \\
ZTF21aaswuea & 2021ahvy & SN~Ia? & 2021nsh & 7 & Host-$z$=0.134. $M_{abs}\sim-19.0$~mag. \\
ZTF21aajtfas & \textit{2021ahvx} & SN~Ia? & 2021tvc & 5 & Well sampled. \\
ZTF18abmofew & \textit{2018mfo} & SN~Ia? & 2021uwa & 5 & Well sampled. \\
ZTF20acuyxki & 2020afes & SN~Ia? & 2021uvx & 1 & Host-$z$=0.121. $M_{abs}\sim-19.2$~mag. \\
ZTF18abmofew & \textit{2018mfo} & SN~Ia? & 2021vku & 4 & Well sampled. \\
\textbf{ZTF18abtjrbt} & 2018mfz & TDE? & 2021vnv & 7 & Blue. SALT3 $x_{1}=+3.00$. Host-$z$=0.065. $M_{abs}\sim-18.4$~mag. \\
\textbf{ZTF18ablqjws} & \textit{2018mga} & SN~II/IIn? & 2021wym & 3 & Well sampled. Tagged by AD model. \\
ZTF20acjoazg & \textit{2020afet} & SN~Ia? & 2021xoh & 6 & Well sampled. \\
ZTF20acthdtq & 2020afeu & SN~Ia? & 2021wwc & 1 & Faint host. \\
ZTF18acbzvzm & \textit{2018mft} & SN~Ia? & 2021xpk & 1 & Host-$z$=0.106. Peak $M_{abs}\sim-19.5$~mag. \\
ZTF18abmofew & \textit{2018mfo} & SN~Ia? & 2021ycy & 5 & Well sampled. \\
ZTF18abuioue & \textit{2018mgb} & SN~Ia? & " & 2 & No visible host. \\
ZTF20acjoazg & \textit{2020afet} & SN~Ia? & 2021ysn & 2 & Well sampled. \\
ZTF18acvgwfq & 2018mgc & SN~Ia? & 2021aafd & 1 & Fast decline. \\
ZTF18acbwdym & \textit{2018mfu} & SN~Ia? & 2021aaem & 2 & Host-$z$=0.091. Peak $M_{abs}\textgreater-19.0$~mag. \\
\textbf{ZTF18abwhsnx} & \textit{2018mgd} & SN~Ib/c? & 2021aaht & 7 & Tagged by AD model. \\
\textbf{ZTF18abwhsnx} & \textit{2018mgd} & SN~Ib/c? & 2021abds & 2 & Tagged by AD model. \\
ZTF18abmofew & \textit{2018mfo} & SN~Ia? & 2021abaq & 7 & Well sampled. \\
ZTF21abdhwhj & 2021ahvz & SN? & 2021abpc & 4 & \nodata \\
ZTF21aamssts & \textit{2021ahvp} & SN~Ia? & 2021acgo & 5 & \nodata \\
ZTF18acpefbb & 2018mge & SN~Ia? & 2017baq & 4 & Host-$z$=0.135. Peak $M_{abs}\sim-19.5$~mag. \\
\textbf{ZTF21aaualdi} & 2021ahwa & SLSN? & 2021acgu & 7 & No visible host. FLEET=46\%(41\%) SLSN-I(II). Tagged by AD model. \\
\hline
\end{tabular}
\end{minipage} \hfill
\newpage
\begin{minipage}{\textwidth}
\footnotesize
\centering
\hspace*{-10mm}
\vspace*{-1mm}
\begin{tabular}{cccccc}

\hline
ZTF ID & IAU Name & Possible Class. & Reference SNe & ANN & Remarks \\
\hline
ZTF18absljwl & \textit{2018mgf} & SN~II? & 2021acfg & 6 & No rise. \\
ZTF18acbzvzm & \textit{2018mft} & SN~Ia? & 2021acwr & 1 & Host-$z$=0.106. Peak $M_{abs}\sim-19.5$~mag. \\
ZTF18acefgee & 2018mgg & SN~I? & 2021aczd & 1 & \nodata \\
ZTF21aabvpub & \textit{2021ahwb} & SN~Ia? & " & 7 & \nodata \\
ZTF21aayngti & 2021ahwc & SN~Ia? & " & 8 & Host-$z$=0.142. Peak $M_{abs}\sim-19.1$~mag.\\
ZTF21abcluco & \textit{2021ahvn} & SN~Ia? & 2021adeq & 2 & \nodata \\
ZTF18acbvijf & \textit{2018mgh} & SN~Ia? & " & 5 & \nodata \\
ZTF18abuioue & \textit{2018mgb} & SN~Ia? & 2021aden & 6 & No visible host. \\
ZTF19aacislx & 2019aatc & SN~Ia? & 2021aefa & 1 & Faint host. \\
ZTF21aamfdzo & \textit{2021ahwd} & SN~Ia? & 2022is & 8 & \nodata \\
ZTF21abghbue & 2021ahwe & SN~Ia? & 2022aby & 4 & Faint host. \\
ZTF18acwward & 2018mgi & SN~Ia? & 2022bcc & 7 & Faint host. \\
\textbf{ZTF20acpgnmw} & 2020afev & TDE? & 2022ccz & 7 & Blue. Nuclear. FLEET=63\% TDE. \\
ZTF18acbvtjm & \textit{2018mfq} & SN? & 2022cmd & 3 & \nodata \\
ZTF21aabvpub & \textit{2021ahwb} & SN~Ia? & 2022csq & 3 & \nodata \\
ZTF18abuluko & 2018mgj & SN~Ia? & " & 4 & No decline. \\
\hline
\multicolumn{6}{c}{\textit{Reference Class: SN~II (243 objects)}} \\
\textbf{ZTF18ablqjws} & \textit{2018mga} & SN~II/IIn? & 2021lmp & 7 & Well sampled. Tagged by AD model. \\
ZTF20acyybza & 2021ahwf & SN~II? & 2018grp & 5 & Well sampled. Faint host. \\
\textbf{ZTF20acbiwfi} & \textit{2020afew} & SLSN? & 2019cob & 4 & FLEET=44\% SLSN-II. \\
\textbf{ZTF20aczhbki} & 2020afex & SLSN? & " & 5 & Host-$z$=0.573. Peak $M_{abs}\sim-23.1$~mag. \\
ZTF20acyzxse & 2020afey & SN~Ia? & " & 7 & \nodata \\
ZTF18abkhqww & 2018mgk & SN~II/IIP? & 2020tab & 2 & Well sampled. Faint host. \\
ZTF18acandwv & 2018mgl & SN~Ia? & 2020tzs & 1 & \nodata \\
ZTF18aczerlj & \textit{2018mfs} & SN~II/IIP? & 2020vcp & 5 & Well sampled. Host-$z$=0.066. Peak $M_{abs}\sim-17.7$~mag. \\
\textbf{ZTF21aaiapis} & 2021ahwg & SN~Ib/c? & 2021adcw & 4 & Tagged by AD model. \\
ZTF18aaxckpt & 2018mgm & SN? & 2020yae & 7 & \nodata \\
ZTF18aczerlj & \textit{2018mfs} & SN~II/IIP? & 2020acjg & 3 & Well sampled. Host-$z$=0.066. Peak $M_{abs}\sim-17.7$~mag. \\
ZTF20acyydhn & 2020afez & SN~II? & 2021crx & 3 & Host-$z$=0.021. Peak $M_{abs}\sim-15.3$~mag. \\
\textbf{ZTF21abasbpe} & 2021ahwh & SN~IIn/SLSN? & 2021uuz & 4 & Faint host. No decline. FLEET=61\% SLSN-II. \\
ZTF18abmofew & \textit{2018mfo} & SN~Ia? & 2021yky & 3 & \nodata \\
ZTF21aajtfas & \textit{2021ahvx} & SN~Ia? & 2021aaev & 8 & Well sampled. \\
\hline
\multicolumn{6}{c}{\textit{Reference Class: SN~IIn (59 objects)}} \\
\textbf{ZTF18ablqjws} & \textit{2018mga} & SN~II/IIn? & 2021lft & 1 & Well sampled. Tagged by AD model. \\
ZTF19aasalud & 2019aatd & SN~Ia? & 2021acnp & 6 & \nodata \\
\hline
\multicolumn{6}{c}{\textit{Reference Class: SN~Ia-91T-like (37 objects)}} \\
ZTF21aalvdng & 2021ahwi & SN~Ia? & 2022cvt & 6 & No decline. \\
ZTF21aamfdzo & \textit{2021ahwd} & SN~Ia? & 2020veg & 7 & \nodata \\
ZTF18abokvkt & \textit{2018mfw} & SN~Ia? & 2020wze & 8 & Faint host. \\
\textbf{ZTF20aclghmy} & \textit{2020affa} & SN~IIn? & 2021hj & 5 & SALT3 $x_{1}=+3.00$. Host-$z$=0.136. Peak $M_{abs}\sim-19.4$~mag. \\
\hline
\multicolumn{6}{c}{\textit{Reference Class: SN~Ib (25 objects)}} \\
ZTF18acbvijf & \textit{2018mgh} & SN~Ia? & 2021lax & 5 & \nodata \\
\textbf{ZTF18abwhsnx} & \textit{2018mgd} & SN~Ib/c? & 2021riw & 6 & Tagged by AD model. \\
ZTF18acrdqvy & 2018mgn & SN~II? & 2021aghp & 3 & Host-$z$=0.049. Peak $M_{abs}\sim-17.0$~mag. \\
ZTF19aabblsx & 2019aate & SN~Ia? & " & 7 & \nodata \\
\hline
\multicolumn{6}{c}{\textit{Reference Class: SN~IIb (21 objects)}} \\
ZTF18aaiuynw & 2018mgo & SN~Ia? & 2018hqu & 3 & \nodata \\
ZTF18acenyrb & 2018mgp & SN~I? & 2020abkp & 4 & \nodata \\
\hline
\multicolumn{6}{c}{\textit{Reference Class: SN~Ic (21 objects)}} \\
ZTF21abvatnb & 2021ahwj & SN~Ia? & 2021lei & 3 & \nodata \\
ZTF21aagyuvz & 2021ahwk & SN~I? & " & 8 & \nodata \\
ZTF19aafmxxd & 2019aatf & SN~Ia? & 2021acwh & 2 & Host-$z$=0.095. Peak $M_{abs}\sim-19.1$~mag. \\
ZTF21aajtfas & \textit{2021ahvx} & SN~Ia? & 2021adgu & 2 & Well sampled. \\
\hline
\end{tabular}
\end{minipage} \hfill
\newpage
\begin{minipage}{\textwidth}
\footnotesize
\centering
\hspace*{-10mm}
\vspace*{-1mm}
\begin{tabular}{cccccc}

\hline
ZTF ID & IAU Name & Possible Class. & Reference SNe & ANN & Remarks \\
\hline
\multicolumn{6}{c}{\textit{Reference Class: TDE (20 objects)}} \\
ZTF20acqyjih & 2020affb & SN~Ia? & 2021lo & 7 & Nuclear. \\
\textbf{ZTF21aasdcgt} & 2021ahwl & TDE? & 2021gje & 6 & FLEET=60\% TDE. Nuclear, blue. Tagged by AD model. \\
ZTF21acbgyai & 2021ahwm & SN~Ia? & 2022rz & 2 & Nuclear. \\
\hline
\multicolumn{6}{c}{\textit{Reference Class: SN~IIP (14 objects)}} \\
ZTF18absljwl & \textit{2018mgf} & SN~II? & 2020tet & 7 & No rise. \\
\hline
\multicolumn{6}{c}{\textit{Reference Class: SLSN-II (14 objects)}} \\
ZTF18abuioue & \textit{2018mgb} & SN~Ia? & 2020vws & 7 & No visible host. \\
ZTF21aajtfas & \textit{2021ahvx} & SN~Ia? & 2021fmu & 4 & Well sampled. \\
ZTF20acllkua & 2020affc & SN~Ia? & 2022akb & 1 & \nodata \\
ZTF18aajmlnp & 2019aatg & SN~Ia? & " & 1 & Host-$z$=0.117. Peak $M_{abs}\sim-19.1$~mag.\\
\hline
\multicolumn{6}{c}{\textit{Reference Class: SN~Ic-BL (14 objects)}} \\
ZTF21aalimtb & 2021ahwn & SN~I? & 2021otc & 3 & \nodata. \\
\hline
\multicolumn{6}{c}{\textit{Reference Class: SLSN-I (11 objects)}} \\
ZTF18abuioue & \textit{2018mgb} & SN~Ia? & 2020xga & 4 & No visible host. \\
ZTF20acteioa & 2020affd & SN~Ia? & 2021rwz & 3 & Faint host. \\
ZTF18abwfhqb & 2018mgq & SN~II? & 2021ybf & 5 & Faint host. \\
ZTF19aaapmth & 2018mgr & SN~Ia? & " & 6 & Faint host. \\
ZTF18abnzney & \textit{2018mfm} & SN~Ia? & 2021ybf & 7 & \nodata \\
\hline
\multicolumn{6}{c}{\textit{Reference Class: SN~Ia-pec (10 objects)}} \\
ZTF19aaptarf & 2019aath & SN~Ia? & 2021wwu & 5 & \nodata \\
ZTF18acvgehz & 2018mgs & SN~Ia? & 2022bbt & 6 & Host-$z$=0.120. Peak $M_{abs}\sim-19.4$~mag. \\
\hline
\multicolumn{6}{c}{\textit{Reference Class: AGN (7 objects)}} \\
ZTF18abmofew & \textit{2018mfo} & SN~Ia? & 2021swi & 4 & Well sampled. \\
\hline
\multicolumn{6}{c}{\textit{Reference Class: SN~Ibn (7 objects)}} \\
ZTF18acbwdym & \textit{2018mfu} & SN~Ia? & 2021bbv & 2 & Host-$z$=0.091. Peak $M_{abs}\textgreater-19.0$~mag. \\
\textbf{ZTF20acbiwfi} & \textit{2020afew} & SLSN? & " & 7 & FLEET=44\% SLSN-II. \\
\hline
\multicolumn{6}{c}{\textit{Reference Class: SN (5 objects)}} \\
\textbf{ZTF18acueeoo} & 2018mgt & SN~II? & 2020uts & 6 & Host-$z$=0.085. Peak $M_{abs}\sim-18.2$~mag. Tagged by AD model.\\
\hline
\multicolumn{6}{c}{\textit{Reference Class: SN~IIn-pec (1 object)}} \\
\textbf{ZTF20aclghmy} & \textit{2020affa} & SN~IIn? & 2021vlu & 5 &  SALT3 $x_{1}=+3.00$. Host-$z$=0.136. Peak $M_{abs}\sim-19.4$~mag.  \\
\hline
\end{tabular}
\end{minipage} \hfill

\newpage

\hspace*{-10mm}
\begin{minipage}{\textwidth}
\footnotesize
\centering \captionof{table}{\textbf{241 Previously Undiscovered SNe Found by Exhaustive Search}}
\vspace*{-1mm}
\label{tab:remaining_no_TNS}
\begin{tabular}{cccc}
\toprule
ZTF ID & IAU Name & Possible Class. & Remarks \\ 
\hline 
\hline
ZTF18aabteyx & 2018mgu & SN? & SN in template thumbnail. \\
\textbf{ZTF18aajgowk} & 2018mgw & SN Ib/c? & Tagged by AD model. \\
ZTF18aajkgtr & 2018mgx & SN Ia? & \nodata \\
ZTF18aamdjfh & 2018mgv & SN? & SN in template thumbnail. \\
ZTF18aaqzdge & 2018mgy & SN? & No rise. \\
ZTF18aathofv & 2018mgz & SN Ia? & No rise. \\
ZTF18aawlezq\footnote{This object is also listed as ZTF18adjzsqt.} & 2018mha & SN II? & \nodata \\
ZTF18aawonvg & 2018mhb & SN Ia? & \nodata \\
ZTF18aaxmjit & 2018mhc & SN? & No rise. \\
ZTF18aaxqwnd & 2018mhd & SN II? & \nodata \\
ZTF18aaxzhzf & 2018mhe & SN Ia? & \nodata \\
ZTF18aayyedm & 2018mhf & SN Ia? & \nodata \\
ZTF18abhttmr & 2018mhg & SN? & No visible host. \\
\textbf{ZTF18abiitmq} & 2018mhh & SN II? & Long ($\sim$70~d) rise. Candidate member of long-rising SN~II class. No decline. \\
ZTF18abjicev & 2018mhi & SN II? & Lasts $\sim$200~d. \\
\textbf{ZTF18abkmfaj} & 2018mhj & SN II? & Long ($\sim$50~d) rise. Candidate member of long-rising SN~II class. No decline. \\
ZTF18ablwatx & 2018mhk & SN I? & Edge-on host galaxy. \\
ZTF18abnysyy & 2018mhl & SN Ia? & Well sampled. \\
ZTF18abolrnr & 2018mhm & SN Ia? & \nodata \\
ZTF18abpucqe & 2018mhn & SN? & \nodata \\
ZTF18abrlhnm & 2018mho & SN Ia? & Faint host. \\
ZTF18abrlurr & 2018mhp & SN I? & \nodata \\
ZTF18abrwqti & 2018mhq & SN Ia? & \nodata \\
ZTF18abscetq & 2018mhr & SN Ia? & \nodata \\
ZTF18abscghc & 2018mhs & SN Ia? & \nodata \\
ZTF18absdgly & 2018mht & SN I? & Faint host. \\
ZTF18abshfmb & 2018mhu & SN II? & Faint host. \\
ZTF18abskrcm & 2018mhv & SN Ia? & \nodata \\
ZTF18absljlb & 2018mhw & SN II/IIP? & Faint host. \\
ZTF18absllcn & 2018mhx & SN Ia? & \nodata \\
ZTF18absloog & 2018mhy & SN Ia? & \nodata \\
ZTF18absmsbm & 2018mhz & SN Ia? & Faint host. \\
ZTF18absopva & 2018mia & SN? & \nodata \\
ZTF18absqohp & 2018mib & SN Ia? & No visible host. \\
ZTF18absquza & 2018mic & SN I? & Edge-on host galaxy. \\
ZTF18absrljp & 2018mid & SN Ia? & \nodata \\
ZTF18abtflop & 2018mie & SN Ia? & \nodata \\
ZTF18abtigie & 2018mif & SN? & \nodata \\
ZTF18abtlxae & 2018mig & SN Ia? & \nodata \\
ZTF18abtmnha & 2018mih & SN Ia? & \nodata \\
ZTF18abtnwpa & 2018mii & SN Ia? & \nodata \\
ZTF18abtotsq & 2018mij & SN Ia? & No visible host. \\
ZTF18abtpgms & 2018mik & SN? & \nodata \\
ZTF18abtptey & 2018mil & SN? & \nodata \\
ZTF18abtqceg & 2018mim & SN Ia? & Faint host. \\
ZTF18abtqidt & 2018min & SN~Ia? & \nodata \\
\textbf{ZTF18abuahio} & 2018mio & SN Ia? & No visible host. Tagged by AD model. \\
ZTF18abubjqi & 2018mip & SN~Ia? & \nodata \\
ZTF18abufwxs & 2018miq & SN I? & Faint host. No decline. \\
ZTF18abuvqgo & 2018mir & SN~Ia? & \nodata \\
ZTF18abuxgmk & 2018mis & SN? & \nodata \\
ZTF18abuyomg & 2018mit & SN~Ia? & \nodata \\
\hline
\end{tabular}
\end{minipage} \hfill
\newpage
\begin{minipage}{\textwidth}
\footnotesize
\centering
\hspace*{-10mm}
\vspace*{-1mm}
\begin{tabular}{cccc}

\hline
ZTF ID & IAU Name & Possible Class. & Remarks \\ 
\hline
ZTF18abvctat & 2018miu & SN? & No visible host. \\
ZTF18abvexiz & 2018miv & SN Ia? & Faint host. \\
ZTF18abvrgjc & 2018miw & SN Ia? & Faint host. \\
ZTF18abvywcl & 2018mix & SN Ia? & Faint host. \\
ZTF18abwblyl & 2018miy & SN~Ia? & \nodata \\
ZTF18abwqgsc & 2018miz & SN~Ia? & \nodata \\
ZTF18abwsfdy & 2018mja & SN~II/IIP? & Faint host. \\
ZTF18abxfott & 2018mjb & SN~Ia? & Faint host. \\
ZTF18abxthsc & 2019aati & SN~I? & \nodata \\
\textbf{ZTF18acajloc} & 2018mjc & SN? & Faint host. Tagged by AD model. \\
ZTF18acbvgtj & 2018mjd & SN? & \nodata \\
ZTF18acbvzsi & 2018mje & SN~Ia? & Faint host. \\
ZTF18acbwele & 2018mjf & SN? & No rise. \\
ZTF18acbxrft & 2018mjg & SN~Ia? & \nodata \\
ZTF18acbzvfj & 2018mjh & SN? & Faint host. \\
ZTF18acbzvow & 2018mji & SN? & \nodata \\
ZTF18acchzlc & 2018mjj & SN? & Faint host. \\
ZTF18accjwbc & 2018mjk & SN? & Edge-on host galaxy. \\
ZTF18accjyim & 2018mjl & SN? & Faint host. \\
ZTF18acckcfr & 2018mjm & SN? & \nodata \\
ZTF18accndre & 2018mjn & SN? & Faint host. \\
ZTF18accvkpt & 2018mjo & SN~Ia? & \nodata \\
ZTF18accwild & 2018mjp & SN? & Faint host. \\
ZTF18acegbeo & 2018mjq & SN~Ia? & Edge-on host galaxy. \\
ZTF18acehtvv & 2018mjr & SN? & \nodata \\
ZTF18aceitqw & 2018mjs & SN? & Faint host. \\
\textbf{ZTF18acevgyz} & 2018mjt & SN~I? & Tagged by AD model. \\
ZTF18achdign & 2018mju & SN? & \nodata \\
ZTF18acmyfyu & 2018mjv & SN~Ia? & \nodata \\
ZTF18acmymdx & 2018mjw & SN? & \nodata \\
ZTF18acnmicp & 2018mjx & SN? & \nodata \\
ZTF18acnneyt & 2018mjy & SN~II/IIP? & Faint host. \\
ZTF18acnnfqr & 2018mjz & SN~Ia? & Host-$z$=0.171. Peak $M_{abs}\sim-19.4$~mag.\\
ZTF18acqzakw & 2018mka & SN~Ia? & Faint host. \\
ZTF18acurkik & 2018mkb & CC~SN? & SALT3 $x1=+3.00$. Host-$z$=0.068. Peak $M_{abs}\sim-17.9$~mag.\\
\textbf{ZTF18acvgmpx} & 2018mkc & SN~II? & Host-$z$=0.069. Peak $M_{abs}\sim-17.6$~mag. In ring galaxy? \\
\textbf{ZTF18acvwkvc} & 2018mkd & TDE? & FLEET=40\% TDE, 27\% SLSN-II. \\
ZTF18acwtrfe & 2018mke & SN? & \nodata \\
ZTF18acwworr & 2018mkf & SN~II? & Host-$z$=0.092. Peak $M_{abs}\sim-18.3$~mag.\\
ZTF18acwyauj & 2018mkg & SN? & \nodata \\
ZTF18acyxxen & 2018mkh & SN~Ia? & \nodata \\
ZTF18aczzvtw & 2018mki & SN? & Faint host. \\
ZTF18aczzwjf & 2018mkj & SN? & Faint host. \\
ZTF18adaksjk & 2018mkk & SN? & Faint host. \\
ZTF18adazgwk & 2018mkl & SN? & \nodata \\
ZTF18adcbkiq & 2018mkm & SN? & \nodata \\
ZTF19aaaefkt & 2018mkn & SN~Ia? & \nodata \\
ZTF19aaaeuxz & 2019aatj & SN~I? & \nodata \\
ZTF19aaafmjk & 2019aatk & SN~II? & Host-$z$=0.062. Peak $M_{abs}\sim-17.6$~mag.\\
ZTF19aaafotc & 2019aatl & SN~Ia? & Host-$z$=0.137. Peak $M_{abs}\sim-19.6$~mag. \\
ZTF19aaajbre & 2018mko & SN~Ia? & \nodata \\
ZTF19aaapdum & 2019aatm & SN~Ia? & Host-$z$=0.171. Peak $M_{abs}\textgreater-19.6$~mag.\\
ZTF19aacwiqw & 2019aatn & SN~Ia? & \nodata \\
ZTF19aacyrdk & 2019aato & SN~Ia? & Faint host. \\
ZTF19aadgceq & 2019aatp & SN~Ia? & \nodata \\
\hline
\end{tabular}
\end{minipage} \hfill
\newpage
\begin{minipage}{\textwidth}
\footnotesize
\centering
\hspace*{-10mm}
\vspace*{-1mm}
\begin{tabular}{cccc}

\hline
ZTF ID & IAU Name & Possible Class. & Remarks \\ 
\hline
\textbf{ZTF19aavoqbe} & 2019aatq & SN~Ia-91bg-like? & Host-$z$=0.13. SALT3 $c=+0.3$. Peak $M_{abs}\sim-18.8$~mag. \\
ZTF19abiinmg & 2019aatr & SN~Ia? & \nodata \\
\textbf{ZTF19abljudj} & 2019aats & SN~II? & In elliptical galaxy? \\
ZTF20abribtl & 2020affe & SN? & \nodata \\
ZTF20abztvjb & 2020afff & SN~II/IIP? & Red. Lasted $\sim$~80~d. Faint host. \\
\textbf{ZTF20acbptqx} & 2020affg & SN~IIn? & Faint host. FLEET=52\% SN~II, 31\% SLSN-II. \\
ZTF20acfamrq & 2020affh & SN~II? & \nodata \\
ZTF20achofuu & 2021ahwo & SN~Ia? & \nodata \\
ZTF20acipeqz & 2020affi & SN~II? & \nodata \\
ZTF20ackjgbt & 2020affj & SN~I? & Faint host. \\
ZTF20acmsesb & 2020affk & SN~II? & \nodata \\
ZTF20acnbhsq & 2020affl & SN~Ia? & \nodata \\
ZTF20acnexgl & 2020affm & SN~Ia? & Host-$z$=0.118. Peak $M_{abs}\sim-19.2$~mag. \\
\textbf{ZTF20acngnvf} & 2020affn & SN~IIn? & Faint host. Lasted 125~d. FLEET=70\% SN~II. \\
ZTF20acnviuc & 2020affo & SN~Ia? & \nodata \\
ZTF20acnvoql & 2020affp & SN~Ia? & \nodata \\
ZTF20acnzpsj & 2020affq & SN~Ia? & \nodata \\
ZTF20acnztaa & 2020affr & SN~II/IIn? & Faint host. \\
ZTF20acnzzlo & 2020affs & SN~Ia? & \nodata \\
ZTF20acotzgs & 2020afft & SN~Ia? & Faint host. \\
ZTF20acpggpe & 2020affu & SN~Ia? & \nodata \\
ZTF20acpskkx & 2020affv & SN? & \nodata \\
ZTF20acpxtkj & 2020affw & SN~Ia? & \nodata \\
\textbf{ZTF20acpzjsk} & 2020affx & TDE? & Blue. Nuclear. FLEET=40\% SN II. \\
ZTF20acrktth & 2020affy & SN~Ia? & \nodata \\
\textbf{ZTF20acrssoi} & 2020affz & SN~IIn? & Faint host. FLEET=74\% SN II. \\
ZTF20acsfdpi & 2020afga & SN? & \nodata \\
\textbf{ZTF20actkqax} & 2020afgb & SN~IIn/SLSN? & Faint host. FLEET=40\% SN II, 36\% SLSN-II. Tagged by AD model. \\
ZTF20acuhlsk & 2020afgc & SN~Ia? & \nodata \\
ZTF20acusjnl & 2020afgd & SN~Ia? & Faint host. \\
\textbf{ZTF20acxyrkk} & 2020afge & SN~IIn/SLSN? & Faint host. FLEET=50\% SN II. Tagged by AD model. \\
\textbf{ZTF20acyroio} & 2021ahwp & SLSN? & Host-$z$=0.545. Peak $M_{abs}\sim-22.8$~mag. \\
ZTF21aaahiba & 2021ahwq & SN~Ia? & \nodata \\
ZTF21aaaolli & 2021ahwr & SN~Ia? & \nodata \\
ZTF21aaaxnnv & 2021ahws & SN~I? & \nodata \\
ZTF21aaayfre & 2021ahwt & SN~Ia? & \nodata \\
\textbf{ZTF21aabyoxk} & 2021ahwu & TDE/SN~IIn? & Blue. Unclear host. Possible host-$z$=0.14. \\
ZTF21aacuckk & 2021ahwv & SN~I? & Host-$z$=0.094. Peak $M_{abs}\sim-19.0$~mag.\\
ZTF21aadplfw & 2021ahww & SN~Ia? & \nodata \\
ZTF21aadrrbf & 2021ahwx & SN~Ia? & \nodata \\
ZTF21aadruuo & 2021ahwy & SN~II? & Edge-on host galaxy. \\
ZTF21aadsosc & 2021ahwz & SN? & No decline. \\
ZTF21aaekerb & 2021ahxa & SN? & \nodata \\
ZTF21aaewjmq & 2021ahxb & SN~II? & \nodata \\
ZTF21aagkynz & 2021ahxc & SN? & \nodata \\
ZTF21aagtspb & 2021ahxd & SN~I? & \nodata \\
ZTF21aagywat & 2021ahxe & SN~Ia? & Faint host. \\
ZTF21aahhegg & 2021ahxf & SN~Ia? & \nodata \\
\textbf{ZTF21aaiahsu} & 2021ahxg & SN~Ia-91bg-like? & Host-$z$=0.088. SALT3 $c=+0.3$. Peak $M_{abs}\sim-18.8$~mag. \\
ZTF21aaiaqvc & 2021ahxh & SN~Ia? & \nodata \\
ZTF21aaiqdbp & 2021ahxi & SN~Ia? & \nodata \\
ZTF21aaiqdkm & 2021ahxj & SN~Ia? & \nodata \\
ZTF21aaiqifj & 2021ahxk & SN~Ia? & No visible host. \\
ZTF21aaitqpu & 2021ahxl & SN~I? & Faint host. \\
ZTF21aakitay & 2021ahxm & SN~Ia? & \nodata \\
\hline
\end{tabular}
\end{minipage} \hfill
\newpage
\begin{minipage}{\textwidth}
\footnotesize
\centering
\hspace*{-10mm}
\vspace*{-1mm}
\begin{tabular}{cccc}

\hline
ZTF ID & IAU Name & Possible Class. & Remarks \\ 
\hline
\textbf{ZTF21aakjxhg} & 2021ahxn & SN~Ia? & Ring host galaxy? \\
ZTF21aalgiex & 2021ahxo & SN~Ia? & \nodata \\
\textbf{ZTF21aalimxp} & 2021ahxp & SN~Ib/c? & Tagged by AD model. \\
ZTF21aantmww & 2021ahxq & SN~II/IIP? & \nodata \\
ZTF21aaoijsw & 2021ahxr & SN~Ia? & \nodata \\
ZTF21aapjzyl & 2021ahxs & SN~Ia? & \nodata \\
ZTF21aaqgmfz & 2021ahxt & SN~Ia? & \nodata \\
ZTF21aarbdjl & 2021ahxu & SN~Ia? & \nodata \\
ZTF21aardvtr & 2021ahxv & SN~Ia? & \nodata \\
ZTF21aarffuz & 2021ahxw & SN~I? & \nodata \\
ZTF21aarhnil & 2021ahxx & SN~I? & \nodata \\
ZTF21aarnxig & 2021ahxy & SN~Ia? & Faint host. \\
ZTF21aarspet & 2021ahxz & SN~Ia? & \nodata \\
ZTF21aarvysz & 2021ahya & SN? & No visible host. \\
ZTF21aasjilo & 2021ahyb & SN? & \nodata \\
ZTF21aasjlxg & 2021ahyc & SN~Ia? & \nodata \\
ZTF21aasttsm & 2021ahyd & SN~Ia? & \nodata \\
ZTF21aatbbjt & 2021ahye & SN~Ia? & \nodata \\
ZTF21aathago & 2021ahyf & SN~II/IIP? & Faint host. \\
ZTF21aatwoxq & 2021ahyg & SN~I? & \nodata \\
ZTF21aavdtcy & 2021ahyh & SN~I? & \nodata \\
ZTF21aawlnxl & 2021ahyi & SN~Ia? & \nodata \\
\textbf{ZTF21aawzmne} & 2021ahyj & SN~Ia? & Blue. Off-nuclear. SALT3 $x=+3.00, c=-0.27$. \\
ZTF21aaxslhj & 2021ahyk & SN? & \nodata \\
ZTF21aaxsqht & 2021ahyl & SN~Ia? & \nodata \\
ZTF21aaxszmx & 2021ahym & SN~Ia? & Host-$z$=0.12. Peak $M_{abs}\sim-19.0$~mag. \\
\textbf{ZTF21aayebcv} & 2021ahyn & SN~Ia? & Tagged by AD model. \\
ZTF21aayebwx & 2021ahyo & SN~Ia? & Host-$z$=0.148. Peak $M_{abs}\sim-19.7$~mag. \\
ZTF21aayfqtb & 2021ahyp & SN~Ia? & \nodata \\
ZTF21aaynwgp & 2021ahyq & SN~Ia? & \nodata \\
ZTF21aayotul & 2021ahyr & SN~Ia? & \nodata \\
ZTF21aazqpgb & 2021ahys & SN~Ia? & \nodata \\
ZTF21aazzciu & 2021ahyt & SN? & \nodata \\
ZTF21abamprv & 2021ahyu & SN~II? & \nodata \\
ZTF21abbwxbf & 2021ahyv & SN~Ia? & \nodata \\
ZTF21abcnhrd & 2021ahyw & SN~Ia? & \nodata \\
ZTF21abcopnn & 2021ahyx & SN~Ia? & \nodata \\
ZTF21abcothc & 2021ahyy & SN~Ia? & \nodata \\
ZTF21abcsifi & 2021ahyz & SN? & Faint host. \\
ZTF21abcskis & 2021ahza & SN? & \nodata \\
ZTF21abcsqmp & 2021ahzb & SN? & \nodata \\
ZTF21abcudam & 2021ahzc & SN~Ia? & \nodata \\
ZTF21abejeyz & 2021ahzd & SN? & No decline. \\
ZTF21abenlno & 2021ahze & SN? & Faint host. \\
ZTF21abezpqc & 2021ahzf & SN~II/IIP? & \nodata \\
ZTF21abfreru & 2021ahzg & SN~Ia? & \nodata \\
ZTF21abfspnh & 2021ahzh & SN~Ia? & \nodata \\
ZTF21abhsttv & 2021ahzi & SN~Ia? & Faint host. \\
ZTF21abhzape & 2021ahzj & SN~Ia? & Host-$z$=0.145. Peak $M_{abs}\sim-19.4$~mag. \\
ZTF21abidgad & 2021ahzk & SN~Ia? & \nodata \\
\textbf{ZTF21abigfjv} & 2021ahzl & SN? & Faint host. Tagged by AD model. \\
ZTF21abjkdli & 2021ahzm & SN~Ia? & \nodata \\
ZTF21abjqcmb & 2021ahzn & SN~Ia? & Host-$z$=0.166. Peak $M_{abs}\sim-19.6$~mag. \\
ZTF21abjtxur & 2021ahzo & SN? & No decline. \\
ZTF21abkapuy & 2021ahzp & SN~II? & \nodata \\
\hline
\end{tabular}
\end{minipage} \hfill
\newpage
\begin{minipage}{\textwidth}
\footnotesize
\centering
\hspace*{-10mm}
\vspace*{-1mm}
\begin{tabular}{cccc}

\hline
ZTF ID & IAU Name & Possible Class. & Remarks \\ 
\hline
ZTF21abowuoq & 2021ahzq & SN~Ia? & Faint host. \\
ZTF21abtoqij & 2021ahzr & SN~I? & \nodata \\
ZTF21abutvjt & 2021ahzs & SN~Ia? & \nodata \\
ZTF21abvqzuu & 2021ahzt & SN~II? & \nodata \\
ZTF21abvtzbv & 2021ahzu & SN~Ia? & \nodata \\
ZTF21abvubre & 2021ahzv & SN~II? & Faint host. \\
ZTF21abvufbg & 2021ahzw & SN~Ia? & Faint host. \\
ZTF21abwulxy & 2021ahzx & SN~II? & \nodata \\
ZTF21abwymva & 2021ahzy & SN? & Faint host. \\
ZTF21abxabce & 2021ahzz & SN~Ia? & Faint host. \\
ZTF21abxayqz & 2021aiaa & SN~II? & \nodata \\
ZTF21abxcqfx & 2021aiab & SN~Ia? & Faint host. \\
ZTF21abxjdtn & 2021aiac & SN~Ia? & Faint host. \\
ZTF21abxkqry & 2021aiad & SN~Ia? & No visible host. \\
ZTF21abxoxhb & 2021aiae & SN~Ia? & \nodata \\
ZTF21abxyydw & 2021aiaf & SN~II/IIP? & \nodata \\
ZTF21abxzzel & 2021aiag & SN~Ia? & Faint host. \\
ZTF21abyaetw & 2021aiah & SN~Ia? & Faint host. \\
ZTF21acblsec & 2021aiai & SN~Ia? & \nodata \\
ZTF21aceylki & 2021aiaj & SN~Ia? & \nodata \\
ZTF21acguaop & 2021aiak & SN~Ia? & Edge-on host. Host-$z$=0.096. Peak $M_{abs}\textgreater-18.5$~mag. No rise. \\
ZTF21achsdxh & 2021aial & SN~Ia? & \nodata \\
ZTF21acjotbc & 2021aiam & SN~Ia? & \nodata \\
ZTF21acjouhg & 2021aian & SN~I? & \nodata \\
\hline
\end{tabular}
\end{minipage} \hfill





\end{document}